\documentclass[onecolumn,txfonts]{aa}
\usepackage{graphicx}
\usepackage{txfonts}
\begin{document}
\title{Transport and mixing in the radiation zones of rotating stars: II-Axisymmetric magnetic field}
\titlerunning{Transport and mixing in  rotating stars: II-Axisymmetric field}

  \author{S. Mathis         
	 \inst{1}          
	 \and              
	 J.-P. Zahn          
	 \inst{1}
	  }

   \offprints{J.-P. Zahn}

   \institute{LUTH, Observatoire de Paris, F-92195 Meudon\\
              \email{stephane.mathis@obspm.fr; jean-paul.zahn@obspm.fr}     
             }

\date{Received January 5; accepted May 11, 2005}

\abstract{
The purpose of this paper is to improve the modeling of the  mixing of chemical elements that occurs in stellar radiation zones.  In addition to the classical rotational mixing considered in our previous paper, which results of the combined action of the thermally-driven meridional circulation and of the turbulence generated by the shear of differential rotation,  we include here  the effect of an axisymmetric magnetic field in a self-consistent way. We treat the advection of the field by the meridional circulation, its Ohmic diffusion, and the production of its toroidal component through the shear of differential rotation. The Lorentz force is assumed not to exceed the centrifugal force; it acts on the baroclinic balance and therefore on the meridional flow, and it has a strong impact on the transport of angular momentum. All variables and governing equations are expanded in spherical or spherical vectorial functions, to arbitrary order: this yields a system of partial differential equations in time and in the radial coordinate, which is ready to be implemented in a stellar structure code. 

}

\maketitle

\section{Motivation}

The need for improved stellar structure models, which go beyond the so-called standard model, is now well recognized. The radiation zones can no longer be treated as motionless regions, where no mixing occurs, but one has to include physical processes which lead to the transport of matter and angular momentum. So far the focus has been on the effect of rotation, which generates a thermally driven meridional circulation; by advecting angular momentum, that circulation renders the rotation non-uniform and prone to shear turbulence. This rotational mixing has been formulated by Zahn (1992) and Maeder \& Zahn (1998), assuming that the angular velocity varies much less in the horizontal than in the vertical direction (shellular rotation). For rapidly rotating stars, models built along these lines are in much better agreement than standard models; this was first shown by Talon et al. (1997), and confirmed by Maeder and Meynet (2000), Maeder and Meynet (2001), who in addition refined the description for the mass loss by taking into account the latitudinal variation of the wind.

However, these models fail to correctly reproduce the flat rotation profile observed in the solar radiative interior (Matias \& Zahn 1997). In solar-type stars, at least once they have been spun down through their magnetized wind, other physical processes are therefore responsible for the transport of angular momentum, and the most plausible candidates are now being investigated. One is the transport by internal gravity waves emitted at the base of the convection zones, associated with turbulence which smoothes the differential rotation; this mechanism was  described by Kumar et al. (1999), Talon et al. (2002) and applied by Talon and Charbonnel (2003, 2004) to stars of various types.
The other possibility is magnetic torquing, as advocated already by Mestel (1953): he showed that even a very modest field could enforce rigid rotation. In the present paper we shall address the latter effect, and examine how rotational mixing is affected by an axisymmetric field.

Moss (1974) was the first to map the thermally-driven meridional circulation in a star containing an axisymmetric poloidal field; when rotation was added, it was assumed constant and uniform. Various generalizations of this work followed, by Mestel and Moss (1977) and by Moss (1977, 1984, 1985, 1987). The {\it a priori} assumption of rigid rotation was relaxed by Tassoul and Tassoul (1986),  who claimed that no magnetic field could render the rotation nearly uniform in presence of meridional circulation; however their treatment failed to correctly represent the generation of toroidal field through the shearing of  poloidal field, which provides the major feed-back in this problem. This was pointed out by Mestel et al.  (1988), who carried out numerical simulations that illustrated how a weak magnetic field could enforce nearly uniform rotation within a few Alfv\'en times; for sake of simplification, both the poloidal field and the meridional circulation were taken as given, and constant in time.

Recently Garaud (2002)  treated the problem in a fully consistent way, and she applied it to the description of the solar tachocline: she let the field adjust through Ohmic diffusion and be advected by the meridional circulation. However, due to numerical limitations, her circulation was driven mainly through Ekman pumping, rather than through thermal diffusion; also, only stationary solutions were constructed.
 
The purpose of the present paper is to model the time-dependent rotational mixing in presence of an axisymmetric magnetic field, thus extending the results obtained by Mathis \& Zahn (2004; from now on referred to as Paper I) in the purely hydrodynamical case. To achieve this goal:
\begin{itemize}
\item we allow for a general axisymmetric magnetic field, including all its effects,
\item we expand all variables in spherical harmonics, formally to unlimited order, 
\item we treat explicitely the departures from shellular rotation, in the linear approximation, in order to capture the tachocline(s),
\item {we filter out the short times, and keep only the time derivative in the heat equation and in that describing the transport of angular momentum.}\footnote {All waves are thus filtered out (acoustic, gravity, inertial, Alfv\'en); the equation for the transport of angular momentum (eq. \ref{AM-diff-adv} below) still formally allows for mixed Alfv\'en-inertial waves, but the timestep of the evolutionary calculation (which uses an implicit scheme) will in general exceed the typical wave travel time, and suppress these waves.}
\end{itemize}
In a forthcoming paper, we shall generalize the present work to a non-axisymmetric magnetic field.

\section{Assumptions and expansions}

We focus here our attention to magnetic fields of moderate strength, whose Alfv\'en speed $V_A$ does not exceed the rotation velocity $R \Omega$, but is larger than the meridional circulation velocity $V_M$. Thus the field will not affect much the hydrostatic balance in the meridional plane, but it may play a major role in the transport of angular momentum, by tending to render the angular velocity $\Omega$ constant along the field lines of the poloidal field (Ferraro 1937). We also assume that the field does not vary on a timescale which is shorter than the Alfv\'en time $R/V_A$; thus we shall deal with fields of primordial origin, and exclude possible dynamo fields produced in the adjacent convection zones, whose penetration would be damped anyway by skin effect (Garaud 1999). 

Our treatment is not able to address the question whether the field is stable or not, because it is axisymmetric, and because we do not resolve the Alfv\'en waves.  We will have to rely on our intuition, guided by 3D simulations such as recently performed  by Braithwaite and Spruit (2004), to choose the magnetic configurations we implement in our code. As for instabilities which may lead to a turbulent state, for instance the magneto-rotational instability when the equator rotates slower than the poles (Balbus \& Hawley 1994), their transport properties will have to be accounted for by a suitable parametrization, as was done previously in the hydrodynamical case.

As in Paper I, we assume that the rotating star is only weakly two-dimensional, and this for two reasons. The first is that the rotation rate and the magnetic field, which is taken axisymmetric, are sufficiently moderate to allow the centrifugal and the Lorentz forces to be considered as perturbations compared to gravity. The second reason rests on the less justified hypothesis that the shear instabilities due to the differential rotation give rise to turbulent motions which are strongly anisotropic due to the stable stratification, with much stronger transport in the horizontal directions than in the vertical. In the radiation zones, we expect it to smooth out the horizontal variations of angular velocity and of chemical composition, a property we shall use to discard certain non-linear terms. Let us emphasize also that the influence of the magnetic field on such turbulence is not taken into account, as seems to be allowed by the condition $V_A  < R \Omega$. 
   
Hence, we consider an axisymmetric star, and assume that the horizontal variations of all quantities are small and smooth enough to allow their linearization and their expansion in a modest number of spherical harmonics. As reference surface, we choose either the sphere or the isobar, and write all scalar quantities either as:
\begin{equation}
X(r,\theta)=X_{0}(r)+\sum_{l}\widehat{X}_{l}(r)P_{l}\left(\cos\theta\right)
\end{equation}
or
\begin{equation}
X(P,\theta)=\overline{X}(P)+\sum_{l>0}\widetilde{X}_{l}(P)P_{l}\left(\cos\theta\right),
\end{equation}
where $\tilde{X}_{l}({r})$ and $\widehat{X}_{l}(r)$ are related by (cf. Paper I, \S2)
\begin{equation}
\tilde{X}_{l}({r})=\widehat{X}_{l}(r)-
\left( \frac{{\rm d} X_{0}}  {{\rm d} P_{0}} \right)  \widehat{P}_{l}(r),
\label{xtilde}
\end{equation}
with $r$ being here the mean radius of an isobar.

Likewise, we expand all axisymmetric vector fields using the method outlined by Rieutord (1987):
\begin{equation}
\vec{u}(r,\theta)=\sum_{l=0}^{\infty}\left\{u_{0}^{l}(r)\vec R_{l}^{0}(\theta)+v_{0}^{l}(r)\vec S_{l}^{0}(\theta)+w_{0}^{l}(r)\vec T_{l}^{0}(\theta)\right\}.
\end{equation}
The orthogonal axisymmetric vectorial spherical harmonics $\vec R_{l}^{0}\left(\theta\right)$, $\vec S_{l}^{0}\left(\theta\right)$, $\vec T_{l}^{0}\left(\theta\right)$ are defined by:
\begin{equation}
\vec R_{l}^{0}(\theta)=Y_{l}^{0}(\theta)\,\vec{\widehat{e}}_{r}\hbox{, }\vec S_{l}^{0}(\theta)=\vec\nabla_{\mathcal{S}}Y_{l}^{0}(\theta)\hbox{ and }\vec T_{l}^{0}(\theta)=\vec\nabla_{\mathcal{S}}\wedge\vec R_{l}^{0}(\theta)
\end{equation}
where $Y_{l}^{0}\left(\theta\right)$ is the classical spherical harmonic with $m=0$ (cf. Edmonds 1974) and $\vec\nabla_{\mathcal{S}}$ is the horizontal gradient
\begin{equation}
\vec\nabla_{\mathcal{S}}=\vec{\widehat{e}}_{\theta}\partial_{\theta}+\vec{\widehat{e}}_{\varphi}\frac{1}{\sin\theta}\partial_{\varphi}.
\end{equation}
Their detailed properties are given in Appendix A.

Using (\ref{def}) and (\ref{axi}), $\vec u$ may be written explicitly on the usual vector basis in spherical coordinates:
\begin{eqnarray}
\vec u&=&\sum_{l=0}^{\infty}\mathcal{N}_{l}^{0}\left[u_{0}^{l}\left(r\right)P_{l}\left(\cos\theta\right)\vec{\widehat e}_{r}+v_{0}^{l}\left(r\right)\partial_{\theta}P_{l}\left(\cos\theta\right)\vec{\widehat e}_{\theta}-w_{0}^{l}\left(r\right)\partial_{\theta}P_{l}\left(\cos\theta\right)\vec{\widehat e}_{\varphi}\right]\nonumber\\
&=&\mathcal{N}_{0}^{0}u_{0}^{0}\left(r\right)\vec{\widehat e}_{r}+\sum_{l=1}^{\infty}\mathcal{N}_{l}^{0}\left[u_{0}^{l}\left(r\right)P_{l}\left(\cos\theta\right)\vec{\widehat e}_{r}-v_{0}^{l}\left(r\right)P_{l}^{1}\left(\cos\theta\right)\vec{\widehat e}_{\theta}+w_{0}^{l}\left(r\right)P_{l}^{1}\left(\cos\theta\right)\vec{\widehat e}_{\varphi}\right];
\end{eqnarray}
$\vec{\widehat e}_{r}$, $\vec{\widehat e}_{\theta}$ and $\vec{\widehat e}_{\varphi}$ are the unit-vectors repectively in the $r$, $\theta$ and $\varphi$ directions.

This expansion of the vector fields allows us to separate explicitly the spatial coordinates $(r, \theta)$ in the vectorial partial differential equations which govern the problem: using this decomposition, we cast the problem into partial differential equations in $t$ and $r$ only. This point is very important numerically, because the existing stellar evolution codes in which we have to introduce the transport equations are all 1-dimensional, and also because we need to achieve much higher accuracy in the radial direction than in the horizontal.

\section{Axisymmetric magnetic field}
We start with expansions and equations involving the magnetic field.
   
\subsection{Definitions}
The magnetic field, being divergenceless, is conveniently split in its poloidal and toroidal parts:
\begin{equation}
\vec B\left(r,\theta\right)=\vec B_{P}\left(r,\theta\right)+\vec B_{T}\left(r,\theta\right) \equiv  \vec\nabla\wedge\vec\nabla\wedge\left(\xi_{P}\left(r,\theta\right)\vec{\widehat{e}}_{r}\right)+\vec\nabla\wedge\left(\xi_{T}\left(r,\theta\right)\vec{\widehat{e}}_{r}\right)\label{Bdef},
\end{equation}
where $\xi_{P}$ and $\xi_{T}$ are respectively the poloidal and the toroidal magnetic stream-functions. Using the classical method introduced by Bullard and Gellman (1954) for the spectral treatment of the geodynamo problem (see also James 1973, 1974 and Serebrianaya 1988), we get the following expansion in radial functions and in spherical harmonics which are defined in \S A.1.1.:
\begin{equation}
\xi_{P}\left(r,\theta\right)=\sum_{l=1}^{\infty}\xi_{0}^{l}\left(r\right)Y_{l}^{0}\left(\theta\right)\hbox{ } \quad \hbox{and} \quad \xi_{T}\left(r,\theta\right)=\sum_{l=1}^{\infty}\chi_{0}^{l}\left(r\right)Y_{l}^{0}\left(\theta\right).
\end{equation}
Since $\vec{R}_{l}^{0}\left(\theta\right)=Y_{l}^{0}\left(\theta\right)\vec{\widehat{e}}_{r}$, as we have seen  in the previous section where we have defined $\vec{R}_{l}^{0}\left(\theta\right), \vec{S}_{l}^{0}\left(\theta\right), \vec{T}_{l}^{0}\left(\theta\right)$, the field may be written as 
\begin{equation}
\vec B_P=\sum_{l=1}^{\infty} \vec\nabla\wedge\vec\nabla\wedge\left(\xi_{0}^{l}\vec{R}_{l}^{0}\left(\theta\right)\right)
\quad \hbox{and} \quad 
\vec B_T=\sum_{l=1}^{\infty} \vec\nabla\wedge\left(\chi_{0}^{l}\vec R_{l}^{0}\left(\theta\right)\right).
\label{bpoltor}
\end{equation}
Next, we develop the curl operator as explained in the appendix  (eqs. (\ref{curl}) and (\ref{curl2p})) to obtain the following expansions for the magnetic field:
\begin{equation}
\vec B\left(r,\theta\right)= \vec B_{P}\left(r,\theta\right) + \vec B_{T}\left(r,\theta\right) =
\sum_{l=1}^{\infty}\left\{\left[l(l+1)\frac{\xi_{0}^{l}}{r^2}\right]\vec R_{l}^{0}\left(\theta\right)+\left[\frac{1}{r}\partial_{r}\xi_{0}^{l}\right]\vec S_{l}^{0}\left(\theta\right)\right\}
+ \sum_{l=1}^{\infty}\left\{\left[\frac{\chi_{0}^{l}}{r}\right]\vec T_{l}^{0}\left(\theta\right)\right\}.
\label{expB}
\end{equation}

We proceed likewise for the current density $\vec j$, which in the framework of magnetohydrodynamics  is related with $\vec B$ through the Maxwell-Amp\`ere equation, neglecting the displacement current:
\begin{equation}
\vec j\left(r,\theta\right)=\frac{1}{\mu_{0}}\vec\nabla\wedge\vec B\left(r,\theta\right),
\end{equation}
$\mu_{0}$ being the magnetic permeability of vacuum.\\

Therefore, using expression (\ref{Bdef}) for the field, we get:
\begin{equation}
\vec j=\frac{1}{\mu_{0}}\vec\nabla\wedge\vec B=\frac{1}{\mu_{0}}\left[\vec\nabla\wedge\vec\nabla\wedge\vec\nabla\wedge\left(\xi_{P}\vec{\widehat{e}}_{r}\right)+\vec\nabla\wedge\vec\nabla\wedge\left(\xi_{T}\vec{\widehat{e}}_{r}\right)\right],
\end{equation}
which,  using the relations (\ref{curl2p}) and (\ref{curl3}), we can project again on $\vec R_{l}^{0}\left(\theta\right)$, $\vec S_{l}^{0}\left(\theta\right)$ and $\vec T_{l}^{0}\left(\theta\right)$, and split into its poloidal and toroidal parts:
\begin{equation}
\vec j_{P}\left(r,\theta\right)=\frac{1}{\mu_{0}}\vec\nabla\wedge\vec B_{T}\left(r,\theta\right)=\frac{1}{\mu_{0}}\vec\nabla\wedge\vec\nabla\wedge\left(\xi_{T}\vec{\widehat{e}}_{r}\right)=\frac{1}{\mu_{0}}\sum_{l=1}^{\infty}\left\{\left[l(l+1)\frac{\chi_{0}^{l}}{r^2}\right]\vec R_{l}^{0}\left(\theta\right)+\left[\frac{1}{r}\partial_{r}\chi_{0}^{l}\right]\vec S_{l}^{0}\left(\theta\right)\right\}
\label{jpol}
\end{equation}
\begin{equation}
\vec j_{T}\left(r,\theta\right)=\frac{1}{\mu_{0}}\vec\nabla\wedge\vec B_{P}\left(r,\theta\right)=\frac{1}{\mu_{0}}\vec\nabla\wedge\vec\nabla\wedge\vec\nabla\wedge\left(\xi_{P}\vec{\widehat{e}}_{r}\right)=-\frac{1}{\mu_{0}}\sum_{l=1}^{\infty}\left\{\left[\Delta_{l}\left(\frac{\xi_{0}^{l}}{r}\right)\right]\vec T_{l}^{0}\left(\theta\right)\right\}.
\label{jtor}
\end{equation}
with $\Delta_{l}$ being the laplacian operator
\begin{equation}
\Delta_{l}=\partial_{r,r}+\frac{2}{r}\partial_{r}-\frac{l(l+1)}{r^2}.
\end{equation}

We are now ready to examine the aspects of transport and mixing in the radiation zones of rotating stars that are related to the presence of an axisymmetric magnetic field. To achieve this, we shall proceed in three steps
\begin{itemize}
\item first, we introduce the Lorentz force;
\item next, we derive the transport equation for $\vec B$, which is the classical induction equation;
\item finally, we take into account the energy losses due to Ohmic heating. 
\end{itemize}

\subsection{Lorentz force}
The Lorentz force $\vec{\mathcal F}\!\!_{\mathcal L}$ plays a crucial role in a rotating star, since it acts to render the angular velocity constant on the field lines of the poloidal field $\vec B_{P}$.
In this section, we establish the formal expression of $\vec{\mathcal F}\!\!_{\mathcal L}$ in terms of vectorial spherical harmonics. 

Starting from the definition of the Lorentz force: 
\begin{equation}
\vec{\mathcal{F}}\!\!_{\mathcal{L}}\left(r,\theta\right)=\vec j\wedge\vec B=\left[\frac{1}{\mu_{0}}(\vec\nabla\wedge\vec B)\right]\wedge\vec B,
\end{equation}
we replace $\vec B$ by its expansions (\ref{expB}), and use the algebra ruling the vector product of two general axisymmetric vectors  in Appendix A, whereby we obtain the following 
formal projection:
\begin{eqnarray}
\vec{\mathcal{F}}\!\!_{\mathcal{L}}\left(r,\theta\right)&=&\mathcal{X}_{{\vec {\mathcal F}}_{\mathcal{L}};0}(r)\vec{\widehat{e}}_{r}+\sum_{l=1}^{\infty}\left\{\mathcal{X}_{{\vec {\mathcal F}}_{\mathcal{L}};l}(r)P_{l}(\cos\theta)\vec{\widehat{e}}_{r}+\mathcal{Y}_{{\vec {\mathcal F}}_{\mathcal{L}};l}(r)P_{l}^{1}(\cos\theta)\vec{\widehat{e}}_{\theta}+\mathcal{Z}_{{\vec {\mathcal F}}_{\mathcal{L}};l}(r)P_{l}^{1}(\cos\theta)\vec{\widehat{e}}_{\varphi}\right\}\nonumber\\
&=&\left[\frac{\mathcal{X}_{{\vec {\mathcal F}}_{\mathcal{L}};0}(r)}{\mathcal{N}_{0}^{0}}\right]\vec R_{0}^{0}(\theta)+\sum_{l=1}^{\infty}\left\{\left[\frac{\mathcal{X}_{{\vec {\mathcal F}}_{\mathcal{L}};l}(r)}{\mathcal{N}_{l}^{0}}\right]\vec R_{l}^{0}(\theta)+\left[-\frac{\mathcal{Y}_{{\vec {\mathcal F}}_{\mathcal{L}};l}(r)}{\mathcal{N}_{l}^{0}}\right]\vec S_{l}^{0}(\theta)+\left[\frac{\mathcal{Z}_{{\vec {\mathcal F}}_{\mathcal{L}};l}(r)}{\mathcal{N}_{l}^{0}}\right]\vec T_{l}^{0}(\theta)\right\}  .\nonumber\\
\label{Lorentzexp}
\end{eqnarray}
In Appendix B, we give the explicit expressions for the radial functions $\mathcal{X}_{\vec{\mathcal F}\!\!_{\mathcal L};l}\left(r\right)$, $\mathcal{Y}_{\vec{\mathcal F}\!\!_{\mathcal L};l}\left(r\right)$, $\mathcal{Z}_{\vec{\mathcal F}\!\!_{\mathcal L};l}\left(r\right)$ in terms of the magnetic stream-functions $\xi_{0}^{l}$ and $\chi_{0}^{l}$,   the normalization coefficient $\mathcal{N}_{l}^{0}$ being given in (\ref{norm}). 

As we have done previously for the magnetic field $\vec B$ and for the current $\vec j$, we split the magnetic force into its poloidal and toroidal parts: 
\begin{eqnarray}
\vec{\mathcal{F}}\!\!_{\mathcal{L},P}&=&\vec j_{T}\wedge\vec B_{P}+\vec j_{P}\wedge\vec B_{T}=\mathcal{X}_{{\vec {\mathcal F}}_{\mathcal{L}};0}\vec{\widehat{e}}_{r}+\sum_{l=1}^{\infty}\left\{\mathcal{X}_{{\vec {\mathcal F}}_{\mathcal{L}};l}P_{l}(\cos\theta)\vec{\widehat{e}}_{r}+\mathcal{Y}_{{\vec {\mathcal F}}_{\mathcal{L}};l}P_{l}^{1}(\cos\theta)\vec{\widehat{e}}_{\theta}\right\}\nonumber\\
&=&\left[\frac{\mathcal{X}_{{\vec {\mathcal F}}_{\mathcal{L}};0}}{\mathcal{N}_{0}^{0}}\right]\vec R_{0}^{0}(\theta)+\sum_{l=1}^{\infty}\left\{\left[\frac{\mathcal{X}_{{\vec {\mathcal F}}_{\mathcal{L}};l}}{\mathcal{N}_{l}^{0}}\right]\vec R_{l}^{0}(\theta)+\left[-\frac{\mathcal{Y}_{{\vec {\mathcal F}}_{\mathcal{L}};l}}{\mathcal{N}_{l}^{0}}\right]\vec S_{l}^{0}(\theta)\right\}\nonumber\\
\end{eqnarray}
\begin{equation}
\vec{\mathcal{F}}\!\!_{\mathcal{L},T}=\vec j_{P}\wedge\vec B_{P}=\sum_{l=1}^{\infty}\mathcal{Z}_{{\vec {\mathcal F}}_{\mathcal{L}};l}P_{l}^{1}(\cos\theta)\vec{\widehat{e}}_{\varphi}=\sum_{l=1}^{\infty}\left\{\left[\frac{\mathcal{Z}_{{\vec {\mathcal F}}_{\mathcal{L}};l}}{\mathcal{N}_{l}^{0}}\right]\vec T_{l}^{0}(\theta)\right\} .
\end{equation}
As it was underlined by Mestel et al. (1988) and as we shall see in \S4, \S5 and \S6, this decomposition is very useful: the poloidal Lorentz force $\vec{\mathcal F}\!\!_{\mathcal L,P}$ operates on the hydrostatic balance and thus contributes to the thermal imbalance, while the toroidal component $\vec{\mathcal F}\!\!_{\mathcal L,T}$ acts on the transport of angular momentum through its torque.
    
The expression of $\vec{\mathcal{F}}\!\!_{\mathcal{L}}$ in the case where only the $l\!=\!\left\{1,2,3\right\}$ terms are kept in the expansion of $\vec B$ can be found in Appendix E. One should note that if the expansion of $\vec B$ is terminated at mode $l_{\rm max}=N_{m}$ in (\ref{expB}), then the expansion of $\vec{\mathcal{F}}\!\!_{\mathcal{L}}$ involves terms up to $l_{\rm max}=2N_{m}$, due to selection rules (see Appendix B).

\subsection{Induction equation}
We now turn to  the evolution of the magnetic field in stellar radiation zones, which is governed by the induction equation: 
\begin{equation}
\partial_{t}\vec B=\vec\nabla\wedge\left(\vec V\wedge\vec B\right)-\vec\nabla\wedge\left(||\eta||\otimes\vec\nabla\wedge\vec B\right) ,
\label{induction}
\end{equation}
where we have allowed for an anisotropic eddy diffusivity $||\eta||$.
We recall (cf. Paper I) that the macroscopic velocity field $\vec V$ is the sum of a zonal flow $\vec{\mathcal U}_{\varphi}  = r\sin\theta \,\Omega\left(r,\theta\right) \vec{\widehat{e}}_{\varphi}$ and of a meridional flow $\vec{\mathcal U}\left(r,\theta\right)$:
\begin{equation}
\vec V\left(r,\theta\right)=r\sin\theta \, \Omega\left(r,\theta\right)\vec{\widehat{e}}_{\varphi}+\vec{\mathcal{U}}\left(r,\theta\right).
\label{vsplit1}
\end{equation}
The latter can be split into a spherically symmetric part, which represents the contractions and dilatations of the star during its evolution, plus the thermally-driven circulation
\begin{equation}
\vec{\mathcal{U}}\left(r,\theta\right)=\dot{r}\,\vec{\widehat{e}}_{r}+\vec{\mathcal{U}}_{M}\left(r,\theta\right) ,
\label{vsplit2}
\end{equation}
the meridional flow being expanded in spherical functions:
\begin{equation}
\vec{\mathcal{U}}_{M}=\sum_{l>0}\left[U_{l}(r)P_{l}\left(\cos\theta\right)\widehat{\vec e}_{r}+V_{l}(r)\frac{{\rm d}P_{l}(\cos\theta)}{{\rm d}\theta}\widehat{\vec e}_{\theta}\right].
\label{meridexp}
\end{equation}
The radial functions $U_{l}(r)$ and $V_{l}(r)$ are related by the continuity equation, i.e. $\vec{\nabla} \cdot (\rho\,\vec {\mathcal{U}}_{M})=0$ in the anelastic approximation:
\begin{equation}
V_{l}=\frac{1}{l(l+1)\rho r}{{\rm d} \over {\rm d}r} \left(\rho r^2U_{l}\right) .
\end{equation}
 
We introduce the expanded form of the velocity field in the induction equation (\ref{induction}):
\begin{equation}
\partial_{t}\vec B-\vec\nabla\wedge\left(\dot{r}\, \vec{\widehat{e}}_{r}\wedge\vec B\right)=\vec\nabla\wedge\left[\left(\vec{\mathcal U}_{\varphi}+\vec{\mathcal U}_{M}\right)\wedge\vec B\right]-\vec\nabla\wedge\left(||\eta||\otimes\vec\nabla\wedge\vec B\right) ,
\label{indufinal}
\end{equation}
and as before we project this equation on the vectorial spherical harmonics  $\vec R_{l}^{0}\left(\theta\right)$, $\vec S_{l}^{0}\left(\theta\right)$, $\vec T_{l}^{0}\left(\theta\right)$.
The time derivative on the left-hand side of (\ref{indufinal}) is readily derived (cf. \ref{expB}):
\begin{equation}
\partial_{t}\vec B=\sum_{l=1}^{\infty}\left\{\left[l(l+1)\frac{\partial_{t}\xi_{0}^{l}}{r^2}\right]\vec R_{l}^{0}(\theta)+\left[\frac{1}{r}\partial_{t,r}\xi_{0}^{l}\right]\vec S_{l}^{0}(\theta)+\left[\frac{\partial_{t}\chi_{0}^{l}}{r}\right]\vec T_{l}^{0}(\theta)\right\}.
\label{i1}
\end{equation}
Then, using the algebra related to the $\vec R_{l}^{0}\left(\theta\right)$, $\vec S_{l}^{0}\left(\theta\right)$, $\vec T_{l}^{0}\left(\theta\right)$ and the identity (\ref{curl}), the following expansion of the contractions/dilatations term is obtained:
\begin{equation}
\vec\nabla\wedge\left(\dot{r}\hbox{ }\vec{\widehat e}_{r}\wedge\vec B\right)=-\sum_{l=1}^{\infty}\left\{\left[l(l+1)\frac{\dot{r}}{r^2}\partial_{r}\xi_{0}^{l}\right]\vec R_{l}^{0}\left(\theta\right)+\left[\frac{1}{r}\partial_{r}\left(\dot{r}\partial_{r}\xi_{0}^{l}\right)\right]\vec S_{l}^{0}\left(\theta\right)+\left[\frac{1}{r}\partial_{r}\left(\dot{r}\chi_{0}^{l}\right)\right]\vec T_{l}^{0}\left(\theta\right)\right\}.
\label{i2}
\end{equation}
In the same way, we get for the dissipation term:
\begin{eqnarray}
\vec\nabla\wedge\left(||\eta||\otimes\vec\nabla\wedge\vec B\right)&=&-\sum_{l=1}^{\infty}\left\{\left[\eta_{h}\frac{1}{r}\Delta_{l}\left(l\left(l+1\right)\frac{\xi_{0}^{l}}{r}\right)\right]\vec R_{l}^{0}\left(\theta\right)+\left[\frac{1}{r}\partial_{r}\left(r\eta_{h}\Delta_{l}\left(\frac{\xi_{0}^{l}}{r}\right)\right)\right]\vec S_{l}^{0}\left(\theta\right)\right.\nonumber\\
&+&{\left.\left[\frac{1}{r}\partial_{r}\left(\eta_{h}\partial_{r}\chi_{0}^{l}\right)-\eta_{v}l\left(l+1\right)\frac{\chi_{0}^{l}}{r^3}\right]\vec T_{l}^{0}\left(\theta\right)\right\}},
\label{i3}
\end{eqnarray}
where we recall that $\Delta_{l}$ is the laplacian operator: $\Delta_{l}=\partial_{r,r}+({2}/{r})\partial_{r}-{l(l+1)}/{r^2}$.
 
The task is more difficult with the non-linear expressions describing the stretching and advection of the magnetic field:  $\vec\nabla\wedge\left[\vec{\mathcal U}_{\varphi}\wedge\vec B\right]+\vec\nabla\wedge\left[\vec{\mathcal U}_{M}\wedge\vec B\right]$. The first term corresponds to the generation of toroidal field trough the shearing of the poloidal component by the differential rotation, while the second one represents the advection of the field by the meridional circulation.
     
The first step is to expand ($\vec{\mathcal U}_{\varphi} + \vec{\mathcal U}_{M})$ on the $\vec R_{l}^{0}\left(\theta\right)$, $\vec S_{l}^{0}\left(\theta\right)$ and $\vec T_{l}^{0}\left(\theta\right)$:
\begin{equation}
\vec{\mathcal U}_{\varphi}+\vec{\mathcal U}_{M}=\sum_{l=0}^{\infty}\left\{u_{0}^{l}\left(r\right)\vec R_{l}^{0}\left(\theta\right)+v_{0}^{l}\left(r\right)\vec S_{l}^{0}\left(\theta\right)+w_{0}^{l}\left(r\right)\vec T_{l}^{0}\left(\theta\right)\right\}.
\label{uproject}
\end{equation}
From (\ref{meridexp}), we get immediately for the meridional part:
\begin{equation}
u_{0}^{l}\left(r\right)=\frac{U_{l}\left(r\right)}{\mathcal{N}_{l}^{0}}\hbox{ and }v_{0}^{l}\left(r\right)=\frac{V_{l}\left(r\right)}{\mathcal{N}_{l}^{0}}.
\end{equation}
Next, we expand the rotation law as in Paper I:
\begin{equation}
\Omega\left(r,\theta\right)=\overline{\Omega}\left(r\right)+\widehat\Omega\left(r, \theta\right)=
\overline{\Omega}\left(r\right)+\sum_{l>0}\Omega_{l}\left(r\right)Q_{l}\left(\theta\right)=\sum_{l=0}^{\infty}\Omega_{l}^{*}\left(r\right)P_{l}\left(\cos\theta\right),
\label{Bomexp}
\end{equation}
where $\overline{\Omega}\left(r\right)\!\!=\!\!\int_{0}^{\pi}\Omega\left(r,\theta\right)\sin^{3}\theta{\rm d}\theta/\int_{0}^{\pi}\sin^{3}\theta {\rm d}\theta$ the horizontal functions are given by $Q_{l}\left(\theta\right)\!\!=\!\!P_{l}\left(\cos\theta\right)-I_{l}$, with \\ $I_{l}={\int_{0}^{\pi}P_{l}\left(\cos\theta\right)\sin^{3}\theta \, d\theta}/{\int_{0}^{\pi}\sin^{3}\theta \,d\theta}=\delta_{l,0}-\frac{1}{5}\delta_{l,2}$.
Thus, we get:
\begin{equation}
\left\{
\begin{array}{l@{\quad}l}
\Omega_{0}^{*}\left(r\right)=\overline{\Omega}\left(r\right)-\sum_{l>0}\Omega_{l}\left(r\right)I_{l}=\Omega_{0}\left(r\right)+\frac{1}{5}\Omega_{2}\left(r\right)\\
\Omega_{l}^{*}\left(r\right)=\Omega_{l}\left(r\right)\hbox{ for }l>0.
\end{array}\right.
\end{equation}
Using the identity (\ref{r2}), we obtain the zonal flow $\vec{\mathcal U}_{\varphi}=\sum_{l>0}w_{0}^{l}\left(r\right){\vec T}_{l}^{0}\left(\theta\right)$, where $w_{0}^{l}\left(r\right)$ is given by
\begin{equation}
w_{0}^{l}(r)=r\left[\frac{D_{l-1}^{0}}{\mathcal{N}_{l-1}^{0}}\Omega_{l-1}^{*}(r)-\frac{C_{l+1}^{0}}{\mathcal{N}_{l+1}^{0}}\Omega_{l+1}^{*}(r)\right].
\label{wOmega}
\end{equation}
The expression of the $w_{0}^{l}$ when we keep only the two first terms of the expansion (\ref{Bomexp}), $\overline{\Omega}$ and $\Omega_{2}$ is given in (\ref{InductionAnn}).

It remains to project  the vector product of the two axisymmetric vectors, ($\vec{\mathcal U}_{\varphi}+\vec{\mathcal U}_{M}$) and $\vec B$, on the vector harmonics $\vec R_{l}^{0}\left(\theta\right)$, $\vec S_{l}^{0}\left(\theta\right)$ and $\vec T_{l}^{0}\left(\theta\right)$. This task is accomplished in \S A.2.3.; applying it to the advection term, we get:
\begin{eqnarray}
\left(\vec{\mathcal U}_{\varphi}+\vec{\mathcal U}_{M}\right)\wedge\vec B&=&\sum_{l=1}^{\infty}\left\{\mathcal{X}_{{\bf Ad};l}(r)P_{l}(\cos\theta)\vec{\widehat{e}}_{r}+\mathcal{Y}_{{\bf Ad};l}(r)P_{l}^{1}(\cos\theta)\vec{\widehat{e}}_{\theta}+\mathcal{Z}_{{\bf Ad};l}(r)P_{l}^{1}(\cos\theta)\vec{\widehat{e}}_{\varphi}\right\}\nonumber\\
&=&\sum_{l=1}^{\infty}\left\{\left[\frac{\mathcal{X}_{{\bf Ad};l}(r)}{\mathcal{N}_{l}^{0}}\right]\vec R_{l}^{0}(\theta)+\left[-\frac{\mathcal{Y}_{{\bf Ad};l}(r)}{\mathcal{N}_{l}^{0}}\right]\vec S_{l}^{0}(\theta)+\left[\frac{\mathcal{Z}_{{\bf Ad};l}(r)}{\mathcal{N}_{l}^{0}}\right]\vec T_{l}^{0}(\theta)\right\} , \nonumber\\
\end{eqnarray}
where the radial functions $\mathcal{X}_{{\bf Ad};l}\left(r\right)$, $\mathcal{Y}_{{\bf Ad};l}\left(r\right)$ and $\mathcal{Z}_{{\bf Ad};l}\left(r\right)$, which are explicit functions of $\overline{\Omega}$, $\Omega_{l}$, $U_{l}$, $\xi_{0}^{l}$ and $\chi_{0}^{l}$, are given in Appendix C. Finally, using once again (\ref{curl}), we obtain:  
\begin{eqnarray}
\vec\nabla\wedge\left[\left(\vec{\mathcal U}_{\varphi}+\vec{\mathcal U}_{M}\right)\wedge\vec B\right]&=&\sum_{l=1}^{\infty}\left\{\left[\frac{l(l+1)}{\mathcal{N}_{l}^{0}}\frac{\mathcal{Z}_{{\bf Ad};l}}{r}\right]\vec R_{l}^{0}(\theta)+\left[\frac{1}{\mathcal{N}_{l}^{0}}\frac{1}{r}\partial_{r}\left(r\mathcal{Z}_{{\bf Ad};l}\right)\right]\vec S_{l}^{0}(\theta)\right.\nonumber\\
&+&{\left.\left[\frac{1}{\mathcal{N}_{l}^{0}}\left(\frac{\mathcal{X}_{{\bf Ad};l}}{r}+\frac{1}{r}\partial_{r}\left(r\mathcal{Y}_{{\bf Ad};l}\right)\right)\right]\vec T_{l}^{0}\left(\theta\right)\right\}}.
\label{i4}
\end{eqnarray}

We are now ready to put in their final form the one-dimensional advection/diffusion equations respectively for the poloidal  $\xi_{0}^{l}$ and for the toroidal magnetic stream-functions $\chi_{0}^{l}$, by collecting  the $\vec R_{l}^{0}\left(\theta\right)$ and the $\vec T_{l}^{0}\left(\theta\right)$  components of (\ref{i1}), (\ref{i2}), (\ref{i3}) and (\ref{i4}):
\begin{equation}
\frac{{\rm d}\xi_{0}^{l}}{{\rm d}t}=\frac{1}{\mathcal{N}_{l}^{0}}r\mathcal{Z}_{{\bf Ad};l}+\eta_{h}r\Delta_{l}\left(\frac{\xi_{0}^{l}}{r}\right)
\label{i5}
\end{equation}
\begin{equation}
\frac{{\rm d}\chi_{0}^{l}}{{\rm d}t}+\partial_{r}\left(\dot{r}\right)\chi_{0}^{l}=\frac{1}{\mathcal{N}_{l}^{0}}\left[\mathcal{X}_{{\bf Ad};l}+\partial_{r}\left(r\mathcal{Y}_{{\bf Ad};l}\right)\right]+\left[\partial_{r}\left(\eta_{h}\partial_{r}\chi_{0}^{l}\right)-\eta_{v}l(l+1)\frac{\chi_{0}^{l}}{r^2}\right].
\label{i6}
\end{equation}
Note that these equations involve the Lagrangian time-derivative, which makes them suitable for their implementation in stellar evolution codes.
They are the equivalent in our formalism of the two classical equations for $\vec{B}_{P}\!=\!\vec\nabla\wedge\vec A$, $\vec A\!=\!A\vec{\widehat e}_{\varphi}$ being the potential vector, and $\vec{B}_{T}\!=\!B_{T}\vec{\widehat e}_{\varphi}$:
\begin{equation}
\partial_{t}A+\frac{1}{s}\vec{\mathcal{U}}_{M}\cdot\vec\nabla\left(sA\right)=\eta\left(\nabla^2 A-\frac{A}{s^2}\right) \quad \hbox{and} \quad
\partial_{t}B_{T}+s\vec{\mathcal{U}}_{M}\cdot\vec\nabla\left(\frac{B_{T}}{s}\right)+B_{T}\vec\nabla\cdot\vec{\mathcal U}_{P}=s\vec B_{P}\cdot\vec\nabla\Omega+\eta\left(\nabla^2 B_{T}-\frac{B_{T}}{s^2}\right) ,
\end{equation}
where $s=r\sin\theta$ and $\eta$ is taken isotropic and uniform (Campbell 1997, Mestel 1999).\\

Finally, we note that in the ideal case, in the absence of shear, meridional circulation and Ohmic diffusion, (\ref{i5}) and (\ref{i6})  reduce to:
\begin{equation}
\frac{{\rm d}\xi_{0}^{l}}{{\rm d}t}=0\hbox{ and }\frac{{\rm d}}{{\rm d}t}\left(\frac{\chi_{0}^{l}}{r^2\rho}\right)=0 \quad \hbox{and thus to} \quad \frac{{\rm d}}{{\rm d}t}\left(r^2 B_{r}\right)=0 \hbox{ and } \; \frac{{\rm d}}{{\rm d}t}\left(\frac{B_{\varphi}}{r\rho}\right)=0.
\end{equation}
These equations express  the Lagrangian flux conservation of respectively 
\begin{itemize}
\item the magnetic field through the sphere of radius $r$, as the star expands or contracts,
\item the toroidal field in a homothetic contraction or expansion where the density varies as $r^{-3}$ (cf. Cowling 1957).
\end{itemize}
In Appendix  E.1. we give the explicit expressions of (\ref{i5}) and (\ref{i6}) for the dipole, the quadrupole and the octupole ($l=\left\{1,2,3\right\}$), retaining the lowest order terms in the rotation law: $\Omega(r,\theta)=\overline{\Omega}\left(r\right)+\Omega_2 \left(r\right)Q_{2}\left(\theta\right)$, and the associated meridional circulation. 

\subsection{Ohmic heating}
The last point we shall consider for completeness is  Ohmic heating, which contributes somewhat to the thermal balance/imbalance. In the case of an anisotropic eddy-magnetic diffusivity, its input rate is given by
\begin{equation}
\mathcal{J}\left(r,\theta\right)=\frac{1}{\mu_{0}}\left[||\eta||\otimes\left(\vec\nabla\wedge\vec B\right)\right]\cdot\left(\vec\nabla\wedge\vec B\right) ,
\end{equation} 
which reduces to the classical result $\mathcal{J}={{\vec j}^{2}}/{\sigma}$, with  $\sigma={1}/\left({\mu_{0}\eta}\right)$ being the conductivity, when the diffusivity is isotropic.

The scalar product of two axisymmetric vectors is treated in the \S A.2.3.; applying the result to $\mathcal{J}$, we get:
\begin{equation}
\mathcal J\left(r,\theta\right)=\sum_{l>0}\mathcal{J}_{l}\left(r\right)P_{l}\left(\cos\theta\right),
\end{equation} 
where the radial functions $\mathcal{J}_{l}(r)$, which are functions of the magnetic stream-functions $\xi_{0}^{l}$ and $\chi_{0}^{l}$, are given in Appendix D, and in Appendix E for the special case where only  the $l=\left\{1,2,3\right\}$ modes are kept. Note again here  that if the expansion of $\vec B$ is cut at mode $l_{\rm max}=N_{m}$ in (\ref{expB}), then the expansion of $\mathcal{J}$ involves terms up to $l_{\rm max}=2N_{m}$.
\medskip

We are now ready to introduce the Lorentz force in the equations governing the transport of momentum and of heat.

\section{Transport of angular momentum}
  
We start from the momentum equation
\begin{equation}
\rho\left[ \partial_{t}\vec V+\left( \vec V \cdot \vec \nabla \right) \vec V \right]=-\vec\nabla P -\rho\vec\nabla\phi+\vec\nabla\cdot ||\tau||+\vec{\mathcal{F}}\!\!_{\mathcal{L}} \, ,
\label{NV-vec} 
\end{equation} 
where $\rho$ is the density, $\phi$ the gravitational potential, $||\tau||$ the turbulent stresses and $\vec{\mathcal F}\!\!_{\mathcal L}$ the Lorentz force. We insert in it the expression for the 
macroscopic velocity $\vec V$ given above in (\ref{vsplit1}, \ref{vsplit2}), and take its azimuthal component, which yields an advection/diffusion equation for the angular momentum density:
\begin{equation}
\rho {{\rm d} \over {\rm d}t} \left( r^{2} \sin^{2} \theta \, \Omega \right) + \vec \nabla \cdot \left( \rho r^{2} \sin^{2} \theta \, \Omega \, \vec{\mathcal{U}_M} \right) 
= \frac{\sin^{2}\theta}{r^2} \partial_{r} \left( \rho \nu_{v}r^{4} \partial_{r} \Omega \right) + \frac{1}{\sin\theta} \partial_{\theta} \left( \rho\nu_{h} \sin^{3} \theta \, \partial_{\theta} \Omega \right)+\Gamma_{\vec{\mathcal F}\!\!_{\mathcal L}}\left(r,\theta\right).
\label{AM-diff-adv}
\end{equation} 
Note that here again we have introduced the Lagrangian time derivative, making use of the anelastic continuity equation. This equation is of course identical to that given in Paper I  (eq. 14), except for the magnetic torque $\Gamma_{\vec{\mathcal F}\!\!_{\mathcal L}}\left(r,\theta\right)= r \sin \theta \, \vec e _\varphi \cdot \vec{\mathcal F}\!\!_{\mathcal L}\left(r,\theta\right)$. As in Zahn (1992), we assume that the effect of the turbulent stresses on the large scale flow is adequately described by an anisotropic eddy viscosity, whose components are $\nu_{v}$ and $\nu_{h}$ respectively in the vertical and horizontal directions.  In the absence of circulation, turbulence and magnetic field, we retrieve the Lagrangian conservation of angular momentum.

It remains to project this equation on spherical harmonics;  the expansions of the meridional circulation and the angular velocity were established in Paper I and they have been recalled above in (\ref{meridexp}, \ref{Bomexp}). We  proceed likewise for the magnetic torque:
\begin{equation}
\Gamma_{\vec{\mathcal F}\!\!_{\mathcal L}}\left(r,\theta\right)=r\sin\theta \, \mathcal{F}_{\mathcal{L},\varphi}=r\sin\theta\sum_{l=0}^{\infty}\mathcal{Z}_{\vec{\mathcal{F}}_{\mathcal L};l}(r)P_{l}^{1}\left(\cos\theta\right) ,
\end{equation}
where $\mathcal{F}_{\mathcal{L},\varphi}$ is the azimuthal component of $\vec{\mathcal F}\!\!_{\mathcal L}$ obtained from (\ref{Lorentzexp}). Then, using the following property of the associated Legendre functions:
\begin{equation}
P_{l}^{1}\left(\cos\theta\right)=-\frac{{\rm d}}{{\rm d}\theta}P_{l}\left(\cos\theta\right)=\sin\theta\frac{{\rm d}}{{\rm d}\mu}P_{l}\left(\mu\right)\hbox{ where }\mu=\cos\theta
\end{equation}
and the expression for ${dP_{l}\left(\mu\right)}/{d\mu}$ which is given in (\ref{Ryzhik}), we put $\Gamma_{\vec{\mathcal F}\!\!_{\mathcal L}}$ in its final form:
\begin{equation}
\Gamma_{\vec{\mathcal F}\!\!_{\mathcal L}}\left(r,\theta\right)=\sum_{l=0}^{\infty}\Gamma_{l}\left(r\right)\sin^2\theta \, P_{l}\left(\cos\theta\right) \quad \hbox{where} \quad\Gamma_{l}\left(r\right)=\sum_{k=0}^{\infty}r\mathcal{Z}_{\vec{\mathcal F}\!\!_{\mathcal L};k}\left(r\right)\left[\sum_{j=0}^{E\left[\frac{k-1}{2}\right]}\left(2k-4j-1\right)\delta_{l,k-2j-1}\right],
\label{Gamma2D}
\end{equation}
$E\left[x\right]$ is the integer part of $x$, and $\mathcal{Z}_{\vec{\mathcal F}\!\!_{\mathcal L};k}$ is defined in (\ref{Lorentzexp}) and spelled out in explicit form in Appendix B.
  
\subsection{Evolution equation for the angular mean angular velocity}
  
Taking the horizontal average of equation (\ref{AM-diff-adv}) over an isobar and using the assumption that $\overline{\Omega}(r) \gg \Omega_{l}(r)$, we obtain the following vertical advection/diffusion equation for the mean angular velocity $\overline{\Omega}$:  
\begin{equation}
\rho {{\rm d} \over {\rm d}t} (r^2\overline{\Omega})=\frac{1}{5r^2}\partial_{r}\left(\rho r^4\overline{\Omega}U_{2}\right)+\frac{1}{r^2}\partial_{r}\left(\rho\nu_{v}r^4\partial_{r}\overline{\Omega}\right)+\overline{\Gamma}_{\vec{\mathcal{F}}_{\mathcal L}}\left(r\right),
\label{mean-AM}
\end{equation}

\begin{equation}
\hbox{where the mean Lorentz torque is} \quad  \overline{\Gamma}_{\vec{\mathcal{F}}_{\mathcal L}}\left(r\right)=\frac{\int_{0}^{\pi}\Gamma_{\vec{\mathcal F}\!\!_{\mathcal L}}\left(r,\theta\right)\sin\theta{\rm d}\theta}{\int_{0}^{\pi}\sin^{3}\theta{\rm d}\theta}=\left[\Gamma_{0}\left(r\right)-\frac{1}{5}\Gamma_{2}\left(r\right)\right].
\end{equation}
Note that only the $l=2$ component of the circulation is able to advect a net amount of angular momentum; the higher order components of $\vec{\mathcal{U}}_{M}$ (for instance those induced in its tachocline by a differentially rotating convection zone) do not contribute to the vertical transport of angular momentum, as was pointed out in Spiegel  and Zahn (1992). The explicit form of (\ref{mean-AM}), keeping only the $l=\left\{1,2,3\right\}$ terms in the expansion of $\vec B$ (cf. eq. (\ref{expB})) is derived in Appendix E.2.1. .
   
\subsection{Evolution equation for the differential rotation in latitude}
  
We establish the equation governing the  horizontal transport of angular momentum by  multiplying eq. (\ref{mean-AM}) through $\sin^{2}\theta$ and subtracting it from the original form (\ref{AM-diff-adv}):   
\begin{eqnarray}
\rho \lefteqn{{{\rm d} \over {\rm d}t}  \left(r^2\sin^2\theta \, \widehat{\Omega}\right)+\vec\nabla\cdot\left(\rho r^{2}\sin^2\theta\, \overline{\Omega} \, \vec{\mathcal{U}}_{M}\right) +
\frac{\sin^2 \theta}{5r^2}\partial_{r}\left(\rho r^4\overline{\Omega} \, U_{2}\right)  = \frac{\sin^2\theta}{r^2}\partial_{r}\left(\rho\nu_{v}r^4\partial_{r}\widehat{\Omega}\right)+\frac{1}{\sin\theta}\partial_{\theta}\left(\rho\nu_{h}\sin^3\theta \, \partial_{\theta}\widehat{\Omega}\right)}\nonumber\\
&+&\Gamma_{\vec{\mathcal F}\!\!_{\mathcal L}}-\sin^2\theta \, \overline{\Gamma}_{\vec{\mathcal F}\!\!_{\mathcal L}}.
\end{eqnarray}
In the advection term we have again neglected the fluctuation $\widehat{\Omega}(r,\theta)$ compared to the mean $\overline{\Omega}$. 

The next step is to replace $\widehat{\Omega}(r,\theta)$ by its expansion (\ref{Bomexp}) in the horizontal functions ${Q}_{l}(\theta)$. For $l=2$, the equation separates neatly into
\begin{equation}
\rho{{\rm d} \over {\rm d}t} \left(r^2 \Omega_{2}\right)-2\rho\overline{\Omega}r\left[2V_{2}-\alpha(r)U_{2}\right]=\frac{1}{r^2}\partial_{r}\left(\rho\nu_{v}r^4\partial_{r}\Omega_{2}\right)-10\rho\nu_{h}\Omega_{2}+\Gamma_{2}
\end{equation}
\begin{equation} 
\hbox{with} \quad V_{2}=\frac{1}{6\rho r}\frac{{\rm d}}{{\rm d}r}\left(\rho r^2 U_{2}\right)
\quad  \hbox{ and } \quad 
\alpha(r)=\frac{1}{2}\frac{{\rm d}\ln\left(r^2\overline{\Omega}\right)}{{\rm d}\ln r}.
\end{equation}
It can be simplified by assuming that the turbulent transport is more efficient in the horizontal than in the vertical direction (i.e.  $\nu_{v} \ll \nu_{h}$):
\begin{equation}
\rho{{\rm d} \over {\rm d}t} \left(r^2 \Omega_{2}\right)-2\rho\overline{\Omega}r\left[2V_{2}-\alpha U_{2}\right]=-10\rho\nu_{h}\Omega_{2}+\Gamma_{2}.
\label{Omega2}
\end{equation}
In the asymptotic regime  $t \gg  {r^2}/{\nu_{h}}$, a stationary state is reached:
\begin{equation}
\nu_{h}\Omega_{2}=\frac{1}{5}r\left[2V_{2}-\alpha U_{2}\right]\overline{\Omega}+\frac{\Gamma_{2}}{10\rho} ,
\end{equation}
where horizontal diffusion balances horizontal advection and the torque due to the Lorentz force.\\ \newline
For $l\!\!>\!\!2$ the situation is more intricate, because there are couplings between terms of different $l$, which prevent a clean separation for each $l$. This is mainly due to the magnetic torque. Indeed, we recall that in the hydrodynamical case the hypothesis of Spiegel \& Zahn (1992) allows such separation (cf. Paper I,  \S3.2.). Therefore, we choose here to stop the expansion of the rotation law at $\Omega_{2}$. The explicit form of (\ref{Omega2}), keeping only the $l=\left\{1,2,3\right\}$ terms in the expansion of $\vec B$ (cf. eq. (\ref{expB})) is derived in Appendix E.2.1. .

\section{Structural properties of the differentially rotating magnetic star}
  
\subsection{Baroclinic relation}
  
As in Paper I, we consider a non-uniform and a non-cylindrical rotation law, and add here a general axisymmetric magnetic field. Then neither the centrifugal force nor  the Lorentz force derive from a potential, and therefore the isobars and the surfaces of constant density in general do not coincide.  In order to determine how the density varies on an isobar, we start from the hydrostatic equation:    
\begin{equation}
\frac{1}{\rho}\vec\nabla P=\vec g= - \vec\nabla \phi+\vec{\mathcal F}_{\mathcal{C}}+\frac{\vec{\mathcal F}_{\mathcal{L},P}}{\rho},
\label{barocl-orig}
\end{equation}
where $\phi$ is the gravitational potential and $\vec g$ the local effective gravity, which includes both the centrifugal force \\ $\vec{\mathcal F}_{\mathcal{C}}= \frac{1}{2}{\Omega}^{2}\vec\nabla\left(r^{2}\sin^{2}\theta\right)$ and the meridional magnetic force $\vec{\mathcal{F}}_{\mathcal L,P}$. Taking the curl of this equation, we get 
\begin{equation}
-\frac{1}{{\rho}^{2}}\vec\nabla\rho\wedge\vec\nabla P=-\frac{1}{\rho}\vec\nabla \rho\wedge\vec g=\frac{1}{2}\vec\nabla (\Omega)^{2}\wedge\vec\nabla(r\sin\theta)^{2}+\vec\nabla\wedge\left(\frac{\vec{\mathcal F}_{\mathcal{L},P}}{\rho}\right),
\label{barocl-origb}
\end{equation}
which to first order reduces to
\begin{equation}
-\frac{\vec\nabla {\rho}^{'}\wedge\vec g}{\overline{\rho}}=\left[\partial_{r}(\Omega^{2})r\cos\theta\sin\theta-\partial_{\theta}(\Omega^{2})\sin^{2}\theta\right]\widehat{e}_{\varphi}+\vec\nabla\wedge\left(\frac{\vec{{\mathcal F}}_{\mathcal L,P}}{\overline{\rho}}\right),
\label{barocline}
\end{equation} 
with $\rho' (r, \theta)$ being the variation of the density on the isobar.

We now expand all terms in spherical harmonics. For the density fluctuation this is readily done:
\begin{equation}
\rho^{'}\left(r,\theta\right)=\sum_{l>0} \widetilde{\rho_{l}}\left(r\right)P_{l}(\cos\theta).
\end{equation}
Then, differentiating the equation of state:
\begin{equation}
{d \rho \over \rho} = \alpha {d P \over P} - \delta {d T \over T} + \varphi {d \mu \over \mu}
\end{equation}
we can write the modal amplitude of the  density  fluctuation as
\begin{equation}
{ \widetilde{\rho_{l}} \over \overline\rho} = \Theta_l = \varphi \Lambda_l - \delta \Psi_l .
\end{equation}
For the centrifugal force, we recall that 
\begin{equation}
\Omega(r,\theta)=\overline{\Omega}(r)+\sum_{l>0}\Omega_{l}(r)\left(P_{l}(\cos\theta)-I_{l}\right) ,
\end{equation} 
which leads us to
\begin{equation}
\Omega^{2}(r,\theta)=\left[{\overline{\Omega}}^{2}-2\overline{\Omega}\sum_{l>0}\Omega_{l}I_{l}\right]+2\overline{\Omega}\sum_{l>0}\Omega_{l}P_{l}(\cos\theta)
\label{omega-sq}
\end{equation}
keeping only the terms linear in $\Omega_l$ (beside $\overline{\Omega}^{2}$).   
Finally, we draw the expansion of the poloidal Lorentz force from (\ref{Lorentzexp}): 
\begin{equation}
\vec{\mathcal{F}}\!\!_{\mathcal{L},P}\left(r,\theta\right)=\sum_{l=1}^{\infty}\left\{\mathcal{X}_{{\vec {\mathcal F}}_{\mathcal{L}};l}(r)P_{l}(\cos\theta)\vec{\widehat{e}}_{r}+\mathcal{Y}_{{\vec {\mathcal F}}_{\mathcal{L}};l}(r)P_{l}^{1}(\cos\theta)\vec{\widehat{e}}_{\theta}\right\} .
\end{equation}

It remains to insert these expansions in eq. \ref{barocline}, noting that it projects itself only on the azimuthal vectorial spherical harmonics $\vec{T}_{l}^{0}\left(\theta\right)$. Using the algebra related with these harmonics, we reach the following expression for the modal amplitudes of the relative fluctuation of density on an isobar:
\begin{equation}
\frac{\widetilde{\rho_{l}}\left(r\right)}{\overline\rho}=\varphi \Lambda_l - \delta \Psi_l =\frac{r}{\overline{g}}\left[\mathcal{D}_{l}(r)+\frac{{\mathcal X}_{\vec{\mathcal F}\!\!_{\mathcal L};l}\left(r\right)}{r\overline{\rho}\left(r\right)}+\frac{1}{r}\frac{{\rm d}}{{\rm d}r}\left(r\frac{\mathcal{Y}_{\vec{\mathcal{F}}_{\mathcal L};l}\left(r\right)}{\overline{\rho}\left(r\right)}\right)\right].
\label{baro1}
\end{equation}
$\overline{g}$ is the horizontal average of the modulus of $\vec g$ and $\mathcal{D}_{l}\left(r\right)$ has been derived in Paper I (eq. 47):
\begin{eqnarray}
\mathcal{D}_{l}(r)&=&\mathcal{N}_{l}^{0}\left\{r\partial_{r}\left[\overline{\Omega}^{2}(r)-2\overline{\Omega}(r)\Omega_{2}(r)I_{2} \right] \frac{1}{3\mathcal{N}_{2}^{0}}\delta_{l,2}\right . \nonumber\\
&+&2r\sum_{s>0}\partial_{r}\left(\overline{\Omega}(r)\Omega_{s}(r)\right)\frac{1}{\mathcal{N}_{s}^{0}}\left[A_{s}^{0}\left(-C_{s-1}^{0}\delta_{l,s-2}+D_{s-1}^{0}\delta_{l,s}\right)+B_{s}^{0}
\left(-C_{s+1}^{0}\delta_{l,s}+D_{s+1}^{0}\delta_{l,s+2}\right)\right]\nonumber\\
&-&{\left. 2\overline{\Omega}(r)\sum_{s>0}\frac{\Omega_{s}(r)}{\mathcal{N}_{s}^{0}}\left[G_{s}^{0}\left(-C_{s+1}^{0}\delta_{l,s}+D_{s+1}^{0}\delta_{l,s+2}\right)-H_{s}^{0}\left(-C_{s-1}^{0}\delta_{l,s-2}+D_{s-1}^{0}\delta_{l,s}\right)\right]\right\}},
\label{baro2}
\end{eqnarray}
where all the numerical coefficients ($A^0_l, B^0_l, C^0_l, D^0_l, G^0_l, H^0_l$) are given in Appendix A. 

This baroclinic equation  (\ref{baro1} - \ref{baro2}) plays a key role in linking the density fluctuation on an isobar with the differential rotation and the magnetic field. It allows us to close the system formed by the induction equation, that for the transport of angular momentum, that for the transport of the chemical species and that for the transport of heat, which we shall establish in \S6.  
   
\subsection{Effective gravity and perturbed potential} \label{effgrav-pot}
  
It remains to examine how the redistribution of mass we just described modifies the local gravity. In the absence of magnetic field, the perturbing force is just the centrifugal force $\vec{\mathcal F}_{\mathcal{C}} =\frac{1}{2}\Omega^{2}\vec\nabla(r^2\sin^2\theta)$, whose $r$ and $\theta$ components are expanded as
\begin{equation}
{\mathcal  F}_{{\mathcal C}, r}(r,\theta)=\sum_{l}a_{l}(r)P_{l}(\cos\theta) \qquad
{\mathcal  F}_{{\mathcal C},\theta}(r,\theta)=-\sum_{l}b_{l}(r)\partial_\theta P_{l}(\cos\theta).
\label{ab}
\end{equation}
where $a_{l}(r)$ and $b_{l}(r)$ are given by: 
\begin{eqnarray}
a_{l}(r)&=&\frac{2}{3}r\left[\overline{\Omega}^{2}(r)-2\overline{\Omega}(r)\Omega_{2}(r)I_{2}\right]\left(\delta_{l,0}-\delta_{l,2}\right)\label{al}\nonumber\\
&+& 2r\overline{\Omega}(r)\sum_{s>0}\Omega_{s} (r)\left[\frac{1}{\mathcal{N}_{s}^{0}} \left\{C_{s}^{0} \left(G_{s-1}^{0}\mathcal{N}_{s}^{0}\delta_{l,s}-H_{s-1}^{0}\mathcal{N}_{s-2}^{0}\delta_{l,s-2}\right)
- D_{s}^{0}\left(G_{s+1}^{0}\mathcal{N}_{s+2}^{0}\delta_{l,s+2}-H_{s+1}^{0}\mathcal{N}_{s}^{0}\delta_{l,s}\right)\right\}\right],
\label{al}
\end{eqnarray} 
\vskip - 1pc
\begin{eqnarray}
b_{l}(r)&=&\frac{1}{3}r\left[\overline{\Omega}^{2}(r)-2\overline{\Omega}(r)\Omega_{2}(r)I_{2}\right]\delta_{l,2}\nonumber\\
&+&2r\overline{\Omega}(r)\sum_{s>0}\frac{\Omega_{s}ñ(r)}{\mathcal{N}_{s}^{0}}\left\{A_{s}^{0}\left(-C_{s-1}^{0}\mathcal{N}_{s-2}^{0}\delta_{l,s-2}+D_{s-1}^{0}\mathcal{N}_{s}^{0}\delta_{l,s}\right)
+ B_{s}^{0}\left(-C_{s+1}^{0}\mathcal{N}_{s}^{0}\delta_{l,s}+D_{s+1}^{0}\mathcal{N}_{s+2}^{0}\delta_{l,s+2}\right)\right\}. 
\label{bl} 
\end{eqnarray}
 
To this we have here to add the meridional components of the Lorentz force
\begin{equation}
{\mathcal F}_{{\mathcal L},P;r}\left(r,\theta\right)=\sum_{l}\mathcal{X}_{\vec{\mathcal{F}}\!\!_{\mathcal{L}};l}\left(r\right)P_{l}\left(\cos\theta\right) \qquad {\mathcal F}_{{\mathcal L},P;\theta}=-\sum_{l}\mathcal{Y}_{\vec{\mathcal{F}}\!\!_{\mathcal{L}};l}\left(r\right)\partial_{\theta}P_{l}\left(\cos\theta\right). 
\end{equation}

In the expressions that we have derived in Paper I, it suffices then replace $\rho_0 a_l$ by $\rho_0 a_l +\mathcal{X}_{\vec{\mathcal{F}}\!\!_{\mathcal{L}};l}$ and $\rho_0 b_l$ by $\rho_0 b_l + \mathcal{Y}_{\vec{\mathcal{F}}\!\!_{\mathcal{L}};l}$ to establish the version including the effect of the magnetic field. Thus for the gravity perturbation along an isobar we now get (cf. Paper I, eq. 72)
\begin{equation}
\frac{\widetilde{g}_{l}}{\overline{g}}=-\left[\frac{{\rm d} g_{0}}{{\rm d} r}\frac{1}{{g_{0}^{2}}}r\left(b_{l}+\frac{\mathcal{Y}_{\vec{\mathcal F}\!\!_{\mathcal L};l}}{\rho_{0}}\right) + \frac{1}{g_{0}}\left(a_{l}+\frac{\mathcal{X}_{\vec{\mathcal F}\!\!_{\mathcal L};l}}{\rho_{0}}\right)\right] + \frac{{\rm d}}{{\rm d}r}\left(\frac{\widehat{\phi}_{l}}{g_{0}}\right) .
\label{geff}
\end{equation}
The last term involves the fluctuation of the gravity field along the sphere, $\widehat{\phi}_{l}$, which is obtained by integrating the Poisson equation  (cf. Paper I, eq. 60), modified along the same rules:
\begin{equation}
\frac{1}{r}\frac{{\rm d}^2}{{\rm d}r^2}\left(r\widehat{\phi}_{l}\right)-\frac{l(l+1)}{r^2}\widehat{\phi}_{l}-\frac{4\pi G}{g_{0}}\frac{{\rm d}\rho_{0}}{{\rm d}r}\widehat{\phi}_{l}=\frac{4\pi G}{g_{0}}\left[\rho_{0}a_{l}+\frac{{\rm d}}{{\rm d}r}\left(r\rho_{0}b_{l}\right)+\mathcal{X}_{\vec{\mathcal{F}}\!\!_{\mathcal{L}};l}+\frac{{\rm d}}{{\rm d}r}\left(r\mathcal{Y}_{\vec{\mathcal{F}}\!\!_{\mathcal{L}};l}\right)\right] .
\label{poisson-pert}
\end{equation}
Let us recall that Sweet (1950) was the first to establish this result for the most general perturbing force.

\section{Thermal imbalance and transport of heat}

It remains to implement the magnetic field in the heat equation:
\begin{equation}
\rho T \left[ {\partial S \over \partial t} + \vec{\mathcal{U}} \cdot \vec\nabla{S} \right] =\vec\nabla\cdot\left(\chi\vec\nabla T\right)+\rho\epsilon-\vec\nabla\cdot{\vec{F}}_{h}+\mathcal{J} ,
\label{entropy-orig}
\end{equation} 
where $S$ is the entropy per unit mass, $\chi$ the thermal conductivity, $\epsilon$ the nuclear energy production rate per unit mass and $\vec F_{h}$ the flux carried in the horizontal direction by the anisotropic turbulence.  Note that in a medium of varying composition, we have to take into account the entropy of mixing (cf. Maeder \& Zahn 1998). In the simplest case, applicable to main-sequence stars, where the stellar material can be approximated by a  mixture of hydrogen and helium with a fixed abundance of metals, it can be expressed in terms of the mean molecular weight only; then we have
\begin{equation}
{\rm d}S=C_{p}\left[\frac{{\rm d}T}{T}-\nabla_{\rm ad}\frac{{\rm d}P}{P}+\Phi\left(P,T,\mu\right)\frac{{\rm d}\mu}{\mu}\right]
\end{equation}  
where  $\nabla_{\rm ad}$ is the adiabatic gradient and $\Phi$ is a function of the metal mass fraction and of $\mu$, the mean molecular weight.

The magnetic field manifests itself in the heat equation (\ref{entropy-orig}) through the Ohmic heating term $\mathcal{J}$, but also in the divergence of the thermal flux, because it involves the divergence of the perturbing force. 
This force includes both the centrifugal force $\vec{\mathcal{F}}_{\mathcal{C}}$ and the Lorentz force per unit volume, ${\vec{\mathcal F}_{\mathcal{L},P}}/{\rho_{0}}$, and their divergence will again be expanded in spherical harmonics as
\begin{equation}
\vec\nabla\cdot\vec{\mathcal{F}}_{\mathcal{C}}=\overline{f}_{\mathcal{C}}+\sum_{l>0}\widetilde{f}_{\mathcal{C},l}P_{l}\left(\cos\theta\right) \quad  \hbox{and}  \quad\vec\nabla\cdot\left( \frac {\vec {\mathcal{F}}\!\!_{\mathcal{L,P}}} {\rho_0} \right)
=\overline{f}_{\mathcal{L}}
+\sum_{l>0}\widetilde{f}_{\mathcal{L},l}P_{l}\left(\cos\theta\right).
\end{equation}
In Paper I (eq. 85) we gave the expressions of $\overline{f}_{\mathcal{C}}$ and $\widetilde{f}_{\mathcal{C},l}$ in terms of $a_l$ and $b_l$:
\begin{equation}
\overline{f}_{\mathcal{C}}=\frac{1}{r^2}\partial_{r}\left(r^2 a_{0}\right)\hbox{ and }\widetilde{f}_{\mathcal{C},l}=\frac{1}{r^2}\partial_{r}\left(r^2a_{l}\right)+l\left(l+1\right)\frac{b_{l}}{r} .
\end{equation}
As above in \S\ref{effgrav-pot}., it suffices to replace $a_l$ by $\mathcal{X}_{\vec{\mathcal{F}}\!\!_{\mathcal{L}};l}/\rho_0$ and $b_l$ by $\mathcal{Y}_{\vec{\mathcal{F}}\!\!_{\mathcal{L}};l}/\rho_0$ to obtain the equivalent expressions for the divergence of the Lorentz force:
\begin{equation}
\overline{f}_{\mathcal{L}}=\frac{1}{r^2}\partial_{r}\left(r^2\frac{\mathcal{X}_{\vec{\mathcal F}\!\!_{\mathcal L};0}}{\rho_{0}}\right)\quad\hbox{ and }\quad\widetilde{f}_{\mathcal{L},l}=\frac{1}{r^2}\partial_{r}\left(r^2\frac{\mathcal{X}_{\vec{\mathcal F}\!\!_{\mathcal L};l}}{\rho_{0}}\right)+l\left(l+1\right)\frac{\mathcal{Y}_{\vec{\mathcal F}_{\mathcal{L};l}}}{r\rho_{0}}.
\end{equation}

This leads us to the modal form of the heat equation, which can be implemented directly into a stellar structure code:
\begin{equation}
\overline{T} C_{p}\left[{{\rm d} \Psi_{l} \over {\rm d} t}+\Phi\frac{{\rm d}\ln\overline{\mu}}{{\rm d}t}\Lambda_{l}+\frac{U_{l}(r)}{H_{p}}\left(\nabla_{ad}-\nabla\right)\right]=\frac{L(r)}{M(r)}{\mathcal T}_{l}(r)+\frac{\mathcal{J}_{l}\left(r\right)}{\overline{\rho}},
\label{heat1}
\end{equation}
where $\mathcal{T}_{l}$ is given by:
\begin{eqnarray}
\mathcal{T}_{l}&=&2\left[1-\frac{\overline{f}_{\mathcal{C}}+\overline{f}_{\mathcal{L}}}{4\pi G\overline{\rho}}-\frac{\left(\overline{\epsilon}+\overline{\epsilon}_{\rm grav}\right)}{\epsilon_{m}}\right]\frac{\widetilde{g}_{l}}{\overline{g}}+\frac{\widetilde{f}_{\mathcal{C},l}+\widetilde{f}_{\mathcal{L},l}}{4\pi G\overline{\rho}}-\frac{\overline{f}_{\mathcal{C}}+\overline{f}_{\mathcal{L}}}{4\pi G\overline{\rho}}\left(-\delta\Psi_{l}+\varphi\Lambda_{l}\right)\nonumber\\
&+&\frac{\rho_{m}}{\overline{\rho}}\left[ \frac{r}{3}\partial_{r}\left(H_{T}\partial_{r}\Psi_{l}-(1-\delta+\chi_{T})\Psi_{l}-(\varphi+\chi_{\mu})\Lambda_{l}\right)-\frac{l(l+1)H_{T}}{3r}\left(1 + \frac{D_{h}}{K}\right){\Psi_l}\right] \nonumber\\
&+&\frac{\left(\overline{\epsilon}+\overline{\epsilon}_{grav}\right)}{\epsilon_{m}}\left\{ \left(H_{T}  \partial_{r}\Psi_{l}-(1-\delta+\chi_{T})\Psi_{l}-(\varphi+\chi_{\mu})\Lambda_{l}\right)+(f_{\epsilon}\epsilon_{T} - f_{\epsilon}\delta + \delta)\Psi_{l}+(f_{\epsilon}\epsilon_{\mu}+f_{\epsilon}\varphi - \varphi)\Lambda_{l}\right\} . \nonumber\\    
\label{tcal-final}
\end{eqnarray}
We recall that $L$ is the luminosity, $M$ the mass, $\overline{T}$ the horizontal average of the temperature, $C_{p}$ the specific heat at constant pressure, and $\nabla$ the radiative gradient. We have also introduced the temperature scale-height $H_{T}=\left|{{\rm d}r}/{{\rm d}\ln \overline{T}}\right|$, the thermal diffusivity $K={\overline{\chi}}/{\overline{\rho}C_{p}}$, the horizontal eddy-diffusivity $D_{h}$ and $f_{\epsilon}={\overline{\epsilon}}/{\left(\overline{\epsilon}+\overline{\epsilon}_{\rm grav}\right)}$, with $\overline{\epsilon}$ and $\overline{\epsilon}_{\rm grav}$ being respectively the mean nuclear and gravitational energy release rates, whereas $\epsilon_{\mu}$ and  $\chi_{\mu}$ are the logaritmic derivatives of $\epsilon$ and of the radiative conductivity $\chi$ with respect to $\mu$, their derivatives with respect to $T$ being noted as $\epsilon_{T}$ and $\chi_{T}$. Moreover, we have $\epsilon_{m}=L\left(r\right)/M\left(r\right)$ and $\rho_{m}$ is the mean density inside the considered level surface.

Three remarks before we conclude this section.
   
\begin{itemize}
\item First, we have noted previously that if the expansion of $\vec B$ is stopped at mode $l_{\rm max}=N_{m}$ in (\ref{expB}), then the expansions of $\vec{\mathcal{F}}\!\!_{\mathcal{L}}$ and of $\mathcal{J}$ must include all terms up to  $l_{\rm max}=2N_{m}$, due to selection rules. Therefore, keeping $\Omega\left(r,\theta\right)=\overline{\Omega}\left(r\right)+\Omega_{2}\left(r\right)Q_{2}\left(\theta\right)$, the expansion of the meridional flow will extend to $l_{\vec{\mathcal U}_{M};{\rm max}}={\rm max}\left\{4,2N_{m}\right\}$.
\item Next, the relative importance of the terms contributed respectively by the magnetic force and by the centrifugal force depends on the ratio $\left({V_{A}}/{V_{\Omega}}\right)^{2}$ where $V_{A}={B}/{\sqrt{\mu_{0}\rho_{0}}}$ is the Alfv\'en speed and $V_{\Omega}=R\Omega$, $R$ being the radius of the star. If  the magnetic field is weak enough, this ratio will be small except just below the surface because of the decay of the density. In this case, for the implementation in existing stellar evolution codes of the equations related to the thermal imbalance and to the structural properties of the star (cf. \S5), we keep only the modes $l=\left\{2,4\right\}$, which allows to describe the tachocline(s) circulation to first order (see Appendix E). However, in the case of a strong field higher order modes of the Lorentz force must be taken into account.
\item Finally, the ratio between  the term representing the Ohmic heating, ${\mathcal{J}_{l}\left(r\right)}/{\overline{\rho}}$, and the highest-order derivative term of $\vec\nabla\cdot\left(\chi\vec\nabla T\right)$ in (\ref{heat1}-\ref{tcal-final}), namely $({L}/{M}) ({\rho_{m}}/{\overline{\rho}}) ({r}/{3}) \partial_{r}\left(H_{T}\partial_{r}\Psi_{l}\right)$, is given by $({\eta}/{K}) \left({{V_{A}}}/{V_{\Omega}}\right)^{2}$. Therefore the Ohmic term can be neglected since $V_A<R\Omega$ and $\eta \ll K$.
\end{itemize}

In conclusion, the link between the circulation and its cause, namely the thermal imbalance due to the rotation, the magnetic field and the chemical composition, is established through (\ref{heat1}-\ref{tcal-final}) where the temperature fluctuation on an isobar is governed by an advection/diffusion equation from which we can derive the radial component of $\vec{\mathcal{U}}_{M}$. These equations have been established for a rotation law which depends  on $r$ and $\theta$, and  that allows us to treat simultaneously the bulk of the radiation zones and the tachoclines in presence of a general axisymmetric magnetic field.

\section{Boundary conditions}

The system of equations is now complete: we have the induction equation, an advection/diffusion equation for the transport of angular momentum (mean and fluctuating) including the action of  the magnetic torque,  another for the temperature (mean and fluctuating), and  the baroclinic relation which allows us to close the system. The equation for the transport of chemical species (or alternatively for the transport of molecular weight) is unchanged. It thus remains to specify the boundaries conditions of this system, to be applied on the limits of the radiation zone.
To be specific, we consider a star with a radiation zone located between a convective core and an upper convection zone. We designate by $r_{b}$ and $r_{t}$ the respective radius of the base and of the top of that radiative zone. Of course, in a solar-type  main-sequence star we have $r_b=0$, whereas for a massive main sequence star $r_t=R$.

The boundary conditions for the equations already present in the hydrodynamic case, which was examined in Paper I, are unchanged, except those to be applied to the equation of angular momentum transport, since it includes the Lorentz torque.
The novelty here is the induction equation, which we have split in two equations, respectively for the
 poloidal and  toroidal magnetic stream-functions $\xi_{0}^{l}$ (eq. \ref{i5}) and $\chi_{0}^{l}$ (eq. \ref{i6}). These equations are of second order in $r$, and therefore each of them requires two boundary conditions.  In principle they can be obtained by solving  the induction equation, together with the equation of momentum, etc. in each adjacent convective zone, but this represents a formidable task which is well beyond our scope here. Instead,  we make the following - simplifying but reasonable - hypothesis, valid only in the convection zone(s): 
\begin{itemize}
\item (i): we do not take into account the meridional velocity field. The only motion we retain  is the zonal flow associated with the differential rotation: $\vec{\mathcal U}_{\varphi}=r\sin\theta \, \Omega_{CZ}\left(r,\theta\right)\vec{\widehat{e}}_{\varphi}$, with $\Omega_{CZ}(r, \theta)$ being the angular velocity in the convection zone;
\item (ii): we put $\eta_{v}=\eta_{h}=\eta_{\rm {turb}}$ where $\eta_{\rm {turb}}$ is the magnetic eddy-diffusivity, which we assume constant  inside the considered convective region;
\item (iii): this eddy-diffusivity is high enough to allow for a stationary state (compared to the slow evolution in the radiation zone). 
\end{itemize}
With these assumptions, the poloidal field is potential, and (\ref{i5}) reduces to:
\begin{equation}
\frac{{\rm d^{2}}\xi_{0}^{l}}{{\rm d}r^{2}}-\frac{l\left(l+1\right)}{r^2}\xi_{0}^{l}=0;
\end{equation}
it admits the general solution:
\begin{equation}
\xi_{0}^{l}=Ar^{l+1}+\frac{B}{r^{l}}.
\label{polCLfin}
\end{equation}
When applying it to a convective core where we must take the non-singular solution at the origin: \begin{equation}
r\frac{{\rm d}\xi_{0}^{l}}{{\rm d}r}-\left(l+1\right)\xi_{0}^{l}=0\quad \hbox{at} \; r=r_{b} ,
\end{equation}
which becomes $\xi_{0}^{l}=0$ if $r_{b}=0$.
In the case of a convective envelope, surrounded by vacuum, we have
\begin{equation}
r\frac{{\rm d}\xi_{0}^{l}}{{\rm d}r}+l\xi_{0}^{l}=0\quad \hbox{at} \; r=r_{t}.
\end{equation}
With the same assumptions, the toroidal field obeys the simplified form of (\ref{i6}):
\begin{equation}
\frac{{\rm d^{2}}\chi_{0}^{l}}{{\rm d}r^{2}}-\frac{l\left(l+1\right)}{r^2}\chi_{0}^{l}={\mathcal{S}_{l}\left(\Omega_{\rm{CZ}}\right)\over \eta_{\rm{turb}}}   \hbox{ } 
\quad \hbox{where} \quad \hbox{ }\mathcal{S}_{l}\left(\Omega_{\rm{CZ}}\right)=-\frac{1}{\mathcal{N}_{0}^{l}}\left[\mathcal{X}_{{\bf Ad};l}\left(\Omega_{\rm{CZ}}\right)+\partial_{r}\left(r\mathcal{Y}_{{\bf Ad};l}\left(\Omega_{\rm{CZ}}\right)\right)\right].
\label{torCL1}
\end{equation}
$\mathcal{X}_{{\bf Ad};l}\left(\Omega_{\rm{CZ}}\right)$ and $\mathcal{Y}_{{\bf Ad};l}\left(\Omega_{\rm{CZ}}\right)$ are derived using the previous multipolar solutions for $\xi_{0}^{l}$ (cf. eq. \ref{polCLfin}), expanding $\Omega_{\rm{CZ}}$ in spherical functions like $\Omega_{\rm{CZ}}\left(r,\theta\right)=\sum_{l}\Omega_{l}\left(r\right)P_{l}\left(\cos\theta\right)$ where the $\Omega_{l}$ are taken from the results of the inversion of helioseismological data (Corbard et al. 2002),  using (\ref{wOmega}) and Appendix~C. Then, the solution of (\ref{torCL1}) inside a convection zone is derived using Green's functions:
\begin{equation}
\chi_{0}^{l}=\frac{1}{2l+1} {1  \over \eta_{\rm{turb}}} \left[r^{l+1}\int_{r}^{R_{\rm{extCZ}}}x^{-l} \mathcal{S}_{l}\left(\Omega_{\rm{CZ}}\right) \,
{\rm d}x+r^{-l}\int_{R_{intCZ}}^{r}x^{l+1}
\mathcal{S}_{l}\left(\Omega_{\rm{CZ}}\right) \,
{\rm d}x\right]
\label{torCLfin}
\end{equation}
where $R_{intCZ}$ and $R_{\rm{extCZ}}$ are respectively the radius at its base and at its top. 
Applying (\ref{torCLfin}) to a convective core where $R_{intCZ}=0$ and $R_{\rm{extCZ}}=r_{b}$, we get:
\begin{equation}
\chi_{0}^{l}=\frac{1}{2l+1} {1  \over \eta_{\rm{turb}}} r_{b}^{-l}\int_{0}^{r_{b}}x^{l+1}
\mathcal{S}_{l} \left(\Omega_{\rm{CZ}}\right) \,
{\rm d}x \quad \hbox{at} \; r=r_{b}.
\end{equation}
Finally, for a convective envelope where  $R_{intCZ}=r_{t}$ and $R_{\rm{extCZ}}=R$, we obtain:
\begin{equation}
\chi_{0}^{l}=\frac{1}{2l+1}{1  \over \eta_{\rm{turb}}} r_{t}^{l+1}\int_{r_{t}}^{R}x^{-l}
\mathcal{S}_{l}\left(\Omega_{\rm{CZ}}\right) \,
{\rm d}x\quad \hbox{at} \; r=r_{t}.
\end{equation}
In the case where $\mathcal{S}_{l}\left(\Omega_{\rm{CZ}}\right)$ is negligible, the previous results leads to:
\begin{equation}
\chi_{0}^{l}=0\hbox{ }\hbox{at}\hbox{ }r=r_{b} \quad \hbox{and} \; r=r_{t}, \quad \hbox{and thus to} \quad \vec B_{T}=\vec 0 .
\end{equation}

It remains to write down the boundary conditions to be applied on the transport of angular momentum.
 Since the equation for the mean angular velocity $\overline \Omega$ (\ref {mean-AM}) is of second order in $r$,  it requires two boundary conditions.  They are obtained by evaluating the budget of angular momentum in each adjacent convective zone: 
    \begin{equation}
 { {\rm d} \over {\rm d}t} \left[\int_{0}^{r_{b}}r^{4}\rho \Omega {\rm d}r \right]={1 \over 5} r^{4}\rho\overline{\Omega}U_{2} - \mathcal{F}_{B}(r_b) \quad \hbox{at} \; r=r_{b} , \qquad
{{\rm d} \over {\rm d}t} \left[\int_{r_{t}}^{R}r^{4}\rho \Omega {\rm d}r\right]=-{1 \over 5} r^{4}\rho\overline{\Omega}U_{2}-\mathcal{F}_{\Omega} + \mathcal{F}_{B}(r_t) \quad \hbox{at} \;  r=r_{t} .
 \label{bc-mean-omega}
  \end{equation}
 $\mathcal{F}_{\Omega}$ is the (signed) flux of angular momentum which is lost at the surface by the stellar wind, and  $\mathcal{F}_{B}(r)$ the flux carried by the magnetic field through the considered surface:
\begin{equation}
\mathcal{F}_{B}(r) = -{1 \over \mu_0} r^3 \int_0^\pi B_r \left(r, \theta\right) \, B_\varphi \left(r, \theta\right) \sin^2 \theta \, {\rm d}\theta=\frac{3}{2\mu_{0}}\sum_{l>0}\left\{l\left(l+1\right)\mathcal{N}_{l}^{0}\xi_{0}^{l}\left[\frac{\left(l+1\right)\left(l+2\right)\mathcal{N}_{l+1}^{0}}{\left(2l+1\right)\left(2l+3\right)}\chi_{0}^{l+1}-\frac{\left(l-1\right)l\mathcal{N}_{l-1}^{0}}{\left(2l-1\right)\left(2l+1\right)}\chi_{0}^{l-1}\right]\right\}.
\end{equation}
Here again the perturbation $\Omega_2$ obeys an evolution equation which does not include any derivative in $r$, and  therefore it needs no boundary condition.

\section{Conclusion}

The work presented here is the continuation of Paper I (Mathis \& Zahn 2004), where the star was assumed without magnetic field. Here we allow for a magnetic field of moderate strength in the radiation zone(s), with an Alfv\'en speed not exceeding the rotational velocity: $(V_A)^2 < (R\Omega)^2$. Such a field has little impact on the hydrostatic balance, since we also assume, as in Paper I, that the centrifugal force is small compared to gravity: $R \Omega^2 \ll g$. But it will compete with the centrifugal force in the baroclinic balance (\ref{barocl-origb}), and therefore it will participate in governing the meridional flow.  Moreover, such a field tends to enforce uniform rotation along the field lines of the poloidal field (Ferraro's law). The question then arises whether or not the poloidal field threads into the convection zone(s), where such differential rotation is probably maintained through the turbulent motions, as observed in the Sun. To answer that question one has to treat consistently the evolution of both angular velocity and magnetic field, with special care for the dynamics in the tachocline(s), and for this reason we expanded  them in spherical harmonics to arbitrary order.

\begin{figure}[h!]
\centering
\resizebox{12cm}{!}{\includegraphics{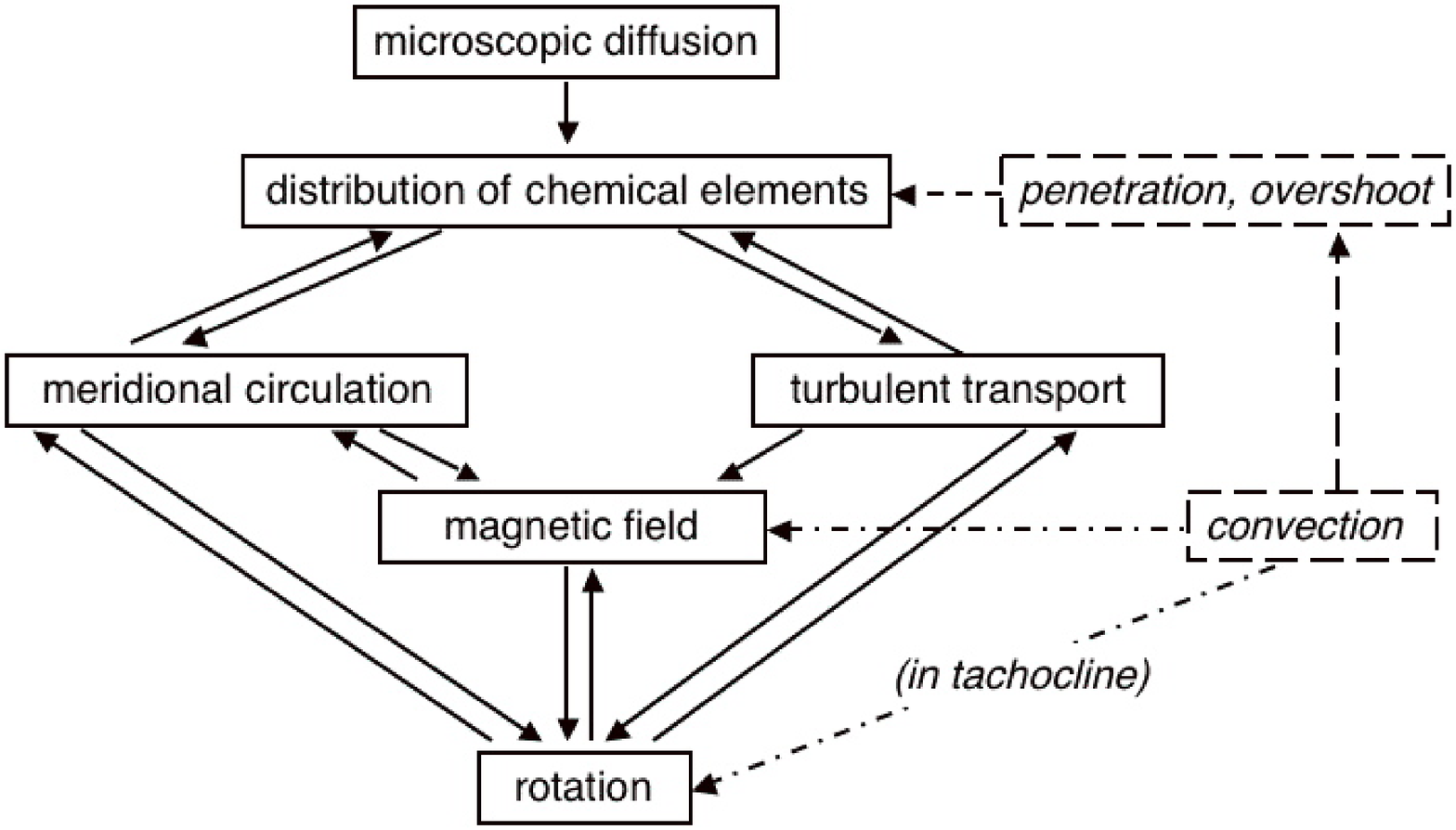}}
\caption{Rotational mixing in stellar interiors with magnetic field: a highly non-linear problem (the action of the convection through penetration and overshoot which enhance the mixing has been added).}
\label{diagram}
\end{figure}

The overall problem of rotational mixing in a magnetized star is highly non-linear, with multiple feed-backs, as illustrated by the diagram displayed in fig.~\ref{diagram}.
Mixing is achieved through the meridional circulation, which is due to the thermal imbalance caused by the centrifugal force and by the Lorentz force (mainly through the baroclinic balance), and also through the turbulence generated by the shear of differential rotation and possibly by the magnetic field. These motions modify both the rotation profile and the magnetic field through large-scale advection and turbulent diffusion. 
As the star evolves, molecular gradients build up through microscopic diffusion (including radiative levitation and gravitational settling), and through nuclear burning. Large-scale advection turns them into horizontal gradients, which react on the meridional circulation and are smoothed by the turbulence.

The boundary conditions play a crucial role: in a non-magnetic star, the circulation is driven by the loss (or gain) of angular momentum, as explained in Zahn (1992); in the absence of such loss, the circulation tends to vanish, as was shown by Busse (1982). It remains to be seen whether this is still the case when angular momentum is transported mainly through magnetic stresses, with the poloidal field anchored in the convection zone(s).

The main weakness of our modeling remains the description of the turbulence.  As in Paper I, it is assumed to be anisotropic, due to the stable stratification, and that it tends to smooth angular velocity and chemical composition on horizontal surfaces. The action of magnetic field on such turbulence is not taken in account, which seems reasonable as long as the Alfv\'en speed does not exceed the rotational velocity. Moreover, we have deliberately ignored here the possibility of MHD instabilities generating their own turbulence (Balbus \& Hawley 1994; Spruit 1999).

From the technical point of view, we adopted here the formalism developed in Paper I and introduced the spherical vectorial harmonics, which obey the Racah-Wigner angular momentum algebra widely used in quantum mechanics. This allowed us to separate the colatitude $\theta$  in the vectorial partial differential equations which govern the problem, and to reduce the problem to solve coupled partial differential equations in $t\hbox{ and }r$ only. It is then straightforward to introduce these equations in existing stellar structure codes, a task we have now undertaken.

Among the first applications we plan to model a solar-type star, to check whether a fossil magnetic field, which is able to extract angular momentum from the radiative interior as the star spins down, yields a profile of angular velocity that is compatible with the helioseismic data. Next we will undertake the modeling of magnetized massive stars, since they are progenitors of neutron stars which are the seat of strong magnetic fields (Maeder \& Meynet 2003, 2004; Heger et al. 2004).

Most of the Appendix is dedicated to the algebra involving the spherical and vectorial spherical harmonics used in this paper. The numerical values of the coupling coefficients given in the Tables have been computed with Mathematica. In Appendix E, we state the equations in a form ready to be implemented in a stellar structure code, when the magnetic field is expanded up to the octupole ($l=3$) and only the first term $\Omega_2$ of the differential rotation is retained.

\begin{acknowledgements}
This work was supported by the Centre National de la Recherche Scientifique (Programme National de Physique Stellaire and GdR Dynamo).     
\end{acknowledgements}

  
\appendix{}

\section{Algebra related to the spherical harmonics}

\subsection{Scalar quantities}

\subsubsection{Definition and basic properties}
The  spherical harmonics are defined by:
\begin{equation}
Y_{l}^{m}(\theta,\varphi)=\mathcal{N}_{l}^{m}P_{l}^{|m|}\left(\cos\theta\right)e^{im\varphi} ,
\label{def}
\end{equation}
where $P_{l}^{|m|}\left(\cos\theta\right)e^{im\varphi}$ is the associated Legendre function, and the normalization coefficient being 
\begin{equation}
\mathcal{N}_{l}^{m}=(-1)^{\frac{\left(m+|m|\right)}{2}}\left[\frac{2l+1}{4\pi}\frac{(l-|m|)!}{(l+|m|)!}\right]^{\frac{1}{2}}.
\label{norm}
\end{equation}
They obey the orthogonality relation:
\begin{equation}
\int_{\Omega}\left(Y_{l_{1}}^{m_{1}}\left(\theta,\varphi\right)\right)^{*}Y_{l_{2}}^{m_{2}}\left(\theta,\varphi\right){\rm d}\Omega=\delta_{l_{1},l_{2}}\delta_{m_{1},m_{2}}
\label{ortho}
\end{equation}
where ${\rm d}\Omega=\sin\theta \, {\rm d}\theta \, {\rm d}\varphi$ and where the complex conjugate spherical harmonic is given by:
\begin{equation}\left(Y_{l}^{m}\left(\theta,\varphi\right)\right)^{*}=\left(-1\right)^{m}Y_{l}^{-m}\left(\theta,\varphi\right).
\label{conj}
\end{equation}
Using these properties, every function $f(\theta,\varphi)$ can be expanded as:
\begin{equation}
f\left(\theta,\varphi\right)=\sum_{l=0}^{\infty}\sum_{m=-l}^{l}f_{m}^{l}Y_{l}^{m}\left(\theta,\varphi\right)
\quad \hbox{where} \quad 
f_{m}^{l}=\int_{\Omega}f(\theta,\varphi)\left(Y_{l}^{m}\left(\theta,\varphi\right)\right)^{*}{\rm d}\Omega.
\label{exp}
\end{equation}

\subsubsection{Special case  of axisymmetric spherical harmonics}
In that case we have:
\begin{equation}
Y_{l}^{0}\left(\theta,\varphi\right)=\mathcal{N}_{l}^{0}P_{l}\left(\cos\theta\right)\hbox{ and }\partial_{\theta}Y_{l}^{0}\left(\theta,\varphi\right)=\mathcal{N}_{l}^{0}\partial_{\theta}P_{l}\left(\cos\theta\right)=\left\{
\begin{array}{l@{\quad}l}
0\hbox{ for }l=0,\\
-\mathcal{N}_{l}^{0}P_{l}^{1}\left(\cos\theta\right)\hbox{ for }l>0.
\end{array}
\right.
\label{axi}
\end{equation}

\subsubsection{Linear differential and recursion relations}
We recall that the spherical harmonics obey to the differential equation:
\begin{equation}
\frac{1}{\sin\theta}\partial_{\theta}\left(\sin\theta\partial_{\theta}Y_{l}^{m}\left(\theta,\varphi\right)\right)+\frac{1}{\sin^{2}\theta}\partial_{\varphi}^{2}Y_{l}^{m}\left(\theta,\varphi\right)=-l\left(l+1\right)Y_{l}^{m}\left(\theta,\varphi\right) .
\label{ode}
\end{equation}
In deriving  the functions which are related to the centrifugal force in \S5 (cf. (\ref{baro2}-\ref{al}-\ref{bl})) we have used the following recursion relations for $m=0$:
\begin{equation}
\cos\theta Y_{l}^{0}(\theta)=A_{l}^{0}Y_{l-1}^{0}(\theta)+B_{l}^{0}Y_{l+1}^{0}(\theta)
\quad \hbox{where} \;
A_{l}^{0}=\frac{l}{\sqrt{(2l+1)(2l-1)}} \quad \hbox{and} \; B_{l}^{0}=\frac{(l+1)}{\sqrt{(2l+3)(2l+1)}} ,
\label{r1}
\end{equation}
\begin{equation}
\sin\theta Y_{l}^{0}(\theta)=C_{l}^{0}\partial_{\theta}Y_{l-1}^{0}(\theta)-D_{l}^{0}\partial_{\theta}Y_{l+1}^{0}(\theta)
\quad \hbox{where} \; C_{l}^{0}=\frac{1}{\sqrt{(2l+1)(2l-1)}} \quad \hbox{and}  \; D_{l}^{0}=\frac{1}{\sqrt{(2l+3)(2l+1)}} ,
\label{r2}
\end{equation}
\begin{equation}
\cos\theta\partial_{\theta}Y_{l}^{0}(\theta)=E_{l}^{0}\partial_{\theta}Y_{l-1}^{0}(\theta)+F_{l}^{0}\partial_{\theta}Y_{l+1}^{0}(\theta)
\quad \hbox{where} \;
E_{l}^{0}=\frac{l+1}{\sqrt{(2l+1)(2l-1)}} \quad \hbox{and} \; F_{l}^{0}=\frac{l}{\sqrt{(2l+3)(2l+1)}} ,
\label{r3}
\end{equation}
\begin{equation}
\sin\theta\partial_{\theta}Y_{l}^{0}(\theta)=G_{l}^{0}Y_{l+1}^{0}(\theta)-H_{l}^{0}Y_{l-1}^{0}(\theta)
\quad \hbox{where} \;
G_{l}^{0}=\frac{l(l+1)}{\sqrt{(2l+3)(2l+1)}} \quad \hbox{and} \; H_{l}^{0}=\frac{l(l+1)}{\sqrt{(2l+1)(2l-1)}} .
\label{r4}
\end{equation}
The identities (\ref{r2}) and (\ref{r3}) have been deduced from the two others (\ref{r1}-\ref{r4}) with the help of (\ref{ode}). 

\subsubsection{Expansion of products of the spherical harmonics}
Using the normalization and the orthogonality of spherical harmonics (cf. (\ref{ortho}-\ref{exp})) and their complex conjugate (cf. (\ref{conj})), we can write: 
\begin{equation}
Y_{l_{1}}^{m_{1}}\left(\theta,\varphi\right)Y_{l_{2}}^{m_{2}}\left(\theta,\varphi\right)=(-1)^{\left(m_{1}+m_{2}\right)}\sum_{l=|l_{1}-l_{2}|}^{l_{1}+l_{2}}\mathcal{I}_{l_{1},l_{2},l}^{m_{1},m_{2},-\left(m_{1}+m_{2}\right)}Y_{l}^{m_{1}+m_{2}}\left(\theta,\varphi\right)
\end{equation}
where we define the integral $\mathcal{I}_{l_{1},l_{2},l}^{m_{1},m_{2},m}$ like in Edmonds (1968) or Varshalovich and al. (1975):
\begin{equation}
\mathcal{I}_{l_{1},l_{2},l}^{m_{1},m_{2},m}=\int_{\Omega}Y_{l_{1}}^{m_{1}}\left(\theta,\varphi\right)Y_{l_{2}}^{m_{2}}\left(\theta,\varphi\right)Y_{l}^{m}\left(\theta,\varphi\right)d\Omega=\sqrt{\frac{(2l_{1}+1)(2l_{2}+1)(2l+1)}{4\pi}}\left(\begin{array}{ccc}
l_{1} & l_{2} & l \\
m_{1} & m_{2} & m
\end{array}\right)
\left(\begin{array}{ccc}
l_{1} & l_{2} & l\\
0 & 0 & 0
\end{array}\right)
\end{equation}
with the 3j-Wigner coefficients which are related to the classical Clebsch-Gordan coefficients by: 
\begin{equation}
\left(\begin{array}{ccc}
l_{1} & l_{2} & l \\
m_{1} & m_{2} & m 
\end{array}\right)=\frac{(-1)^{l_{1}-l_{2}-m}}{\sqrt{2l+1}}C_{l_{1},m_{1},l_{2},m_{2}}^{l,-m}.
\end{equation}
So finally, we get the following expansion for the product of two spherical harmonics:
\begin{equation}
Y_{l_{1}}^{m_{1}}\left(\theta,\varphi\right)Y_{l_{2}}^{m_{2}}\left(\theta,\varphi\right)=\sum_{l=|l_{1}-l_{2}|}^{l_{1}+l_{2}}c_{l_{1},m_{1},l_{2},m_{2}}^{l}Y_{l}^{m_{1}+m_{2}}\left(\theta,\varphi\right)
\end{equation}
where the coefficient $c_{l_{1},m_{1},l_{2},m_{2}}^{l}$ is given by:
\begin{equation}
c_{l_{1},m_{1},l_{2},m_{2}}^{l}=\left(-1\right)^{\left(m_{1}+m_{2}\right)}\sqrt{\frac{(2l_{1}+1)(2l_{2}+1)(2l+1)}{4\pi}}\left(\begin{array}{ccc}
l_{1} & l_{2} & l \\
m_{1} & m_{2} & -\left(m_{1}+m_{2}\right)
\end{array}\right)
\left(\begin{array}{ccc}
l_{1} & l_{2} & l\\
0 & 0 & 0
\end{array}\right).
\end{equation}
Then, using the initial definition of spherical harmonics (cf. \ref{def}-\ref{norm}), we deduce the expansion for the product of two associated Legendre functions: 
\begin{equation}
P_{l_{1}}^{m_{1}}\left(\cos\theta\right)P_{l_{2}}^{m_{2}}\left(\cos\theta\right)=\sum_{l=|l_{1}-l_{2}|}^{l_{1}+l_{2}}d_{l_{1},m_{1},l_{2},m_{2}}^{l}P_{l}^{m_{1}+m_{2}}\left(\cos\theta\right)
\end{equation}
where
\begin{equation}
d_{l_{1},m_{1},l_{2},m_{2}}^{l}=(-1)^{\left(m_{1}+m_{2}\right)}\left(2l+1\right)\sqrt{\frac{\left(l_{1}+m_{1}\right)!\left(l_{2}+m_{2}\right)!\left(l-\left(m_{1}+m_{2}\right)\right)!}{\left(l_{1}-m_{1}\right)!\left(l_{2}-m_{2}\right)!\left(l+\left(m_{1}+m_{2}\right)\right)!}}
\left(\begin{array}{ccc}
l_{1} & l_{2} & l \\
m_{1} & m_{2} & -\left(m_{1}+m_{2}\right)
\end{array}\right)
\left(\begin{array}{ccc}
l_{1} & l_{2} & l\\
0 & 0 & 0
\end{array}\right).
\label{coupling}
\end{equation}
Note that the previous expression is symmetric; therefore
\begin{equation}
d_{l_{1},m_{1},l_{2},m_{2}}^{l}=d_{l_{2},m_{2},l_{1},m_{1}}^{l}.
\end{equation}

\subsection{Vector fields}

\subsubsection{Definitions and basic properties}
Following Rieutord (1987), we expand any vector field $\vec{u}(r,\theta,\phi)$ in vectorial spherical harmonics as
\begin{equation}
\vec{u}(r,\theta,\phi)=\sum_{l=0}^{\infty}\sum_{m=-l}^{l}\left\{u_{m}^{l}(r)\vec R_{l}^{m}(\theta,\varphi)+v_{m}^{l}(r)\vec S_{l}^{m}(\theta,\varphi)+w_{m}^{l}(r)\vec T_{l}^{m}(\theta,\varphi)\right\},
\label{vec}
\end{equation}
where  the vectorial spherical harmonics $\vec R_{l}^{m}\left(\theta,\varphi\right)$, $\vec S_{l}^{m}\left(\theta,\varphi\right)$, $\vec T_{l}^{m}\left(\theta,\varphi\right)$ are defined as:
\begin{equation}
\vec R_{l}^{m}(\theta,\varphi)=Y_{l}^{m}(\theta,\varphi)\vec{\widehat e}_{r}\hbox{, }\vec S_{l}^{m}(\theta,\varphi)=\vec\nabla_{\mathcal{S}}Y_{l}^{m}(\theta,\varphi)\hbox{ and }\vec T_{l}^{m}(\theta,\varphi)=\vec\nabla_{\mathcal{S}}\wedge\vec R_{l}^{m}(\theta,\varphi) ,
\end{equation}
\begin{equation}
\hbox{with the horizontal gradient} \; \vec\nabla_{\mathcal{S}}=\vec{\widehat e}_{\theta}\partial_{\theta}+\vec{\widehat e}_{\varphi}\frac{1}{\sin\theta}\partial_{\varphi}.
\end{equation}
These vector functions obey the following orthogonality relations:
\begin{equation}
\int_{\Omega}\vec R_{l_{1}}^{m_{1}}\cdot\vec S_{l_{2}}^{m_{2}}d\Omega=\int_{\Omega}\vec R_{l_{1}}^{m_{1}}\cdot\vec T_{l_{2}}^{m_{2}}d\Omega=\int_{\Omega}\vec S_{l_{1}}^{m_{1}}\cdot\vec T_{l_{2}}^{m_{2}}d\Omega=0,
\end{equation}
\begin{equation}
\int_{\Omega}\vec R_{l_{1}}^{m_{1}}\cdot\left(\vec R_{l_{2}}^{m_{2}}\right)^{*}d\Omega=\delta_{l_{1},l_{2}}\delta_{m_{1},m_{2}} \; \hbox{ and } \; \int_{\Omega}\vec S_{l_{1}}^{m_{1}}\cdot\left(\vec S_{l_{2}}^{m_{2}}\right)^{*}d\Omega=\int_{\Omega}\vec T_{l_{1}}^{m_{1}}\cdot\left(\vec T_{l_{2}}^{m_{2}}\right)^{*}d\Omega=l_{1}(l_{1}+1)\delta_{l_{1},l_{2}}\delta_{m_{1},m_{2}}.
\end{equation}
The vector function $\vec u$ may also be projected on the classical spherical vectorial basis:
\begin{equation}
\vec u=\sum_{l=0}^{\infty}\sum_{m=-l}^{l}\left\{u_{m}^{l}(r)Y_{l}^{m}(\theta,\varphi)\vec{\widehat e}_{r}
+\left[v_{m}^{l}(r)\partial_{\theta}Y_{l}^{m}(\theta,\varphi)+w_{m}^{l}(r)\frac{im}{\sin\theta}Y_{l}^{m}(\theta,\varphi)\right]\vec{\widehat e}_{\theta}
+\left[v_{m}^{l}(r)\frac{im}{\sin\theta}Y_{l}^{m}(\theta,\varphi) -w_{m}^{l}(r)\partial_{\theta}Y_{l}^{m}(\theta,\varphi)\right]\vec{\widehat e}_{\varphi}\right\} ,
\end{equation}
$\vec{\widehat e}_{r}$, $\vec{\widehat e}_{\theta}$ and $\vec{\widehat e}_{\varphi}$ beeing the unit-vectors respectively in the $r$, $\theta$ and $\varphi$ directions.
These expansions of  vector fields allow us to separate explicitly the angular variables $\theta$ and $\varphi$  in the vectorial partial differential equations which govern the problem. We thus reduce the problem to solve partial differential equations in $t\hbox{ and }r$ only, which are easy to implement in existing stellar structure codes. Note that  $\vec R_{l}^{m}\left(\theta,\varphi\right)$ and $\vec S_{l}^{m}\left(\theta,\varphi\right)$ represent the poloidal part, and $\vec T_{l}^{m}\left(\theta,\varphi\right)$ the toroidal part of $\vec u$.
  
\subsubsection{Expansions of differential operators}
As stated in the previous section, the expansion of the vector fields in $\vec R_{l}^{m}\left(\theta,\varphi\right)$, $\vec S_{l}^{m}\left(\theta,\varphi\right)$ and $\vec T_{l}^{m}\left(\theta,\varphi\right)$ allows us to separate the variables in the vectorial partial differential equations which govern the problem. This will prove particularly useful when dealing with magnetic field, and with the non-linear expressions where it is involved. We start by expanding  the classical linear vectorial operators: gradient, divergence, curl and laplacian (scalar or vectorial). \\

\noindent
{\bf Gradient:}

\medskip
Taking a scalar function $f(r,\theta,\varphi)$ expanded in the $Y_{l}^{m}\left(\theta,\varphi\right)$:
\begin{equation}
f(r,\theta,\varphi)=\sum_{l=0}^{\infty}\sum_{m=-l}^{l}f_{m}^{l}(r)Y_{l}^{m}\left(\theta,\varphi\right),
\end{equation}
it is straightforward to derive its gradient:
\begin{equation}
\vec\nabla f(r,\theta,\varphi)=\sum_{l=0}^{\infty}\sum_{m=-l}^{l}\left\{\partial_{r}f_{m}^{l}\vec R_{l}^{m}(\theta,\varphi)+\frac{f_{m}^{l}}{r}\vec S_{l}^{m}(\theta,\varphi)\right\}.
\end{equation}

\noindent{\bf Divergence:}

 \medskip
Taking a vector field $\vec u$ expanded as in (\ref{vec}), its divergence is given by:
\begin{equation}
\vec\nabla\cdot\vec u(r,\theta,\varphi)=\sum_{l=0}^{\infty}\sum_{m=-l}^{l}\left[\frac{1}{r^{2}}\partial_{r}\left(r^2 u_{m}^{l}\right)-l(l+1)\frac{v_{m}^{l}}{r}\right]Y_{l}^{m}(\theta,\varphi),
\end{equation}
where whe note that, if $\vec u$ is divergence-free, such as the magnetic field $\vec B$ or the momentum density $\rho\vec{\mathcal{U}_{M}}$ in the anelastic approximation, we have the following relation between $u_{m}^{l}$ and $v_{m}^{l}$:
\begin{equation}
v_{m}^{l}=\frac{1}{l(l+1)}\frac{1}{r}\partial_{r}\left(r^2 u_{m}^{l}\right).
\label{divfree}
\end{equation}

\noindent{\bf Laplacian of a scalar function:}

\medskip
With these expressions for the gradient of a scalar quantity and for the divergence of a vector field, one can easily derive the laplacian of a scalar function. Using the well-known property $\nabla^{2}f=\vec\nabla\cdot\left(\vec\nabla f\right)$, we get:
\begin{equation}
\nabla^{2}f=\vec\nabla\cdot(\vec\nabla f)=\sum_{l=0}^{\infty}\sum_{m=-l}^{l}\left[\frac{1}{r}\partial_{r^2}\left(r f_{m}^{l}\right)-l(l+1)\frac{f_{m}^{l}}{r^2}\right]Y_{l}^{m}(\theta,\varphi)=\sum_{l=0}^{\infty}\sum_{m=-l}^{l}\Delta_{l}f_{m}^{l}Y_{l}^{m}(\theta,\varphi)
\end{equation}
where $\Delta_{l}$ is the laplacian operator:
\begin{equation}
\Delta_{l}=\partial_{r,r}+\frac{2}{r}\partial_{r}-\frac{l\left(l+1\right)}{r^2}.
\end{equation}

\noindent{\bf Curl:}

\medskip
This operator is found in the expression of the magnetic field in terms of a stream function, in that of the current density, etc. If we take $\vec u$ expanded as in (\ref{vec}), we retrieve Rieutord's  (1987) result: 
\begin{equation}
\vec\nabla\wedge\vec u=\sum_{l=0}^{\infty}\sum_{m=-l}^{l}\left\{\left[l(l+1)\frac{w_{m}^{l}}{r}\right]\vec R_{l}^{m}(\theta,\varphi)+\left[\frac{1}{r}\partial_{r}(r w_{m}^{l})\right]\vec S_{l}^{m}(\theta,\varphi)+\left[\frac{u_{m}^{l}}{r}-\frac{1}{r}\partial_{r}\left(r v_{m}^{l}\right)\right]\vec T_{l}^{m}(\theta,\varphi)\right\}.
\label{curl}
\end{equation}
Taking this result, the expansion for the laplacian of a vector field is derived. First, we have:
\begin{equation}
\vec\nabla\wedge\left(\vec\nabla\wedge\vec u\right)=\sum_{l=0}^{\infty}\sum_{m=-l}^{l}\left\{\left[\frac{l(l+1)}{r}\left(\frac{u_{m}^{l}}{r}-\frac{1}{r}\partial_{r}\left(r v_{m}^{l}\right)\right)\right]\vec R_{l}^{m}(\theta,\varphi)+\left[\frac{1}{r}\partial_{r}u_{m}^{l}-\frac{1}{r}\partial_{r,r}(r v_{m}^{l})\right]\vec S_{l}^{m}(\theta,\varphi)+\left[-\Delta_{l}w_{m}^{l}\right]\vec T_{l}^{m}(\theta,\varphi)\right\}
\label{curl2p}
\end{equation}
and therefore
\begin{eqnarray}
\lefteqn{\nabla^{2}\vec u=\vec\nabla\left(\vec\nabla\cdot\vec u\right)-\vec\nabla\wedge\left(\vec\nabla\wedge\vec u\right)}\nonumber\\
&=&\sum_{l=0}^{\infty}\sum_{m=-l}^{l}\left\{\left[\Delta_{l}u_{m}^{l}-\frac{2}{r^2}\left(u_{m}^{l}-l(l+1)v_{m}^{l}\right)\right]\vec R_{l}^{m}(\theta,\varphi)+\left[\Delta_{l}v_{m}^{l}+2\frac{u_{m}^{l}}{r^2}\right]\vec S_{l}^{m}(\theta,\varphi)+\left[\Delta_{l}w_{m}^{l}\right]\vec T_{l}^{m}(\theta,\varphi)\right\},
\label{curl2}
\end{eqnarray}
result which becomes in the case where $\vec u$ is divergence-free (cf. (\ref{divfree}) ):
\begin{eqnarray}
\lefteqn{\nabla^{2}\vec u=-\vec\nabla\wedge\left(\vec\nabla\wedge\vec u\right)}\nonumber\\
&=&\sum_{l=0}^{\infty}\sum_{m=-l}^{l}\left\{\left[\frac{1}{r}\Delta_{l}(ru_{m}^{l})\right]\vec R_{l}^{m}(\theta,\varphi)+\left[\frac{1}{r}\partial_{r}\left(r\frac{\Delta_{l}(r u_{m}^{l})}{l(l+1)}\right)\right]\vec S_{l}^{m}(\theta,\varphi)+\left[\Delta_{l}w_{m}^{l}\right]\vec T_{l}^{m}(\theta,\varphi)\right\}.
\label{curl2b}
\end{eqnarray}
Finally, using (\ref{curl}) once again, we get the `triple curl' operator:
\begin{equation}
\left(\vec\nabla\wedge\right)^{3}\vec u=\sum_{l=0}^{\infty}\sum_{m=-l}^{l}\left\{\left[-l(l+1)\frac{\Delta_{l}w_{m}^{l}}{r}\right]\vec R_{l}^{m}(\theta,\varphi)+\left[-\frac{1}{r}\partial_{r}\left(r\Delta_{l}w_{m}^{l}\right)\right]\vec S_{l}^{m}(\theta,\varphi)+\left[\Delta_{l}z_{m}^{l}\right]\vec T_{l}^{m}(\theta,\varphi)\right\}
\label{curl3}
\end{equation}
where:
\begin{equation}
z_{m}^{l}=\frac{1}{r}\partial_{r}(r v_{m}^{l})-\frac{u_{m}^{l}}{r},
\end{equation}
which becomes in the case where $\vec u$ is divergence-free (cf. (\ref{divfree}) ):
\begin{eqnarray}
\left(\vec\nabla\wedge\right)^{3}\vec u=\sum_{l=0}^{\infty}\sum_{m=-l}^{l}\left\{\left[-l(l+1)\frac{\Delta_{l}w_{m}^{l}}{r}\right]\vec R_{l}^{m}(\theta,\varphi)\left[-\frac{1}{r}\partial_{r}\left(r\Delta_{l}w_{m}^{l}\right)\right]\vec S_{l}^{m}(\theta,\varphi)+\left[\frac{1}{l(l+1)}\Delta_{l}\Delta_{l}(r u_{m}^{l})\right]\vec T_{l}^{m}(\theta,\varphi)\right\}.\nonumber\\
\label{curl3b}
\end{eqnarray}

\subsubsection{Products of axisymmetric vectorial spherical harmonics}
\label{products}
Before we turn to the non-linear terms involving the magnetic field: the advection term in the induction equation and the Lorentz force, we shall first perform the projections of  the scalar product of two general axisymmetric vectors and of their vector product, and derive the associated coupling coefficients.\\

\noindent{\bf Scalar product:}

\medskip
We take two general axisymmetric vectors $\vec X_{1}\left(r,\theta\right)$ and $\vec X_{2}\left(r,\theta\right)$, which we expand as in (\ref{vec}):
\begin{equation}
\left\{
\begin{array}{l@{\quad}l}
\vec X_{1}(r,\theta)=\sum_{l_{1}=0}^{\infty}\left\{\mathcal{A}_{0}^{l_{1}}(r)\vec R_{l_{1}}^{0}\left(\theta\right)+\mathcal{B}_{0}^{l_{1}}(r)\vec S_{l_{1}}^{0}\left(\theta\right)+\mathcal{C}_{0}^{l_{1}}(r)\vec T_{l_{1}}^{0}\left(\theta\right)\right\}\\
\vec X_{2}(r,\theta)=\sum_{l_{2}=0}^{\infty}\left\{\mathcal{D}_{0}^{l_{2}}(r)\vec R_{l_{2}}^{0}\left(\theta\right)+\mathcal{E}_{0}^{l_{2}}(r)\vec S_{l_{2}}^{0}\left(\theta\right)+\mathcal{F}_{0}^{l_{2}}(r)\vec T_{l_{2}}^{0}\left(\theta\right)\right\},
\end{array}
\label{2axi}
\right.
\end{equation}
and perform their scalar product.  After some algebra involving the expansion of products of spherical harmonics (cf. \S A.1.4.), we obtain
\begin{equation}
\vec X_{1}\left(r,\theta\right)\cdot\vec X_{2}\left(r,\theta\right)=\sum_{l=0}^{\infty}\mathcal{P}_{\left(\vec X_{1}\cdot\vec X_{2}\right);l}\left(r\right)P_{l}\left(\cos\theta\right)
\label{sp1}
\end{equation}
with the following expression for $\mathcal{P}_{\left(\vec X_{1}\cdot\vec X_{2}\right);l}(r)$:
\begin{eqnarray}
\mathcal{P}_{\left(\vec X_{1}\cdot\vec X_{2}\right);l}(r)&=&\sum_{l_{1}=0}^{\infty}\sum_{l_{2}=0}^{\infty}\mathcal{N}_{l_{1}}^{0}\mathcal{N}_{l_{2}}^{0}\left\{\left[\mathcal{A}_{0}^{l_{1}}\left(r\right)\mathcal{D}_{0}^{l_{2}}\left(r\right)\right]\sum_{j=I\left(l_{1},0,l_{2},0\right)}^{l_{1}+l_{2}}d_{l_{1},0,l_{2},0}^{j}\delta_{l,j}+\frac{2}{3}\left[\mathcal{B}_{0}^{l_{1}}\left(r\right)\mathcal{E}_{0}^{l_{2}}\left(r\right)+\mathcal{C}_{0}^{l_{1}}\left(r\right)\mathcal{F}_{0}^{l_{2}}\left(r\right)\right]X_{l_{1},l_{2}}^{l}\right\} .
\label{sp2}
\end{eqnarray}
Here $X_{l_{1},l_{2}}^{l}$is defined in terms of the coupling coefficients $d_{l_{1},m_{1},l_{2},m_{2}}^{l}$ (\ref{coupling}):  
\begin{equation}
X_{l_{1},l_{2}}^{l}=\sum_{j=I(l_{1},1,l_{2},1)}^{l_{1}+l_{2}}\left(d_{l_{1},1,l_{2},1}^{j}\sum_{p=0}^{E\left[\frac{j-1}{2}\right]}\left[\left(2j-4p-1\right)\cdot X\right]\right)
\label{xm}
\end{equation}
with
\begin{equation}
X=\sum_{q=0}^{E\left[\frac{\left(j-2p-1\right)-1}{2}\right]}\left\{\left[2\left(j-2p-1\right)-4q-1\right]\cdot\left(\delta_{l,\left[(j-2p-1)-2q-1\right]}-\sum_{r=I(2,0,\left[(j-2p-1)-2q-1\right],0)}^{\left[(j-2p-1)-2q-1\right]+2}\left[d_{2,0,\left[(j-2p-1)-2q-1\right],0}^{r}\delta_{l,r}\right]\right)\right\} .
\end{equation}
We have used the classical notation $\delta_{i,j}$ for the usual Kronecker symbol, $E\left[x\right]$ is the integer part of $x$ and $I(l_{1},m_{1},l_{2},m_{2})=\max\left(|l_{1}-l_{2}|,m_{1}+m_{2}\right)$. \\

\noindent{\bf Vector product:}

\medskip
We operate likewise for the vector product of two general axisymmetric vectors $\vec X_{1}\left(r,\theta\right)$ and $\vec X_{2}\left(r,\theta\right)$, again expanded as in (\ref{vec}).
We reach the following result:
\begin{eqnarray}
\vec X_{1}(r,\theta)\wedge\vec X_{2}(r,\theta)
&=&\mathcal{X}_{\left(\vec{X}_{1}\wedge\vec{X}_{2}\right);0}(r)\widehat{e}_{r}+\sum_{l=1}^{\infty}\left\{\mathcal{X}_{\left(\vec{X}_{1}\wedge\vec{X}_{2}\right);l}(r)P_{l}\left(\cos\theta\right)\widehat{e}_{r}+\mathcal{Y}_{\left(\vec{X}_{1}\wedge\vec{X}_{2}\right);l}(r)P_{l}^{1}\left(\cos\theta\right)\widehat{e}_{\theta}+\mathcal{Z}_{\left(\vec{X}_{1}\wedge\vec{X}_{2}\right);l}(r)P_{l}^{1}\left(\cos\theta\right)\widehat{e}_{\varphi}\right\}\nonumber\\
&=&\sum_{l=0}^{\infty}\left\{\left[\frac{\mathcal{X}_{\left(\vec{X}_{1}\wedge\vec{X}_{2}\right);l}(r)}{\mathcal{N}_{l}^{0}}\right]\vec R_{l}^{0}\left(\theta\right)+\left[\frac{-\mathcal{Y}_{\left(\vec{X}_{1}\wedge\vec{X}_{2}\right);l}(r)}{\mathcal{N}_{l}^{0}}\right]\vec S_{l}^{0}\left(\theta\right)+\left[\frac{\mathcal{Z}_{\left(\vec{X}_{1}\wedge\vec{X}_{2}\right);l}(r)}{\mathcal{N}_{l}^{0}}\right]\vec{T}_{l}^{0}\left(\theta\right)\right\} ,
\label{vp}
\end{eqnarray}
where we have also used the following property of the Legendre function (Gradshteyn \& Ryzhik 1967):
\begin{equation}
\frac{{\rm d}P_{l}(\mu)}{{\rm d}\mu}=\sum_{k=0}^{E\left(\frac{l-1}{2}\right)}\left(2l-4k-1\right)P_{l-2k-1}(\mu).
\label{Ryzhik}
\end{equation}
The radial function $\mathcal{X}_{\left(\vec{X}_{1}\wedge\vec{X}_{2}\right);l}\left(r\right)$ is given by:
\begin{equation}
\mathcal{X}_{\left(\vec{X}_{1}\wedge\vec{X}_{2}\right);l}(r)=\frac{2}{3}\sum_{l_{1}=0}^{\infty}\sum_{l_{2}=0}^{\infty}\mathcal{N}_{l_{1}}^{0}\mathcal{N}_{l_{2}}^{0}\left\{\left[\mathcal{C}_{0}^{l_{1}}(r)\mathcal{E}_{0}^{l_{2}}(r)-\mathcal{B}_{0}^{l_{1}}(r)\mathcal{F}_{0}^{l_{2}}(r)\right]X_{l_{1},l_{2}}^{l}\right\}\label{cp1} ,
\label{xl}
\end{equation}
with the same coefficient $X_{l_{1},l_{2}}^{l}$ derived above in (\ref{xm}).
The two other radial functions $\mathcal{Y}_{\left(\vec{X}_{1}\wedge\vec{X}_{2}\right);l}\left(r\right)$ and $\mathcal{Z}_{\left(\vec{X}_{1}\wedge\vec{X}_{2}\right);l}\left(r\right)$ are given by
\begin{equation}
\mathcal{Y}_{\left(\vec{X}_{1}\wedge\vec{X}_{2}\right);l}(r)=\sum_{l_{1}=0}^{\infty}\sum_{l_{2}=0}^{\infty}\mathcal{N}_{l_{1}}^{0}\mathcal{N}_{l_{2}}^{0}\left\{\left[\mathcal{C}_{0}^{l_{1}}(r)\mathcal{D}_{0}^{l_{2}}(r)\right]\sum_{j=I(l_{1},1,l_{2},0)}^{l_{1}+l_{2}}d_{l_{1},1,l_{2},0}^{j}\delta_{l,j}-\left[\mathcal{A}_{0}^{l_{1}}(r)\mathcal{F}_{0}^{l_{2}}(r)\right]\sum_{j=I(l_{1},0,l_{2},1)}^{l_{1}+l_{2}}d_{l_{1},0,l_{2},1}^{j}\delta_{l,j}\right\}
\label{yl}
\end{equation}
and
\begin{equation}
\mathcal{Z}_{\left(\vec{X}_{1}\wedge\vec{X}_{2}\right);l}(r)=\sum_{l_{1}=0}^{\infty}\sum_{l_{2}=0}^{\infty}\mathcal{N}_{l_{1}}^{0}\mathcal{N}_{l_{2}}^{0}\left\{\left[\mathcal{B}_{0}^{l_{1}}(r)\mathcal{D}_{0}^{l_{2}}(r)\right]\sum_{j=I(l_{1},1,l_{2},0)}^{l_{1}+l_{2}}d_{l_{1},1,l_{2},0}^{j}\delta_{l,j}-\left[\mathcal{A}_{0}^{l_{1}}(r)\mathcal{E}_{0}^{l_{2}}(r)\right]\sum_{j=I(l_{1},0,l_{2},1)}^{l_{1}+l_{2}}d_{l_{1},0,l_{2},1}^{j}\delta_{l,j}\right\}.
\label{zl}
\end{equation}

\section{Lorentz force}
In this section we shall project the Lorentz force, 
\begin{equation}
\vec{\mathcal{F}}\!\!_{\mathcal{L}}=\vec j\wedge\vec B=\left[\frac{1}{\mu_{0}}(\vec\nabla\wedge\vec B)\right]\wedge\vec B ,
\label{flap}
\end{equation}
on spherical vectorial harmonics, and write the result in terms of the poloidal and the toroidal magnetic stream-functions, respectively $\xi_{0}^{l}$ and $\chi_{0}^{l}$. We have just seen in \S A.2.3. how to expand  the vector product of two axisymmetric vectors in the $\vec R_{l}^{0}(\theta)$, $\vec S_{l}^{0}(\theta)$, $\vec T_{l}^{0}(\theta)$. We apply the method to (\ref{flap}), with (cf. (\ref{jpol}), (\ref{jtor}))
\begin{eqnarray}
\vec X_{1}\left(r,\theta\right)&=&\vec j\left(r,\theta\right)=\frac{1}{\mu_{0}}\left(\vec\nabla\wedge\vec B\left(r,\theta\right)\right)=\sum_{l_{1}=1}^{\infty}\left\{\mathcal{A}_{0}^{l_{1}}(r)\vec R_{l_{1}}^{0}(\theta)+\mathcal{B}_{0}^{l_{1}}(r)\vec S_{l_{1}}^{0}(\theta)+\mathcal{C}_{0}^{l_{1}}(r)\vec T_{l_{1}}^{0}(\theta)\right\}\nonumber\\&=&\frac{1}{\mu_{0}}\sum_{l_{1}=1}^{\infty}\left\{\left[l_{1}(l_{1}+1)\frac{\chi_{0}^{l_{1}}}{r^2}\right]\vec R_{l_{1}}^{0}(\theta)+\left[\frac{1}{r}\partial_{r}\chi_{0}^{l_{1}}\right]\vec S_{l_{1}}^{0}(\theta)+\left[-\Delta_{l_{1}}\left(\frac{\xi_{0}^{l_{1}}}{r}\right)\right]\vec T_{l_{1}}^{0}(\theta)\right\}\nonumber\\
\end{eqnarray}
and   (cf. (\ref{expB}))
\begin{eqnarray}
\vec X_{2}\left(r,\theta\right)&=&\vec B\left(r,\theta\right)=\sum_{l_{2}=1}^{\infty}\left\{\mathcal{D}_{0}^{l_{2}}(r)\vec R_{l_{2}}^{0}(\theta)+\mathcal{E}_{0}^{l_{2}}(r)\vec S_{l_{2}}^{0}(\theta)+ \mathcal{F}_{0}^{l_{2}}(r)\vec T_{l_{2}}^{0}(\theta)\right\}\nonumber\\ 
&=&\sum_{l_{2}=1}^{\infty}\left\{\left[l_{2}(l_{2}+1)\frac{\xi_{0}^{l_{2}}}{r^2}\right]\vec R_{l_{2}}^{0}(\theta)+\left[\frac{1}{r}\partial_{r}\xi_{0}^{l_{2}}\right]\vec S_{l_{2}}^{0}(\theta)+\left[\frac{\chi_{0}^{l_{2}}}{r}\right]\vec T_{l_{2}}^{0}(\theta)\right\}.\nonumber\\
\end{eqnarray}
So, we have, using the definition of \S A.2.3.:
\begin{equation}
\left\{
\begin{array}{l@{\quad}l}
\mathcal{A}_{0}^{l_{1}}=\frac{1}{\mu_{0}}\left[l_{1}(l_{1}+1)\frac{\chi_{0}^{l_{1}}}{r^2}\right]\\
\mathcal{B}_{0}^{l_{1}}=\frac{1}{\mu_{0}}\left[\frac{1}{r}\partial_{r}\chi_{0}^{l_{1}}\right]\\
\mathcal{C}_{0}^{l_{1}}=\frac{1}{\mu_{0}}\left[-\Delta_{l_{1}}\left(\frac{\xi_{0}^{l_{1}}}{r}\right)\right]
\end{array}
\right.\hbox{and}\quad\left\{
\begin{array}{l@{\quad}l}
\mathcal{D}_{0}^{l_{2}}=\left[l_{2}(l_{2}+1)\frac{\xi_{0}^{l_{2}}}{r^2}\right]\\
\mathcal{E}_{0}^{l_{2}}=\left[\frac{1}{r}\partial_{r}\xi_{0}^{l_{2}}\right]\\
\mathcal{F}_{0}^{l_{2}}=\left[\frac{\chi_{0}^{l_{2}}}{r}\right].
\end{array}
\right.
\end{equation}
Therefore, we get using (\ref{vp}-\ref{xl}-\ref{yl}-\ref{zl}): 
\begin{eqnarray}
\lefteqn{\vec{\mathcal{F}}\!\!_{\mathcal{L}}(r,\theta)=\vec X_{1}(r,\theta)\wedge\vec X_{2}(r,\theta)}\nonumber\\
&=&\mathcal{X}_{\vec{\mathcal{F}}\!\!_{\mathcal{L}};0}(r)\widehat{e}_{r}+\sum_{l=1}^{\infty}\left\{\mathcal{X}_{\vec{\mathcal{F}}\!\!_{\mathcal{L}};l}(r)P_{l}\left(\cos\theta\right)\widehat{e}_{r}+\mathcal{Y}_{\vec{\mathcal{F}}\!\!_{\mathcal{L}};l}(r)P_{l}^{1}\left(\cos\theta\right)\widehat{e}_{\theta}+\mathcal{Z}_{\vec{\mathcal{F}}\!\!_{\mathcal{L}};l}(r)P_{l}^{1}\left(\cos\theta\right)\widehat{e}_{\varphi}\right\}\nonumber\\
&=&\left[\frac{\mathcal{X}_{{\vec {\mathcal F}}_{\mathcal{L}};0}}{\mathcal{N}_{0}^{0}}\right]\vec R_{0}^{0}(\theta)+\sum_{l=1}^{\infty}\left\{\left[\frac{\mathcal{X}_{\vec{\mathcal{F}}\!\!_{\mathcal{L}};l}(r)}{\mathcal{N}_{l}^{0}}\right]\vec R_{l}^{0}\left(\theta\right)+\left[\frac{-\mathcal{Y}_{\vec{\mathcal{F}}\!\!_{\mathcal{L}};l}(r)}{\mathcal{N}_{l}^{0}}\right]\vec S_{l}^{0}\left(\theta\right)+\left[\frac{\mathcal{Z}_{\vec{\mathcal{F}}\!\!_{\mathcal{L}};l}(r)}{\mathcal{N}_{l}^{0}}\right]\vec{T}_{l}^{0}\left(\theta\right)\right\},
\end{eqnarray}
where the radial functions $\mathcal{X}_{\vec{\mathcal{F}}\!\!_{\mathcal{L}};l}\left(r\right)$, $\mathcal{Y}_{\vec{\mathcal{F}}\!\!_{\mathcal{L}};l}\left(r\right)$ and $\mathcal{Z}_{\vec{\mathcal{F}}\!\!_{\mathcal{L}};l}\left(r\right)$ are:
\begin{equation}
\mathcal{X}_{\vec{\mathcal{F}}\!\!_{\mathcal{L}};l}=\frac{2}{3}\frac{1}{\mu_{0}}\sum_{l_{1}=1}^{\infty}\sum_{l_{2}=1}^{\infty}\mathcal{N}_{l_{1}}^{0}\mathcal{N}_{l_{2}}^{0}\left\{\left(\left[-\Delta_{l_{1}}\left(\frac{\xi_{0}^{l_{1}}}{r}\right)\right]\left[\frac{1}{r}\partial_{r}\xi_{0}^{l_{2}}\right]-\left[\frac{1}{r}\partial_{r}\chi_{0}^{l_{1}}\right]\left[\frac{\chi_{0}^{l_{2}}}{r}\right]\right)X_{l_{1},l_{2}}^{l}\right\}
\end{equation}
\begin{equation}
\mathcal{Y}_{\vec{\mathcal{F}}\!\!_{\mathcal{L}};l}=\frac{1}{\mu_{0}}\sum_{l_{1}=1}^{\infty}\sum_{l_{2}=1}^{\infty}\mathcal{N}_{l_{1}}^{0}\mathcal{N}_{l_{2}}^{0}\left\{\left(\left[-\Delta_{l_{1}}\left(\frac{\xi_{0}^{l_{1}}}{r}\right)\right]\left[l_{2}(l_{2}+1)\frac{\xi_{0}^{l_{2}}}{r^2}\right]\right)\sum_{j=I(l_{1},1,l_{2},0)}^{l_{1}+l_{2}}d_{l_{1},1,l_{2},0}^{j}\delta_{l,j}-\left(\left[l_{1}(l_{1}+1)\frac{\chi_{0}^{l_{1}}}{r^2}\right]\left[\frac{\chi_{0}^{l_{2}}}{r}\right]\right)\sum_{j=I(l_{1},0,l_{2},1)}^{l_{1}+l_{2}}d_{l_{1},0,l_{2},1}^{j}\delta_{l,j}\right\}
\end{equation}
\begin{equation}
\mathcal{Z}_{\vec{\mathcal{F}}\!\!_{\mathcal{L}};l}=\frac{1}{\mu_{0}}\sum_{l_{1}=1}^{\infty}\sum_{l_{2}=1}^{\infty}\mathcal{N}_{l_{1}}^{0}\mathcal{N}_{l_{2}}^{0}\left\{\left(\left[\frac{1}{r}\partial_{r}\chi_{0}^{l_{1}}\right]\left[l_{2}(l_{2}+1)\frac{\xi_{0}^{l_{2}}}{r^2}\right]\right)\sum_{j=I(l_{1},1,l_{2},0)}^{l_{1}+l_{2}}d_{l_{1},1,l_{2},0}^{j}\delta_{l,j}-\left(\left[l_{1}(l_{1}+1)\frac{\chi_{0}^{l_{1}}}{r^2}\right]\left[\frac{1}{r}\partial_{r}\xi_{0}^{l_{2}}\right]\right)\sum_{j=I(l_{1},0,l_{2},1)}^{l_{1}+l_{2}}d_{l_{1},0,l_{2},1}^{j}\delta_{l,j}\right\}.
\end{equation}
Explicit values for $X_{l_{1},l_{2}}^{l}$ (\ref{xm}) are given in Table~\ref{table1}, and in Table~\ref{table2} for the coupling coefficients $d_{l_{1},1,l_{2},0}^{l}$.

\begin{table}[!]
\caption{Values of the $X_{l_{1},l_{2}}^{l}$ coefficient  involved in the $r$-component of the Lorentz force and in the Ohmic heating}          
\centering          
\fbox{\begin{tabular}{ c c c c c c c c c l l}  
\hline      
\multicolumn{4}{c}{} & \vline &\multicolumn{4}{c}{}\\                    
\multicolumn{4}{c}{$X_{l_{1},l_{2}}^{2}$} & \vline &\multicolumn{4}{c}{$X_{l_{1},l_{2}}^{4}$}\\ 
\multicolumn{4}{c}{} & \vline &\multicolumn{4}{c}{}\\  
\hline
 & & & & \vline & & & &\\                
  $l_{1} \; \diagdown \; l_{2}$ & 1 & 2 & 3 & \vline &  $l_{1} \; \diagdown \; l_{2}$ & 1 & 2 & 3\\  
 & & & &\vline & & & & \\   
 1 & $-1$ & 0 & $\frac{26}{7}$ & \vline & 1 & 0 & 0 & $-\frac{18}{35}$\\
 & & & & \vline & & & &\\             
 2 & 0 & $\frac{93}{35}$ & 0 & \vline & 2 & 0 & $-\frac{108}{175}$ & 0\\
 & & & & \vline & & & &\\     
 3 & $\frac{26}{7}$ & 0 & $\frac{478}{49}$ & \vline & 3 & $-\frac{18}{35}$ & 0 & $\frac{28898}{2695}$\\
 & & & & \vline & & & &\\      
\hline                  
\end{tabular}}
\label{table1}
\end{table}

\begin{table}[!]
\caption{Values of the $d_{l_{1},1,l_{2},0}^{l}$ coefficients involved in the $\theta-$ and in the $\varphi-$ components of the Lorentz force and in the Ohmic heating}           
\centering          
\fbox{\begin{tabular}{c c c c c c c c c c c cl l l}  
\hline      
\multicolumn{4}{c}{} & \vline &\multicolumn{4}{c}{} & \vline &\multicolumn{4}{c}{}\\                    
\multicolumn{4}{c}{$d_{l_{1},1,l_{2},0}^{1}$} & \vline &\multicolumn{4}{c}{$d_{l_{1},1,l_{2},0}^{2}$}  & \vline &\multicolumn{4}{c}{$d_{l_{1},1,l_{2},0}^{3}$}\\ 
\multicolumn{4}{c}{} & \vline &\multicolumn{4}{c}{} & \vline &\multicolumn{4}{c}{}\\  
\hline    
 & & & & \vline & & & & & \vline & & & &\\                
  $l_{1} \; \diagdown \; l_{2}$ & 1 & 2 & 3 & \vline &  $l_{1} \; \diagdown \; l_{2}$ & 1 & 2 & 3 & \vline &  $l_{1} \; \diagdown \; l_{2}$ & 1 & 2 & 3\\  
 & & & & \vline & & & & & \vline & & & &\\   
 1 & 0 & $-\frac{1}{5}$ & 0 & \vline & 1 & $\frac{1}{3}$ & 0 & $-\frac{1}{7}$ & \vline & 1 & 0 & $\frac{1}{5}$ & 0\\
 & & & & \vline & & & & & \vline & & & &\\             
 2 & $\frac{3}{5}$ & 0 & $-\frac{9}{35}$ & \vline & 2 & 0 & $\frac{1}{7}$ & 0 & \vline & 2 & $\frac{2}{5}$ & 0 & $\frac{1}{15}$\\
 & & & & \vline & & & & & \vline & & & &\\     
 3 & 0 & $\frac{18}{35}$ & 0 & \vline & 3 & $\frac{4}{7}$ & 0 & $\frac{2}{21}$ & \vline & 3 & 0 & $\frac{1}{5}$ & 0\\
 & & & & \vline & & & & & \vline & & & &\\      
\hline\hline 
\multicolumn{4}{c}{} & \vline &\multicolumn{4}{c}{} & \vline &\multicolumn{4}{c}{}\\                    
\multicolumn{4}{c}{$d_{l_{1},1,l_{2},0}^{4}$} & \vline &\multicolumn{4}{c}{$d_{l_{1},1,l_{2},0}^{5}$}  & \vline &\multicolumn{4}{c}{$d_{l_{1},1,l_{2},0}^{6}$}\\ 
\multicolumn{4}{c}{} & \vline &\multicolumn{4}{c}{} & \vline &\multicolumn{4}{c}{}\\  
\hline    
 & & & & \vline & & & & & \vline & & & &\\                
  $l_{1} \; \diagdown \; l_{2}$ & 1 & 2 & 3 & \vline &  $l_{1} \; \diagdown \; l_{2}$ & 1 & 2 & 3 & \vline &  $l_{1} \; \diagdown \; l_{2}$ & 1 & 2 & 3\\  
 & & & & \vline & & & & & \vline & & & &\\   
 1 & 0 & 0 & $\frac{1}{7}$ & \vline & 1 & 0 & 0 & 0 & \vline & 1 & 0 & 0 & 0\\
 & & & & \vline & & & & & \vline & & & &\\             
 2 & 0 & $\frac{9}{35}$ & 0 & \vline & 2 & 0 & 0 & $\frac{4}{21}$ & \vline & 2 & 0 & 0 & 0\\
 & & & & \vline & & & & & \vline & & & &\\     
 3 & $\frac{3}{7}$ & 0 & $\frac{9}{77}$ & \vline & 3 & 0 & $\frac{2}{7}$ & 0 & \vline & 3 & 0 & 0 & $\frac{50}{231}$\\
 & & & & \vline & & & & & \vline & & & &\\     
\hline                  
\end{tabular}}
\label{table2}
\end{table}

\section{Advection term of the induction equation}
We deal likewise with the advection term in the induction equation:
\begin{equation}
\left(\vec{\mathcal{U}}_{\varphi}+\vec{\mathcal{U}}_{M}\right)\wedge\vec B,
\label{advap}
\end{equation}
where $\vec{\mathcal{U}}_{\varphi}(r,\theta)=r\sin\theta\,\Omega(r,\theta)\vec{\widehat e}_{\varphi}$ and $\vec{\mathcal{U}}_M$ represents the meridional flow, whose expansions are found in (\ref{uproject}). Therefore 
\begin{eqnarray}
\vec X_{1}\left(r,\theta\right)&=&\vec{\mathcal{U}}_{\varphi}(r,\theta)+\vec{\mathcal{U}}_{M}(r,\theta)=\sum_{l_{1}=0}^{\infty}\left\{\mathcal{A}_{0}^{l_{1}}(r)\vec R_{l_{1}}^{0}(\theta)+\mathcal{B}_{0}^{l_{1}}(r)\vec S_{l_{1}}^{0}(\theta)+\mathcal{C}_{0}^{l_{1}}(r)\vec T_{l_{1}}^{0}(\theta)\right\}\nonumber\\&=&\sum_{l_{1}=0}^{\infty}\left\{u_{0}^{l_{1}}\left(r\right)\vec R_{l_{1}}^{0}(\theta)+v_{0}^{l_{1}}\left(r\right)\vec S_{l_{1}}^{0}(\theta)+w_{0}^{l_{1}}\left(r\right)\vec T_{l_{1}}^{0}(\theta)\right\}\nonumber\\
\end{eqnarray}
and (cf. (\ref{expB}))
\begin{eqnarray}
\vec X_{2}\left(r,\theta\right)&=&\vec B\left(r,\theta\right)=\sum_{l_{2}=1}^{\infty}\left\{\mathcal{D}_{0}^{l_{2}}(r)\vec R_{l_{2}}^{0}(\theta)+\mathcal{E}_{0}^{l_{2}}(r)\vec S_{l_{2}}^{0}(\theta)+ \mathcal{F}_{0}^{l_{2}}(r)\vec T_{l_{2}}^{0}(\theta)\right\}\nonumber\\ 
&=&\sum_{l_{2}=1}^{\infty}\left\{\left[l_{2}(l_{2}+1)\frac{\xi_{0}^{l_{2}}}{r^2}\right]\vec R_{l_{2}}^{0}(\theta)+\left[\frac{1}{r}\partial_{r}\xi_{0}^{l_{2}}\right]\vec S_{l_{2}}^{0}(\theta)+\left[\frac{\chi_{0}^{l_{2}}}{r}\right]\vec T_{l_{2}}^{0}(\theta)\right\}.\nonumber\\
\end{eqnarray}
So, we have, expliciting the velocity field:
\begin{equation}
\left\{
\begin{array}{l@{\quad}l}
\mathcal{A}_{0}^{l_{1}}=u_{0}^{l_{1}}=\frac{U_{l_{1}}}{\mathcal{N}_{l_{1}}^{0}}\\
\mathcal{B}_{0}^{l_{1}}=v_{0}^{l_{1}}=\frac{V_{l_{1}}}{\mathcal{N}_{l_{1}}^{0}}\\
\mathcal{C}_{0}^{l_{1}}=w_{0}^{l_{1}}=r\left[\frac{D_{l_{1}-1}^{0}}{\mathcal{N}_{l_{1}-1}^{0}}\Omega_{l_{1}-1}^{*}-\frac{C_{l_{1}+1}^{0}}{\mathcal{N}_{l_{1}+1}^{0}}\Omega_{l_{1}+1}^{*}\right]
\end{array}
\right.\hbox{where }\left\{
\begin{array}{l@{\quad}l}
\Omega_{0}^{*}\left(r\right)=\Omega_{0}\left(r\right)+\frac{1}{5}\Omega_{2}\left(r\right)\\
\Omega_{l}^{*}\left(r\right)=\Omega_{l}\left(r\right)\hbox{ for }l>0
\end{array}\right.
\hbox{and }\left\{
\begin{array}{l@{\quad}l}
\mathcal{D}_{0}^{l_{2}}=\left[l_{2}(l_{2}+1)\frac{\xi_{0}^{l_{2}}}{r^2}\right]\\
\mathcal{E}_{0}^{l_{2}}=\left[\frac{1}{r}\partial_{r}\xi_{0}^{l_{2}}\right]\\
\mathcal{F}_{0}^{l_{2}}=\left[\frac{\chi_{0}^{l_{2}}}{r}\right].
\end{array}
\right.
\end{equation}
Hence we get, using (\ref{vp}-\ref{xl}-\ref{yl}-\ref{zl}): 
\begin{eqnarray}
\lefteqn{\left(\vec{\mathcal{U}}_{\varphi}(r,\theta)+\vec{\mathcal{U}}_{M}(r,\theta)\right)\wedge\vec B(r,\theta)=\vec X_{1}(r,\theta)\wedge\vec X_{2}(r,\theta)}\nonumber\\
&=&\sum_{l=1}^{\infty}\left\{\mathcal{X}_{\rm{\bf Ad};l}(r)P_{l}\left(\cos\theta\right)\widehat{e}_{r}+\mathcal{Y}_{\rm{\bf Ad};l}(r)P_{l}^{1}\left(\cos\theta\right)\widehat{e}_{\theta}+\mathcal{Z}_{\rm{\bf Ad};l}(r)P_{l}^{1}\left(\cos\theta\right)\widehat{e}_{\varphi}\right\}\nonumber\\
&=&\sum_{l=1}^{\infty}\left\{\left[\frac{\mathcal{X}_{\rm{\bf Ad};l}(r)}{\mathcal{N}_{l}^{0}}\right]\vec R_{l}^{0}\left(\theta\right)+\left[\frac{-\mathcal{Y}_{\rm{\bf Ad};l}(r)}{\mathcal{N}_{l}^{0}}\right]\vec S_{l}^{0}\left(\theta\right)+\left[\frac{\mathcal{Z}_{\rm{\bf Ad};l}(r)}{\mathcal{N}_{l}^{0}}\right]\vec{T}_{l}^{0}\left(\theta\right)\right\},
\end{eqnarray}
where  $\mathcal{X}_{\rm{\bf Ad};l}\left(r\right)$, $\mathcal{Y}_{\rm{\bf Ad};l}\left(r\right)$ and $\mathcal{Z}_{\rm{\bf Ad};l}\left(r\right)$:
\begin{equation}
\mathcal{X}_{\rm{\bf Ad};l}=\frac{2}{3}\sum_{l_{1}=0}^{\infty}\sum_{l_{2}=1}^{\infty}\mathcal{N}_{l_{1}}^{0}\mathcal{N}_{l_{2}}^{0}\left\{\left(w_{0}^{l_{1}}\left[\frac{1}{r}\partial_{r}\xi_{0}^{l_{2}}\right]-v_{0}^{l_{1}}\left[\frac{\chi_{0}^{l_{2}}}{r}\right]\right)X_{l_{1},l_{2}}^{l}\right\},
\end{equation}
\begin{equation}
\mathcal{Y}_{{\rm{\bf Ad}};l}=\sum_{l_{1}=0}^{\infty}\sum_{l_{2}=1}^{\infty}\mathcal{N}_{l_{1}}^{0}\mathcal{N}_{l_{2}}^{0}\left\{\left(w_{0}^{l_{1}}\left[l_{2}(l_{2}+1)\frac{\xi_{0}^{l_{2}}}{r^2}\right]\right)\sum_{j=I(l_{1},1,l_{2},0)}^{l_{1}+l_{2}}d_{l_{1},1,l_{2},0}^{j}\delta_{l,j}-\left(u_{0}^{l_{1}}\left[\frac{\chi_{0}^{l_{2}}}{r}\right]\right)\sum_{j=I(l_{1},0,l_{2},1)}^{l_{1}+l_{2}}d_{l_{1},0,l_{2},1}^{j}\delta_{l,j}\right\},
\end{equation}
\begin{eqnarray}
\mathcal{Z}_{{\rm{\bf Ad}};l}&=&\sum_{l_{1}=0}^{\infty}\sum_{l_{2}=1}^{\infty}\mathcal{N}_{l_{1}}^{0}\mathcal{N}_{l_{2}}^{0}\left\{\left(v_{0}^{l_{1}}\left[l_{2}(l_{2}+1)\frac{\xi_{0}^{l_{2}}}{r^2}\right]\right)\sum_{j=I(l_{1},1,l_{2},0)}^{l_{1}+l_{2}}d_{l_{1},1,l_{2},0}^{j}\delta_{l,j}-\left(u_{0}^{l_{1}}\left[\frac{1}{r}\partial_{r}\xi_{0}^{l_{2}}\right]\right)\sum_{j=I(l_{1},0,l_{2},1)}^{l_{1}+l_{2}}d_{l_{1},0,l_{2},1}^{j}\delta_{l,j}\right\};
\end{eqnarray}
Explicit values for $X_{l_{1},l_{2}}^{l}$ (\ref{xm}) are given in Table~\ref{table3}, and in Tables~\ref{table4}, \ref{table5}, \ref{table6} for the coupling coefficients $d_{l_{1},1,l_{2},0}^{l}$. 

\begin{table}[]
\caption{Values of the $X_{l_{1},l_{2}}^{l}$ coefficients which are involved in the $r$-component of the induction equation}          
\centering          
\fbox{\begin{tabular}{c c c c c c c c c c c cl l l}  
\hline      
\multicolumn{4}{c}{} & \vline &\multicolumn{4}{c}{} & \vline &\multicolumn{4}{c}{}\\                    
\multicolumn{4}{c}{$X_{l_{1},l_{2}}^{1}$} & \vline &\multicolumn{4}{c}{$X_{l_{1},l_{2}}^{2}$}  & \vline &\multicolumn{4}{c}{$X_{l_{1},l_{2}}^{3}$}\\ 
\multicolumn{4}{c}{} & \vline &\multicolumn{4}{c}{} & \vline &\multicolumn{4}{c}{}\\  
\hline    
 & & & & \vline & & & & & \vline & & & &\\                
  $l_{1} \; \diagdown \; l_{2}$ & 1 & 2 & 3 & \vline &  $l_{1} \; \diagdown \; l_{2}$ & 1 & 2 & 3 & \vline &  $l_{1} \; \diagdown \; l_{2}$ & 1 & 2 & 3\\  
 & & & & \vline & & & & & \vline & & & &\\   
 1 & 0 & $\frac{13}{5}$ & 0 & \vline & 1 & $-1$ & 0 & $\frac{26}{7}$ & \vline & 1 & 0 & $-\frac{3}{5}$ & 0\\
 & & & & \vline & & & & & \vline & & & &\\             
 2 & $\frac{13}{5}$ & 0 & $\frac{366}{49}$ & \vline & 2 & 0 & $\frac{93}{35}$ & 0 & \vline & 2 & $-\frac{3}{5}$ & 0 & $\frac{48}{7}$\\
 & & & & \vline & & & & & \vline & & & &\\     
 3 & 0 & $\frac{366}{49}$ & 0 & \vline & 3 & $\frac{26}{7}$ & 0 & $\frac{478}{49}$ & \vline & 3 & 0 & $\frac{48}{7}$ & 0\\
 & & & & \vline & & & & & \vline & & & &\\    
  4 & $\frac{82}{35}$ & 0 & $\frac{3594}{245}$ & \vline & 4 & 0 & $\frac{226}{21}$ & 0 & \vline & 4 & $\frac{92}{15}$ & 0 & $\frac{774082}{38115}$\\
 & & & & \vline & & & & & \vline & & & &\\     
\hline                  
\end{tabular}}
\label{table3}
\end{table}

\begin{table}[]
\caption{Values of the coupling coefficients which are involved in the $\theta$- and in the $\varphi$- components of the induction equation for $l=1$}         
\centering          
\fbox{\begin{tabular}{c c c c c c c c l l}  
\hline      
\multicolumn{4}{c}{} & \vline &\multicolumn{4}{c}{}\\                    
\multicolumn{4}{c}{$d_{l_{1},1,l_{2},0}^{1}$} & \vline &\multicolumn{4}{c}{$d_{l_{1},0,l_{2},1}^{1}$}\\ 
\multicolumn{4}{c}{} & \vline &\multicolumn{4}{c}{}\\  
\hline    
 & & & & \vline & & & &\\                
  $l_{1} \; \diagdown \; l_{2}$ & 1 & 2 & 3 & \vline &  $l_{1} \; \diagdown \; l_{2}$ & 1 & 2 & 3\\  
 & & & & \vline & & & & \\   
 1 & 0 & $-\frac{1}{5}$ & 0 & \vline & 1 & 0 & $\frac{3}{5}$ & 0\\
 & & & & \vline & & & &\\             
 2 & $\frac{3}{5}$ & 0 & $-\frac{9}{35}$ & \vline & 2 & $-\frac{1}{5}$ & 0 & $\frac{18}{35}$\\
 & & & & \vline & & & &\\     
 3 & 0 & $\frac{18}{35}$ & 0 & \vline & 3 & 0 & $-\frac{9}{35}$ & 0\\
 & & & & \vline & & & &\\    
  4 & 0 & 0 & $\frac{10}{21}$ & \vline & 4 & 0 & 0 & -$\frac{2}{7}$\\
 & & & & \vline & & & &\\     
\hline                  
\end{tabular}}
\label{table4}
\end{table}

\begin{table}[]
\caption{Values of the coefficients involved in the $\theta$- and in the $\varphi$- components of the induction equation for $l=2$}     
\centering          
\fbox{\begin{tabular}{c c c c c c c c l l}  
\hline      
\multicolumn{4}{c}{} & \vline &\multicolumn{4}{c}{}\\                    
\multicolumn{4}{c}{$d_{l_{1},1,l_{2},0}^{2}$} & \vline &\multicolumn{4}{c}{$d_{l_{1},0,l_{2},1}^{2}$}\\ 
\multicolumn{4}{c}{} & \vline &\multicolumn{4}{c}{}\\  
\hline    
 & & & & \vline & & & &\\                
  $l_{1} \; \diagdown \; l_{2}$ & 1 & 2 & 3 & \vline &  $l_{1} \; \diagdown \; l_{2}$ & 1 & 2 & 3\\  
 & & & & \vline & & & & \\   
 1 & $\frac{1}{3}$ & 0 & $-\frac{1}{7}$ & \vline & 1 & $\frac{1}{3}$ & 0 & $\frac{4}{7}$\\
 & & & & \vline & & & &\\             
 2 & 0 & $\frac{1}{7}$ & 0 & \vline & 2 & 0 & $\frac{1}{7}$ & 0\\
 & & & & \vline & & & &\\     
 3 & $\frac{4}{7}$ & 0 & $\frac{2}{21}$ & \vline & 3 & $-\frac{1}{7}$ & 0 & $\frac{2}{21}$\\
 & & & & \vline & & & &\\    
  4 & 0 & $\frac{10}{21}$ & 0 & \vline & 4 & 0 & $-\frac{4}{21}$ & 0\\
 & & & & \vline & & & &\\     
\hline                  
\end{tabular}}
\label{table5}
\end{table}

\begin{table}[]
\caption{Same as in Table~\ref{table5}, but  for $l=3$}               
\centering          
\fbox{\begin{tabular}{c c c c c c c c l l}  
\hline      
\multicolumn{4}{c}{} & \vline &\multicolumn{4}{c}{}\\                    
\multicolumn{4}{c}{$d_{l_{1},1,l_{2},0}^{3}$} & \vline &\multicolumn{4}{c}{$d_{l_{1},0,l_{2},1}^{3}$}\\ 
\multicolumn{4}{c}{} & \vline &\multicolumn{4}{c}{}\\  
\hline    
 & & & & \vline & & & &\\                
  $l_{1} \; \diagdown \; l_{2}$ & 1 & 2 & 3 & \vline &  $l_{1} \; \diagdown \; l_{2}$ & 1 & 2 & 3\\  
 & & & & \vline & & & & \\   
 1 & 0 & $\frac{1}{5}$ & 0 & \vline & 1 & 0 & $\frac{2}{5}$ & 0\\
 & & & & \vline & & & &\\             
 2 & $\frac{2}{5}$ & 0 & $\frac{1}{15}$ & \vline & 2 & $\frac{1}{5}$ & 0 & $\frac{1}{5}$\\
 & & & & \vline & & & &\\     
 3 & 0 & $\frac{1}{5}$ & 0 & \vline & 3 & 0 & $\frac{1}{15}$ & 0\\
 & & & & \vline & & & &\\    
 4 & $\frac{5}{9}$ & 0 & $\frac{5}{33}$ & \vline & 4 & $-\frac{1}{9}$ & 0 & $\frac{1}{33}$\\
 & & & & \vline & & & &\\     
\hline                  
\end{tabular}}
\label{table6}
\end{table}

\section{Ohmic heating}

For sake of completeness, we also calculate the Ohmic heating rate, which we shall express in terms of the magnetic stream functions $\xi_{0}^{l}$ and $\chi_{0}^{l}$. From \S2.4, we get:
\begin{equation}
\mathcal{J}=\frac{1}{\mu_{0}}\left[||\eta||\otimes\left(\vec\nabla\wedge\vec B\right)\right]\cdot\left(\vec\nabla\wedge\vec B\right) ,
\end{equation}
where we allow for different eddy-diffusivities ($\eta_{h}, \eta_{v}$) respectively in the horizontal and vertical direction.
We apply again the method of \S A.2.3., and identify the two vector functions which enter the scalar product:
\begin{eqnarray}
\vec X_{1}\left(r,\theta\right)&=&\frac{1}{\mu_{0}}\left[||\eta||\otimes\left(\vec\nabla\wedge\vec B(r,\theta)\right)\right]=\sum_{l_{1}=1}^{\infty}\left\{\mathcal{A}_{0}^{l_{1}}(r)\vec R_{l_{1}}^{0}\left(\theta\right)+\mathcal{B}_{0}^{l_{1}}\left(r\right)\vec S_{l_{1}}^{0}\left(\theta\right)+\mathcal{C}_{0}^{l_{1}}\left(r\right)\vec T_{l_{1}}^{0}\left(\theta\right)\right\}\nonumber\\
&=&\sum_{l_{1}=1}^{\infty}\left\{\frac{\eta_{v}}{\mu_{0}}\left[l_{1}(l_{1}+1)\frac{\chi_{0}^{l_{1}}}{r^2}\right]\vec{R}_{l_{1}}^{0}\left(\theta\right)+\frac{\eta_{h}}{\mu_{0}}\left[\frac{1}{r}\partial_{r}\chi_{0}^{l_{1}}\right]\vec S_{l_{1}}^{0}\left(\theta\right)+\frac{\eta_{h}}{\mu_{0}}\left[-\Delta_{l_{1}}\left(\frac{\xi_{0}^{l_{1}}}{r}\right)\right]\vec T_{l_{1}}^{0}\left(\theta\right)\right\}\nonumber\\
\end{eqnarray}
and
\begin{eqnarray}
\vec X_{2}\left(r,\theta\right)&=&\vec\nabla\wedge\vec B\left(r,\theta\right)=\sum_{l_{2}=1}^{\infty}\left\{\mathcal{D}_{0}^{l_{2}}(r)\vec R_{l_{2}}^{0}\left(\theta\right)+\mathcal{E}_{0}^{l_{2}}\left(r\right)\vec S_{l_{2}}^{0}\left(\theta\right)+\mathcal{F}_{0}^{l_{2}}\left(r\right)\vec T_{l_{2}}^{0}\left(\theta\right)\right\}\nonumber\\
&=&\sum_{l_{2}=1}^{\infty}\left\{\left[l_{2}(l_{2}+1)\frac{\chi_{0}^{l_{2}}}{r^2}\right]\vec{R}_{l_{2}}^{0}\left(\theta\right)+\left[\frac{1}{r}\partial_{r}\chi_{0}^{l_{2}}\right]\vec S_{l_{2}}^{0}\left(\theta\right)+\left[-\Delta_{l_{2}}\left(\frac{\xi_{0}^{l_{2}}}{r}\right)\right]\vec T_{l_{2}}^{0}\left(\theta\right)\right\}.\nonumber\\
\end{eqnarray}
We thus obtain, using (\ref{sp1}-\ref{sp2}): 
\begin{equation}
\mathcal{J}\left(r,\theta\right)=\vec X_{1}\left(r,\theta\right)\cdot\vec X_{2}\left(r,\theta\right)=\sum_{l=0}^{\infty}\mathcal{J}_{l}\left(r\right)P_{l}\left(\cos\theta\right)
\end{equation}
with the following expression for $\mathcal{J}_{l}\left(r\right)$:
\begin{eqnarray}
\mathcal{J}_{l}&=&\frac{1}{\mu_{0}}\sum_{l_{1}=1}^{\infty}\sum_{l_{2}=1}^{\infty}\mathcal{N}_{l_{1}}^{0}\mathcal{N}_{l_{2}}^{0}\left\{\left(\frac{\eta_{v}}{\mu_{0}}\left[l_{1}(l_{1}+1)\frac{\chi_{0}^{l_{1}}}{r^2}\right]\left[l_{2}(l_{2}+1)\frac{\chi_{0}^{l_{2}}}{r^2}\right]\right)\sum_{j=I\left(l_{1},0,l_{2},0\right)}^{l_{1}+l_{2}}d_{l_{1},0,l_{2},0}^{j}\delta_{l,j}\right.\nonumber\\
&+&{\left. \frac{2}{3}\frac{\eta_{h}}{\mu_{0}}\left(\left[\frac{1}{r}\partial_{r}\chi_{0}^{l_{1}}\right]\left[\frac{1}{r}\partial_{r}\chi_{0}^{l_{2}}\right]+\left[\Delta_{l_{1}}\left(\frac{\xi_{0}^{l_{1}}}{r}\right)\right]\left[\Delta_{l_{2}}\left(\frac{\xi_{0}^{l_{2}}}{r}\right)\right]\right)X_{l_{1},l_{2}}^{l}\right\}} , \nonumber\\
\label{ohmproject}
\end{eqnarray}
recalling that $X_{l_{1},l_{2}}^{l}$ is given in (\ref{xm}), and in explicit form in Table~\ref{table1}.

\section{Equations to be implemented in stellar evolution codes}

In this section we shall give the equations ready to be implemented in a stellar structure code, with all coupling coefficients being replaced by their explicit value, in the special case where only the dipole ($l=1$), quadrupole ($l=2$) and octupole ($l=3$) are kept in the magnetic field, and where the differental rotation is reduced to its first term (l=2).

\subsection{Induction equation}

The linear terms of this equation readily separate into their poloidal and toroidal components, and they project on a single multipole, whereas the advection term is the result of various couplings. 

\subsubsection{Equations for the dipole}

The induction equation translates into two evolution equations for the magnetic stream functions: 
\begin{equation}
\left\{
\begin{array}{l@{\quad}l}
\frac{{\rm d}\xi_{0}^{1}}{{\rm d}t}=2\sqrt{\frac{\pi}{3}}r\mathcal{Z}_{{\bf Ad};1}+\eta_{h}r\Delta_{1}\left(\frac{\xi_{0}^{1}}{r}\right)\\
\frac{{\rm d}\chi_{0}^{1}}{{\rm d}t}+\partial_{r}\left(\dot{r}\right)\chi_{0}^{1}=2\sqrt{\frac{\pi}{3}}\left[\mathcal{X}_{{\bf Ad};1}+\partial_{r}\left(r\mathcal{Y}_{{\bf Ad};1}\right)\right]+\left[\partial_{r}\left(\eta_{h}\partial_{r}\chi_{0}^{1}\right)-2\eta_{v}\frac{\chi_{0}^{1}}{r^2}\right]
\end{array}
\right.\hbox{ }\hbox{where}\hbox{ }\Delta_{1}=\partial_{r,r}+\frac{2}{r}\partial_{r}-\frac{2}{r^2} ,
\end{equation}
and where the advective terms are given by:
\begin{equation}
\left\{
\begin{array}{l@{\quad}l}
\mathcal{X}_{{\bf Ad};1}=\frac{2}{3}\frac{1}{2\pi}\left[\frac{13}{2}\sqrt{\frac{3}{5}}\left(\mathcal{C}_{0}^{1}\mathcal{E}_{0}^{2}-\mathcal{B}_{0}^{2}\mathcal{F}_{0}^{1}\right)+\frac{183}{7}\sqrt{\frac{5}{7}}\left(\mathcal{C}_{0}^{3}\mathcal{E}_{0}^{2}-\mathcal{B}_{0}^{2}\mathcal{F}_{0}^{3}\right)-\frac{1}{35}\left(123\sqrt{3}\mathcal{B}_{0}^{4}\mathcal{F}_{0}^{1}+\frac{5391}{\sqrt{7}}\mathcal{B}_{0}^{4}\mathcal{F}_{0}^{3}\right)\right]\\
\mathcal{Y}_{{\bf Ad};1}=\frac{1}{4\pi}\left[\sqrt{\frac{3}{5}}\left(\mathcal{A}_{0}^{2}\mathcal{F}_{0}^{1}-\mathcal{C}_{0}^{1}\mathcal{D}_{0}^{2}\right)+\frac{18}{\sqrt{35}}\left(\mathcal{C}_{0}^{3}\mathcal{D}_{0}^{2}-\mathcal{A}_{0}^{2}\mathcal{F}_{0}^{3}\right)+\frac{6}{\sqrt{7}}\mathcal{A}_{0}^{4}\mathcal{F}_{0}^{3}\right]\\
\mathcal{Z}_{{\bf Ad};1}=\frac{1}{4\pi}\left[\sqrt{\frac{3}{5}}\left(3\mathcal{B}_{0}^{2}\mathcal{D}_{0}^{1}+\mathcal{A}_{0}^{2}\mathcal{E}_{0}^{1}\right)-\frac{9}{\sqrt{35}}\left(\mathcal{B}_{0}^{2}\mathcal{D}_{0}^{3}+2\mathcal{A}_{0}^{2}\mathcal{E}_{0}^{3}\right)+\frac{2}{\sqrt{7}}\left(5\mathcal{B}_{0}^{4}\mathcal{D}_{0}^{3}+3\mathcal{A}_{0}^{4}\mathcal{E}_{0}^{3}\right)\right]
\end{array}
\right.
\end{equation}
with
\begin{equation}
\left\{
\begin{array}{l@{\quad}l}
\mathcal{A}_{0}^{2}=2\sqrt{\frac{\pi}{5}}U_{2}\\
\mathcal{A}_{0}^{4}=\frac{2}{3}\sqrt{\pi}U_{4}\\
\mathcal{B}_{0}^{2}=2\sqrt{\frac{\pi}{5}}V_{2}\\
\mathcal{B}_{0}^{4}=\frac{2}{3}\sqrt{\pi}V_{4}\\
\mathcal{C}_{0}^{1}=w_{0}^{1}=2\sqrt{\frac{\pi}{3}}r\overline{\Omega}\\
\mathcal{C}_{0}^{3}=w_{0}^{3}=\frac{2}{5}\sqrt{\frac{\pi}{7}}r\Omega_{2}
\end{array}
\right.
\hbox{,}\hbox{ }
\left\{
\begin{array}{l@{\quad}l}
\mathcal{D}_{0}^{1}=\left[2\frac{\xi_{0}^{1}}{r^2}\right]\\
\mathcal{E}_{0}^{1}=\left[\frac{1}{r}\partial_{r}\xi_{0}^{1}\right]\\
\mathcal{F}_{0}^{1}=\left[\frac{\chi_{0}^{1}}{r}\right]
\end{array}
\right.
\hbox{,}\hbox{ }
\left\{
\begin{array}{l@{\quad}l}
\mathcal{D}_{0}^{2}=\left[6\frac{\xi_{0}^{2}}{r^2}\right]\\
\mathcal{E}_{0}^{2}=\left[\frac{1}{r}\partial_{r}\xi_{0}^{2}\right]\\
\mathcal{F}_{0}^{2}=\left[\frac{\chi_{0}^{2}}{r}\right]
\end{array}
\right.
\hbox{and}\hbox{ }\hbox{ }
\left\{
\begin{array}{l@{\quad}l}
\mathcal{D}_{0}^{3}=\left[12\frac{\xi_{0}^{3}}{r^2}\right]\\
\mathcal{E}_{0}^{3}=\left[\frac{1}{r}\partial_{r}\xi_{0}^{3}\right]\\
\mathcal{F}_{0}^{3}=\left[\frac{\chi_{0}^{3}}{r}\right].
\end{array}
\right. 
\label{InductionAnn}
\end{equation}

\subsubsection{Equations for the quadrupole}
Likewise
\begin{equation}
\left\{
\begin{array}{l@{\quad}l}
\frac{{\rm d}\xi_{0}^{2}}{{\rm d}t}=2\sqrt{\frac{\pi}{5}}r\mathcal{Z}_{{\bf Ad};2}+\eta_{h}r\Delta_{2}\left(\frac{\xi_{0}^{2}}{r}\right)\\
\frac{{\rm d}\chi_{0}^{2}}{{\rm d}t}+\partial_{r}\left(\dot{r}\right)\chi_{0}^{2}=2\sqrt{\frac{\pi}{5}}\left[\mathcal{X}_{{\bf Ad};2}+\partial_{r}\left(r\mathcal{Y}_{{\bf Ad};2}\right)\right]+\left[\partial_{r}\left(\eta_{h}\partial_{r}\chi_{0}^{2}\right)-6\eta_{v}\frac{\chi_{0}^{2}}{r^2}\right]
\end{array}
\right. \hbox{ }\hbox{where}\hbox{ }\Delta_{2}=\partial_{r,r}+\frac{2}{r}\partial_{r}-\frac{6}{r^2}
\end{equation}
and
\begin{equation}
\left\{
\begin{array}{l@{\quad}l}
\mathcal{X}_{{\bf Ad};2}=\frac{2}{3}\frac{1}{2\pi}\left[13\sqrt{\frac{3}{7}}\left(\mathcal{C}_{0}^{3}\mathcal{E}_{0}^{1}+\mathcal{C}_{0}^{1}\mathcal{E}_{0}^{3}\right)-\frac{3}{2}\mathcal{C}_{0}^{1}\mathcal{E}_{0}^{1}+\frac{239}{7}\mathcal{C}_{0}^{3}\mathcal{E}_{0}^{3}-\frac{93}{14}\mathcal{B}_{0}^{2}\mathcal{F}_{0}^{2}-\frac{113}{7}\sqrt{5}\mathcal{B}_{0}^{4}\mathcal{F}_{0}^{2}\right]\\
\mathcal{Y}_{{\bf Ad};2}=\frac{1}{4\pi}\left[\sqrt{\frac{3}{7}}\left(4\mathcal{C}_{0}^{3}\mathcal{D}_{0}^{1}-\mathcal{C}_{0}^{1}\mathcal{D}_{0}^{3}\right)+\mathcal{C}_{0}^{1}\mathcal{D}_{0}^{1}-\frac{5}{7}\mathcal{A}_{0}^{2}\mathcal{F}_{0}^{2}+\frac{4}{7}\sqrt{5}\mathcal{A}_{0}^{4}\mathcal{F}_{0}^{2}+\frac{2}{3}\mathcal{C}_{0}^{3}\mathcal{D}_{0}^{3}\right]\\
\mathcal{Z}_{{\bf Ad};2}=\frac{1}{4\pi}\left[\frac{5}{7}\left(\mathcal{B}_{0}^{2}\mathcal{D}_{0}^{2}-\mathcal{A}_{0}^{2}\mathcal{E}_{0}^{2}\right)+\frac{2}{7}\sqrt{5}\left(5\mathcal{B}_{0}^{4}\mathcal{D}_{0}^{2}+2\mathcal{A}_{0}^{4}\mathcal{E}_{0}^{2}\right)\right].
\end{array}
\right.
\end{equation}
All functions $\mathcal{A}_{0}^{2}$ . . .  $\mathcal{F}_{0}^{3}$ have the same meaning as in (\ref{InductionAnn}). 

\subsubsection{Equations for the octupole}
The result is similar for the octupole:
\begin{equation}
\left\{
\begin{array}{l@{\quad}l}
\frac{{\rm d}\xi_{0}^{3}}{{\rm d}t}=2\sqrt{\frac{\pi}{7}}r\mathcal{Z}_{{\bf Ad};3}+\eta_{h}r\Delta_{3}\left(\frac{\xi_{0}^{3}}{r}\right)\\
\frac{{\rm d}\chi_{0}^{3}}{{\rm d}t}+\partial_{r}\left(\dot{r}\right)\chi_{0}^{3}=2\sqrt{\frac{\pi}{7}}\left[\mathcal{X}_{{\bf Ad};3}+\partial_{r}\left(r\mathcal{Y}_{{\bf Ad};3}\right)\right]+\left[\partial_{r}\left(\eta_{h}\partial_{r}\chi_{0}^{3}\right)-12\eta_{v}\frac{\chi_{0}^{2}}{r^2}\right]
\end{array}
\right. \hbox{ }\hbox{where}\hbox{ }\Delta_{3}=\partial_{r,r}+\frac{2}{r}\partial_{r}-\frac{12}{r^2}
\end{equation}
and
\begin{equation}
\left\{
\begin{array}{l@{\quad}l}
\mathcal{X}_{{\bf Ad};3}=\frac{2}{3}\frac{1}{\pi}\left[\frac{3}{4}\sqrt{\frac{3}{5}}\left(\mathcal{C}_{0}^{1}\mathcal{E}_{0}^{2}+\mathcal{B}_{0}^{2}\mathcal{F}_{0}^{1}\right)+12\sqrt{\frac{5}{7}}\left(\mathcal{C}_{0}^{3}\mathcal{E}_{0}^{2}-\mathcal{B}_{0}^{2}\mathcal{F}_{0}^{3}\right)-\frac{23}{5}\sqrt{3}\mathcal{B}_{0}^{4}\mathcal{F}_{0}^{1}-\frac{387041}{3630}\frac{1}{\sqrt{7}}\mathcal{B}_{0}^{4}\mathcal{F}_{0}^{3}\right]\\
\mathcal{Y}_{{\bf Ad};3}=\frac{1}{4\pi}\left[\sqrt{\frac{3}{5}}\left(\mathcal{C}_{0}^{1}\mathcal{D}_{0}^{2}-\mathcal{A}_{0}^{2}\mathcal{F}_{0}^{1}\right)+\sqrt{\frac{7}{5}}\left(\mathcal{C}_{0}^{3}\mathcal{D}_{0}^{2}-\mathcal{A}_{0}^{2}\mathcal{F}_{0}^{3}\right)+\frac{1}{\sqrt{3}}\mathcal{A}_{0}^{4}\mathcal{F}_{0}^{1}-\frac{\sqrt{7}}{11}\mathcal{A}_{0}^{4}\mathcal{F}_{0}^{3}\right]\\
\mathcal{Z}_{{\bf Ad};3}=\frac{1}{4\pi}\left[\sqrt{\frac{3}{5}}\left(2\mathcal{B}_{0}^{2}\mathcal{D}_{0}^{1}-\mathcal{A}_{0}^{2}\mathcal{E}_{0}^{1}\right)+\frac{1}{\sqrt{3}}\left(5\mathcal{B}_{0}^{4}\mathcal{D}_{0}^{1}+\mathcal{A}_{0}^{4}\mathcal{E}_{0}^{1}\right)+\sqrt{\frac{7}{5}}\left(\frac{1}{3}\mathcal{B}_{0}^{2}\mathcal{D}_{0}^{3}-\mathcal{A}_{0}^{2}\mathcal{E}_{0}^{3}\right)+\frac{\sqrt{7}}{11}\left(5\mathcal{B}_{0}^{4}\mathcal{D}_{0}^{3}-\mathcal{A}_{0}^{4}\mathcal{E}_{0}^{3}\right)\right].
\end{array}
\right.
\end{equation}

\subsection{Mean equations}
\subsubsection{Mean rotation rate}
We recall the transport equation for the mean angular momentum (\ref{mean-AM}):
\begin{equation}
\rho {{\rm d} \over {\rm d}t} (r^2\overline{\Omega})=\frac{1}{5r^2}\partial_{r}\left(\rho r^4\overline{\Omega}U_{2}\right)+\frac{1}{r^2}\partial_{r}\left(\rho\nu_{v}r^4\partial_{r}\overline{\Omega}\right)+\overline{\Gamma}_{\vec{\mathcal{F}}_{\mathcal L}}
\label{OmVert}
\end{equation}
where the mean magnetic torque is given by:
\begin{equation}
\overline{\Gamma}_{\vec{\mathcal{F}}_{\mathcal L}}=\Gamma_{0}-\frac{1}{5}\Gamma_{2}
\hbox{ }\hbox{with}\hbox{ }
\left\{
\begin{array}{l@{\quad}l}
\Gamma_{0}=r\left(\mathcal{Z}_{\vec{\mathcal{F}}\!\!_{\mathcal{L}};1}+\mathcal{Z}_{\vec{\mathcal{F}}\!\!_{\mathcal{L}};3}+\mathcal{Z}_{\vec{\mathcal{F}}\!\!_{\mathcal{L}};5}\right)\\
\Gamma_{2}=5r\left(\mathcal{Z}_{\vec{\mathcal{F}}\!\!_{\mathcal{L}};3}+\mathcal{Z}_{\vec{\mathcal{F}}\!\!_{\mathcal{L}};5}\right)
\end{array}
\right.
\label{TorqueMagn}
\end{equation}
with
\begin{equation}
\left\{
\begin{array}{l@{\quad}l}
\mathcal{Z}_{\vec{\mathcal{F}}\!\!_{\mathcal{L}};1}=\frac{1}{4\pi}\left\{\sqrt{\frac{3}{5}}\left[3\left(\mathcal{B}_{0}^{2}\mathcal{D}_{0}^{1}-\mathcal{A}_{0}^{1}\mathcal{E}_{0}^{2}\right)-\left(\mathcal{B}_{0}^{1}\mathcal{D}_{0}^{2}-\mathcal{A}_{0}^{2}\mathcal{E}_{0}^{1}\right)\right]+\frac{9}{\sqrt{35}}\left[2\left(\mathcal{B}_{0}^{3}\mathcal{D}_{0}^{2}-\mathcal{A}_{0}^{2}\mathcal{E}_{0}^{3}\right)-\left(\mathcal{B}_{0}^{2}\mathcal{D}_{0}^{3}-\mathcal{A}_{0}^{3}\mathcal{E}_{0}^{2}\right)\right]\right\}\\
\mathcal{Z}_{\vec{\mathcal{F}}\!\!_{\mathcal{L}};3}=\frac{1}{4\pi}\left\{\sqrt{\frac{3}{5}}\left[2\left(\mathcal{B}_{0}^{2}\mathcal{D}_{0}^{1}-\mathcal{A}_{0}^{1}\mathcal{E}_{0}^{2}\right)+\left(\mathcal{B}_{0}^{1}\mathcal{D}_{0}^{2}-\mathcal{A}_{0}^{2}\mathcal{E}_{0}^{1}\right)\right]+\sqrt{\frac{7}{5}}\left[\left(\mathcal{B}_{0}^{3}\mathcal{D}_{0}^{2}-\mathcal{A}_{0}^{2}\mathcal{E}_{0}^{3}\right)+\frac{1}{3}\left(\mathcal{B}_{0}^{2}\mathcal{D}_{0}^{3}-\mathcal{A}_{0}^{3}\mathcal{E}_{0}^{2}\right)\right]\right\}\\
\mathcal{Z}_{\vec{\mathcal{F}}\!\!_{\mathcal{L}};5}=\frac{1}{2\pi}\sqrt{\frac{5}{7}}\left[\left(\mathcal{B}_{0}^{3}\mathcal{D}_{0}^{2}-\mathcal{A}_{0}^{2}\mathcal{E}_{0}^{3}\right)+\frac{2}{3}\left(\mathcal{B}_{0}^{2}\mathcal{D}_{0}^{3}-\mathcal{A}_{0}^{3}\mathcal{E}_{0}^{2}\right)\right].
\end{array}
\right.
\end{equation}
The $\mathcal{A}_{0}^{l}$, $\mathcal{B}_{0}^{l}$, $\mathcal{C}_{0}^{l}$, $\mathcal{D}_{0}^{l}$, $\mathcal{E}_{0}^{l}$ and $\mathcal{F}_{0}^{l}$ with $l=\left\{1,2,3\right\}$ are given by:
\begin{equation}
\left\{
\begin{array}{l@{\quad}l}
\mathcal{A}_{0}^{1}=\frac{1}{\mu_{0}}\left[2\frac{\chi_{0}^{1}}{r^2}\right]\\
\mathcal{B}_{0}^{1}=\frac{1}{\mu_{0}}\left[\frac{1}{r}\partial_{r}\chi_{0}^{1}\right]\\
\mathcal{C}_{0}^{1}=\frac{1}{\mu_{0}}\left[-\Delta_{1}\left(\frac{\xi_{0}^{1}}{r}\right)\right]\\
\mathcal{D}_{0}^{1}=\left[2\frac{\xi_{0}^{1}}{r^2}\right]\\
\mathcal{E}_{0}^{1}=\left[\frac{1}{r}\partial_{r}\xi_{0}^{1}\right]\\
\mathcal{F}_{0}^{1}=\left[\frac{\chi_{0}^{1}}{r}\right]
\end{array}
\right.
\hbox{}\hbox{,}\hbox{ }
\left\{
\begin{array}{l@{\quad}l}
\mathcal{A}_{0}^{2}=\frac{1}{\mu_{0}}\left[6\frac{\chi_{0}^{2}}{r^2}\right]\\
\mathcal{B}_{0}^{2}=\frac{1}{\mu_{0}}\left[\frac{1}{r}\partial_{r}\chi_{0}^{2}\right]\\
\mathcal{C}_{0}^{2}=\frac{1}{\mu_{0}}\left[-\Delta_{2}\left(\frac{\xi_{0}^{2}}{r}\right)\right]\\
\mathcal{D}_{0}^{2}=\left[6\frac{\xi_{0}^{2}}{r^2}\right]\\
\mathcal{E}_{0}^{2}=\left[\frac{1}{r}\partial_{r}\xi_{0}^{2}\right]\\
\mathcal{F}_{0}^{2}=\left[\frac{\chi_{0}^{2}}{r}\right]
\end{array}
\right.
\hbox{}\hbox{and}\hbox{ }
\left\{
\begin{array}{l@{\quad}l}
\mathcal{A}_{0}^{3}=\frac{1}{\mu_{0}}\left[12\frac{\chi_{0}^{3}}{r^2}\right]\\
\mathcal{B}_{0}^{3}=\frac{1}{\mu_{0}}\left[\frac{1}{r}\partial_{r}\chi_{0}^{3}\right]\\
\mathcal{C}_{0}^{3}=\frac{1}{\mu_{0}}\left[-\Delta_{3}\left(\frac{\xi_{0}^{3}}{r}\right)\right]\\
\mathcal{D}_{0}^{3}=\left[12\frac{\xi_{0}^{3}}{r^2}\right]\\
\mathcal{E}_{0}^{3}=\left[\frac{1}{r}\partial_{r}\xi_{0}^{3}\right]\\
\mathcal{F}_{0}^{3}=\left[\frac{\chi_{0}^{3}}{r}\right].
\end{array}
\right.
\label{LFcoeff}
\end{equation}
From (\ref{TorqueMagn}), the final form of (\ref{OmVert}) is immediately derived:
\begin{equation}
\rho {{\rm d} \over {\rm d}t} (r^2\overline{\Omega})=\frac{1}{5r^2}\partial_{r}\left(\rho r^4\overline{\Omega}U_{2}\right)+\frac{1}{r^2}\partial_{r}\left(\rho\nu_{v}r^4\partial_{r}\overline{\Omega}\right)+r\mathcal{Z}_{\vec{\mathcal{F}}\!\!_{\mathcal{L}};1}.
\end{equation}

\subsubsection{Mean chemical composition}
We restate the mean transport equation for the concentration of chemical species, which remains unchanged by the introduction of the magnetic field (Paper I, eq. B.2):
\begin{equation}  \rho {{\rm d} \over {\rm d} t } \overline{c_{i}}
+\frac{1}{r^2}\partial_{r}\left[r^2\rho \overline{c_{i}} U_i^{\rm diff} \right]
=\frac{1}{r^2}\partial_{r}\left[r^2\rho (D_{v}+D_{\rm eff})\partial_{r}\overline{c_{i}}\right] .
\end{equation}

\subsection{System for $l=2$}

\subsubsection{Meridional circulation}
We split (\ref{heat1}-\ref{tcal-final}) in two first order equations as:
\begin{equation}
U_{2}=\frac{L}{M\overline{g}}\left(\frac{P}{\overline{\rho}C_{p}\overline{T}}\right)\frac{1}{\nabla_{\hbox{ad}}-\nabla}\mathcal{B}_{2}
\end{equation}
where 
\begin{eqnarray}
\mathcal{B}_{2}&=&2\left[1-\frac{\overline{f}_{\mathcal{C}}+\overline{f}_{\mathcal{L}}}{4\pi G\overline{\rho}}-\frac{\left(\overline{\epsilon}+\overline{\epsilon}_{grav}\right)}{\epsilon_{m}}\right]\frac{\widetilde{g}_{2}}{\overline{g}}+\frac{\widetilde{f}_{\mathcal{C},2}+\widetilde{f}_{\mathcal{L},2}}{4\pi G\overline{\rho}}-\frac{\overline{f}_{\mathcal{C}}+\overline{f}_{\mathcal{L}}}{4\pi G\overline{\rho}}\left(-\delta\Psi_{2}+\varphi\Lambda_{2}\right)+\frac{\rho_{m}}{\overline{\rho}}\left[ \frac{r}{3}\partial_{r}\mathcal{A}_{2}-\frac{2H_{T}}{r}\left(1 + \frac{D_{h}}{K}\right){\Psi_2}\right]\nonumber\\
&+&\frac{\left(\overline{\epsilon}+\overline{\epsilon}_{grav}\right)}{\epsilon_{m}}\left\{ \mathcal{A}_{2}+(f_{\epsilon}\epsilon_{T} - f_{\epsilon}\delta + \delta)\Psi_{2}+(f_{\epsilon}\epsilon_{\mu}+f_{\epsilon}\varphi - \varphi)\Lambda_{2}\right\}+\frac{M}{L}\left[\frac{\mathcal{J}_{2}}{\overline{\rho}}-C_{p}\overline{T}\left(\frac{{\rm d}\Psi_{2}}{{\rm d}t}+\Phi\frac{{\rm d}\ln\overline{\mu}}{{\rm d}t}\Lambda_{2}\right)\right]\nonumber\\    
\end{eqnarray}
and
\begin{equation}
\mathcal{A}_{2}=H_{T}\partial_{r}\Psi_{2}-(1-\delta+\chi_{T})\Psi_{2}-(\varphi+\chi_{\mu})\Lambda_{2}.
\end{equation}
The coefficients related respectively to the centrifugal force and to the Lorentz force are given by:
\begin{equation}
\overline{f}_{\mathcal{C}}=\frac{1}{r^2}\partial_{r}\left(r^2 a_{0}\right)\hbox{ }\hbox{,}\hbox{ }\widetilde{f}_{\mathcal{C},2}=\frac{1}{r^2}\partial_{r}\left(r^2a_{2}\right)+6\frac{b_{2}}{r}\hbox{,}
\end{equation}
\begin{equation}
\overline{f}_{\mathcal{L}}=\frac{1}{r^2}\partial_{r}\left(r^2\frac{\mathcal{X}_{\vec{\mathcal F}\!\!_{\mathcal L};0}}{\rho_{0}}\right)\quad\hbox{ and }\quad\widetilde{f}_{\mathcal{L},2}=\frac{1}{r^2}\partial_{r}\left(r^2\frac{\mathcal{X}_{\vec{\mathcal F}\!\!_{\mathcal L};2}}{\rho_{0}}\right)+6\frac{\mathcal{Y}_{\vec{\mathcal F}_{\mathcal{L};2}}}{r\rho_{0}}
\end{equation}
where
\begin{equation}
\left\{
\begin{array}{l@{\quad}l}
a_{0}=\frac{2}{3}r\overline{\Omega}^{2}\\
a_{2}=-\frac{2}{3}r\overline{\Omega}^{2}+\frac{24}{35}r\overline{\Omega}\Omega_{2}\\
b_{2}=\frac{1}{3}r\overline{\Omega}^{2}+\frac{8}{35}r\overline{\Omega}\Omega_{2}
\end{array}
\right.
\end{equation}
and
\begin{equation}
\mathcal{X}_{\vec{\mathcal{F}}\!\!_{\mathcal{L}};0}=\frac{2}{3}\frac{1}{\pi}\left\{\frac{\sqrt{21}}{5}\left[\left(\mathcal{C}_{0}^{3}\mathcal{E}_{0}^{1}-\mathcal{B}_{0}^{3}\mathcal{F}_{0}^{1}\right)+\left(\mathcal{C}_{0}^{1}\mathcal{E}_{0}^{3}-\mathcal{B}_{0}^{1}\mathcal{F}_{0}^{3}\right)\right]+\frac{3}{4}\left(\mathcal{C}_{0}^{1}\mathcal{E}_{0}^{1}-\mathcal{B}_{0}^{1}\mathcal{F}_{0}^{1}\right)+\frac{69}{20}\left(\mathcal{C}_{0}^{2}\mathcal{E}_{0}^{2}-\mathcal{B}_{0}^{2}\mathcal{F}_{0}^{2}\right)+\frac{93}{10}\left(\mathcal{C}_{0}^{3}\mathcal{E}_{0}^{3}-\mathcal{B}_{0}^{3}\mathcal{F}_{0}^{3}\right)\right\}
\end{equation}
\begin{eqnarray}
\mathcal{X}_{\vec{\mathcal{F}}\!\!_{\mathcal{L}};2}&=&\frac{2}{3}\frac{1}{2\pi}\left\{13\sqrt{\frac{3}{7}}\left[\left(\mathcal{C}_{0}^{3}\mathcal{E}_{0}^{1}-\mathcal{B}_{0}^{3}\mathcal{F}_{0}^{1}\right)+\left(\mathcal{C}_{0}^{1}\mathcal{E}_{0}^{3}-\mathcal{B}_{0}^{1}\mathcal{F}_{0}^{3}\right)\right]-\frac{3}{2}\left(\mathcal{C}_{0}^{1}\mathcal{E}_{0}^{1}-\mathcal{B}_{0}^{1}\mathcal{F}_{0}^{1}\right)+\frac{93}{14}\left(\mathcal{C}_{0}^{2}\mathcal{E}_{0}^{2}-\mathcal{B}_{0}^{2}\mathcal{F}_{0}^{2}\right)\right. \nonumber\\
&+&{\left. \frac{239}{7}\left(\mathcal{C}_{0}^{3}\mathcal{E}_{0}^{3}-\mathcal{B}_{0}^{3}\mathcal{F}_{0}^{3}\right)\right\}}
\label{XFL2}
\end{eqnarray}
\begin{equation}
\mathcal{Y}_{\vec{\mathcal{F}}\!\!_{\mathcal{L}};2}=\frac{1}{4\pi}\left\{\sqrt{\frac{3}{7}}\left[4\left(\mathcal{C}_{0}^{3}\mathcal{D}_{0}^{1}-\mathcal{A}_{0}^{1}\mathcal{F}_{0}^{3}\right)-\left(\mathcal{C}_{0}^{1}\mathcal{D}_{0}^{3}-\mathcal{A}_{0}^{3}\mathcal{F}_{0}^{1}\right)\right]+\left(\mathcal{C}_{0}^{1}\mathcal{D}_{0}^{1}-\mathcal{A}_{0}^{1}\mathcal{F}_{0}^{1}\right)+\frac{5}{7}\left(\mathcal{C}_{0}^{2}\mathcal{D}_{0}^{2}-\mathcal{A}_{0}^{2}\mathcal{F}_{0}^{2}\right)+\frac{2}{3}\left(\mathcal{C}_{0}^{3}\mathcal{D}_{0}^{3}-\mathcal{A}_{0}^{3}\mathcal{F}_{0}^{3}\right)\right\}.
\label{YFL2}
\end{equation}
The $\mathcal{A}_{0}^{l}$, $\mathcal{B}_{0}^{l}$, $\mathcal{C}_{0}^{l}$, $\mathcal{D}_{0}^{l}$, $\mathcal{E}_{0}^{l}$ and $\mathcal{F}_{0}^{l}$ with $l=\left\{1,2,3\right\}$ are given in (\ref{LFcoeff}).\\
\newline
The relative fluctuation of the effective gravity is given in (\ref{geff}); we apply it here to $l=2$:
\begin{equation}
\frac{\widetilde{g}_{2}}{\overline{g}}=-\left[\frac{{\rm d} g_{0}}{{\rm d} r}\frac{1}{{g_{0}^{2}}}r\left(b_{2}+\frac{\mathcal{Y}_{\vec{\mathcal F}\!\!_{\mathcal L};2}}{\rho_{0}}\right) + \frac{1}{g_{0}}\left(a_{2}+\frac{\mathcal{X}_{\vec{\mathcal F}\!\!_{\mathcal L};2}}{\rho_{0}}\right)\right] + \frac{{\rm d}}{{\rm d}r}\left(\frac{\widehat{\phi}_{2}}{g_{0}}\right) 
\end{equation}
where $\widehat{\phi}_{2}$ is solution of the Poisson equation (cf. eq. \ref{poisson-pert}):
\begin{equation}
\frac{1}{r}\frac{{\rm d}^2}{{\rm d}r^2}\left(r\widehat{\phi}_{2}\right)-\frac{6}{r^2}\widehat{\phi}_{2}-\frac{4\pi G}{g_{0}}\frac{{\rm d}\rho_{0}}{{\rm d}r}\widehat{\phi}_{2}=\frac{4\pi G}{g_{0}}\left[\rho_{0}a_{2}+\frac{{\rm d}}{{\rm d}r}\left(r\rho_{0}b_{2}\right)+\mathcal{X}_{\vec{\mathcal{F}}\!\!_{\mathcal{L}};2}+\frac{{\rm d}}{{\rm d}r}\left(r\mathcal{Y}_{\vec{\mathcal{F}}\!\!_{\mathcal{L}};2}\right)\right].
\end{equation}
The Ohmic heating term is derived from (\ref{ohmproject}):
\begin{eqnarray}
\mathcal{J}_{2}&=&\frac{1}{4\pi}\left\{\left[\mathcal{A}_{0}^{1}{D}_{0}^{1}-\left(\mathcal{B}_{0}^{1}\mathcal{E}_{0}^{1}+\mathcal{C}_{0}^{1}\mathcal{F}_{0}^{1}\right)\right]+\left[3\sqrt{\frac{3}{7}}\left(\mathcal{A}_{0}^{3}\mathcal{D}_{0}^{1}+\mathcal{A}_{0}^{1}\mathcal{D}_{0}^{3}\right)+\frac{52}{\sqrt{21}}\left(\mathcal{B}_{0}^{3}\mathcal{E}_{0}^{1}+\mathcal{C}_{0}^{3}\mathcal{F}_{0}^{1}+\mathcal{B}_{0}^{1}\mathcal{E}_{0}^{3}+\mathcal{C}_{0}^{1}\mathcal{F}_{0}^{3}\right)\right]\right.\nonumber\\
&+&{\left. \frac{2}{7}\left[5\mathcal{A}_{0}^{2}\mathcal{D}_{0}^{2}+31\left(\mathcal{B}_{0}^{2}\mathcal{E}_{0}^{2}+\mathcal{C}_{0}^{2}\mathcal{F}_{0}^{2}\right)\right]+\frac{4}{3}\left[\mathcal{A}_{0}^{3}{D}_{0}^{3}+\frac{239}{7}\left(\mathcal{B}_{0}^{3}\mathcal{E}_{0}^{3}+\mathcal{C}_{0}^{3}\mathcal{F}_{0}^{3}\right)\right]\right\}}
\end{eqnarray}
where the $\mathcal{A}_{0}^{l}$, $\mathcal{B}_{0}^{l}$, $\mathcal{C}_{0}^{l}$, $\mathcal{D}_{0}^{l}$, $\mathcal{E}_{0}^{l}$ and $\mathcal{F}_{0}^{l}$ with $l=\left\{1,2,3\right\}$ are given by:
\begin{equation}
\left\{
\begin{array}{l@{\quad}l}
\mathcal{A}_{0}^{1}=\frac{\eta_{v}}{\mu_{0}}\left[2\frac{\chi_{0}^{1}}{r^2}\right]\\
\mathcal{B}_{0}^{1}=\frac{\eta_{h}}{\mu_{0}}\left[\frac{1}{r}\partial_{r}\chi_{0}^{1}\right]\\
\mathcal{C}_{0}^{1}=\frac{\eta_{h}}{\mu_{0}}\left[-\Delta_{1}\left(\frac{\xi_{0}^{1}}{r}\right)\right]\\
\mathcal{D}_{0}^{1}=\left[2\frac{\chi_{0}^{1}}{r^2}\right]\\
\mathcal{E}_{0}^{1}=\left[\frac{1}{r}\partial_{r}\chi_{0}^{1}\right]\\
\mathcal{F}_{0}^{1}=\left[-\Delta_{1}\left(\frac{\xi_{0}^{1}}{r}\right)\right]
\end{array}
\right.
\hbox{ }\hbox{,}\hbox{ }
\left\{
\begin{array}{l@{\quad}l}
\mathcal{A}_{0}^{2}=\frac{\eta_{v}}{\mu_{0}}\left[6\frac{\chi_{0}^{2}}{r^2}\right]\\
\mathcal{B}_{0}^{2}=\frac{\eta_{h}}{\mu_{0}}\left[\frac{1}{r}\partial_{r}\chi_{0}^{2}\right]\\
\mathcal{C}_{0}^{2}=\frac{\eta_{h}}{\mu_{0}}\left[-\Delta_{2}\left(\frac{\xi_{0}^{2}}{r}\right)\right]\\
\mathcal{D}_{0}^{2}=\left[6\frac{\chi_{0}^{2}}{r^2}\right]\\
\mathcal{E}_{0}^{2}=\left[\frac{1}{r}\partial_{r}\chi_{0}^{2}\right]\\
\mathcal{F}_{0}^{2}=\left[-\Delta_{2}\left(\frac{\xi_{0}^{2}}{r}\right)\right]
\end{array}
\right.
\hbox{ }\hbox{and}\hbox{ }
\left\{
\begin{array}{l@{\quad}l}
\mathcal{A}_{0}^{3}=\frac{\eta_{v}}{\mu_{0}}\left[12\frac{\chi_{0}^{3}}{r^2}\right]\\
\mathcal{B}_{0}^{3}=\frac{\eta_{h}}{\mu_{0}}\left[\frac{1}{r}\partial_{r}\chi_{0}^{3}\right]\\
\mathcal{C}_{0}^{3}=\frac{\eta_{h}}{\mu_{0}}\left[-\Delta_{3}\left(\frac{\xi_{0}^{3}}{r}\right)\right]\\
\mathcal{D}_{0}^{3}=\left[12\frac{\chi_{0}^{3}}{r^2}\right]\\
\mathcal{E}_{0}^{3}=\left[\frac{1}{r}\partial_{r}\chi_{0}^{3}\right]\\
\mathcal{F}_{0}^{3}=\left[-\Delta_{3}\left(\frac{\xi_{0}^{3}}{r}\right)\right].
\end{array}
\right.
\label{OhmCoeff}
\end{equation}

\subsubsection{Baroclinic relation}
We apply (\ref{baro1}) to $l=2$:
\begin{equation}
\varphi\Lambda_{2}-\delta\Psi_{2}=\frac{r}{\overline{g}}\left[\mathcal{D}_{2}+\frac{\mathcal{X}_{\vec{\mathcal{F}}\!\!_{\mathcal{L}};2}}{r\overline{\rho}}+\frac{1}{r}\frac{{\rm d}}{{\rm d}r}\left(r\frac{\mathcal{Y}_{\vec{\mathcal{F}}\!\!_{\mathcal{L}};2}}{\overline{\rho}}\right)\right]
\end{equation}
where $\mathcal{D}_{2}=\frac{1}{3}\left[r\partial_{r}\overline{\Omega}^{2}\right]+\frac{8}{35}\left[r\partial_{r}\left(\overline{\Omega}\Omega_{2}\right)\right]+\frac{8}{7}\overline{\Omega}\Omega_{2}$ and where $\mathcal{X}_{\vec{\mathcal{F}}\!\!_{\mathcal{L}};2}$ and $\mathcal{Y}_{\vec{\mathcal{F}}\!\!_{\mathcal{L}};2}$ have been given in (\ref{XFL2}-\ref{YFL2}).

\subsubsection{Horizontal fluctuation of the molecular weight}
The evolution equation for the mean molecular weight  (Paper I, eq. 39) is unchanged by the introduction of the magnetic field:
\begin{equation}
\frac{{\rm d}\Lambda_{2}}{{\rm d}t}-\frac{{\rm d}\ln \overline{\mu}}{{\rm d}t}\Lambda_{2}=\frac{U_2}{H_p}\nabla_{\mu}-\frac{6}{r^2}D_{h}\Lambda_{2}.
\end{equation}

\subsection{System for $l=4$}
\subsubsection{Horizontal shear}
We have:
\begin{equation}
\rho{{\rm d} \over {\rm d}t} \left(r^2 \Omega_{2}\right)-2\rho\overline{\Omega}r\left[\frac{1}{3\rho r}\partial_{r}\left(\rho r^2 U_{2} \right)-\alpha U_{2}\right]=-10\rho\nu_{h}\Omega_{2}+\Gamma_{2} 
\end{equation}
where the expression for $\Gamma_{2}$ has been given in (\ref{TorqueMagn}).

\subsubsection{Meridional circulation}
In the same way as for $l=2$, we  recast (\ref{heat1}-\ref{tcal-final}) in two first order equations as
\begin{equation}
U_{4}=\frac{L}{M\overline{g}}\left(\frac{P}{\overline{\rho}C_{p}\overline{T}}\right)\frac{1}{\nabla_{\hbox{ad}}-\nabla}\mathcal{B}_{4}
\end{equation}
where
\begin{eqnarray}
\mathcal{T}_{4}&=&2\left[1-\frac{\overline{f}_{\mathcal{C}}+\overline{f}_{\mathcal{L}}}{4\pi G\overline{\rho}}-\frac{\left(\overline{\epsilon}+\overline{\epsilon}_{grav}\right)}{\epsilon_{m}}\right]\frac{\widetilde{g}_{4}}{\overline{g}}+\frac{\widetilde{f}_{\mathcal{C},4}+\widetilde{f}_{\mathcal{L},4}}{4\pi G\overline{\rho}}-\frac{\overline{f}_{\mathcal{C}}+\overline{f}_{\mathcal{L}}}{4\pi G\overline{\rho}}\left(-\delta\Psi_{4}+\varphi\Lambda_{4}\right)+\frac{\rho_{m}}{\overline{\rho}}\left[ \frac{r}{3}\partial_{r}\mathcal{A}_{4}-\frac{20 H_{T}}{3r}\left(1 + \frac{D_{h}}{K}\right){\Psi_4}\right]\nonumber\\
&+&\frac{\left(\overline{\epsilon}+\overline{\epsilon}_{grav}\right)}{\epsilon_{m}}\left\{ \mathcal{A}_{4}+(f_{\epsilon}\epsilon_{T} - f_{\epsilon}\delta + \delta)\Psi_{4}+(f_{\epsilon}\epsilon_{\mu}+f_{\epsilon}\varphi - \varphi)\Lambda_{4}\right\}+\frac{M}{L}\left[\frac{\mathcal{J}_{4}}{\overline{\rho}}-C_{p}\overline{T}\left(\frac{{\rm d}\Psi_{4}}{{\rm d}t}+\Phi\frac{{\rm d}\ln\overline{\mu}}{{\rm d}t}\Lambda_{4}\right)\right]\nonumber   
\end{eqnarray}
and
\begin{equation}
\mathcal{A}_{4}=H_{T}\partial_{r}\Psi_{4}-(1-\delta+\chi_{T})\Psi_{4}-(\varphi+\chi_{\mu})\Lambda_{4}.
\end{equation}
The coefficients related respectively to the centrifugal force and to the Lorentz force are given by:
\begin{equation}
\widetilde{f}_{\mathcal{C},4}=\frac{1}{r^2}\partial_{r}\left(r^2a_{4}\right)+20\frac{b_{4}}{r}\hbox{ }\hbox{and}\hbox{ }
\widetilde{f}_{\mathcal{L},4}=\frac{1}{r^2}\partial_{r}\left(r^2\frac{\mathcal{X}_{\vec{\mathcal F}\!\!_{\mathcal L};4}}{\rho_{0}}\right)+20\frac{\mathcal{Y}_{\vec{\mathcal F}_{\mathcal{L}};4}}{r\rho_{0}}
\end{equation}
where:
\begin{equation}
\left\{
\begin{array}{l@{\quad}l}
a_{4}=-\frac{24}{35}r\overline{\Omega}\Omega_{2}\\
b_{4}=\frac{6}{35}r\overline{\Omega}\Omega_{2}
\end{array}
\right.
\end{equation}
and
\begin{equation}
\left\{
\begin{array}{l@{\quad}l}
\mathcal{X}_{\vec{\mathcal{F}}\!\!_{\mathcal{L}};4}=\frac{2}{3}\frac{1}{5\pi}\left\{\frac{9}{2}\sqrt{\frac{3}{7}}\left[\left(\mathcal{C}_{0}^{1}\mathcal{E}_{0}^{3}-\mathcal{B}_{0}^{1}\mathcal{F}_{0}^{3}\right)-\left(\mathcal{C}_{0}^{3}\mathcal{E}_{0}^{1}-\mathcal{B}_{0}^{3}\mathcal{F}_{0}^{1}\right)\right]+\frac{27}{7}\left(\mathcal{C}_{0}^{2}\mathcal{E}_{0}^{2}-\mathcal{B}_{0}^{2}\mathcal{F}_{0}^{2}\right)+\frac{14449}{154}\left(\mathcal{C}_{0}^{3}\mathcal{E}_{0}^{3}-\mathcal{B}_{0}^{3}\mathcal{F}_{0}^{3}\right)\right\}\\
\mathcal{Y}_{\vec{\mathcal{F}}\!\!_{\mathcal{L}};4}=\frac{1}{4\pi}\left\{\sqrt{\frac{3}{7}}\left(3\mathcal{C}_{0}^{3}\mathcal{D}_{0}^{1}-\mathcal{A}_{0}^{3}\mathcal{F}_{0}^{1}+\mathcal{C}_{0}^{1}\mathcal{D}_{0}^{3}-3\mathcal{A}_{0}^{1}\mathcal{F}_{0}^{3}\right)+\frac{9}{7}\left(\mathcal{C}_{0}^{2}\mathcal{D}_{0}^{2}-\mathcal{A}_{0}^{2}\mathcal{F}_{0}^{2}\right)+\frac{9}{11}\left(\mathcal{C}_{0}^{3}\mathcal{D}_{0}^{3}-\mathcal{A}_{0}^{3}\mathcal{F}_{0}^{3}\right)\right\}.
\end{array}
\right.
\label{XYFL4}
\end{equation}
We get the relative fluctuation of the effective gravity from (\ref{geff}):
\begin{equation}
\frac{\widetilde{g}_{4}}{\overline{g}}=-\left[\frac{{\rm d} g_{0}}{{\rm d} r}\frac{1}{{g_{0}^{2}}}r\left(b_{4}+\frac{\mathcal{Y}_{\vec{\mathcal F}\!\!_{\mathcal L};4}}{\rho_{0}}\right) + \frac{1}{g_{0}}\left(a_{4}+\frac{\mathcal{X}_{\vec{\mathcal F}\!\!_{\mathcal L};4}}{\rho_{0}}\right)\right] + \frac{{\rm d}}{{\rm d}r}\left(\frac{\widehat{\phi}_{4}}{g_{0}}\right) 
\end{equation}
where $\widehat{\phi}_{4}$ obeys the Poisson equation:
\begin{equation}
\frac{1}{r}\frac{{\rm d}^2}{{\rm d}r^2}\left(r\widehat{\phi}_{4}\right)-\frac{20}{r^2}\widehat{\phi}_{4}-\frac{4\pi G}{g_{0}}\frac{{\rm d}\rho_{0}}{{\rm d}r}\widehat{\phi}_{4}=\frac{4\pi G}{g_{0}}\left[\rho_{0}a_{4}+\frac{{\rm d}}{{\rm d}r}\left(r\rho_{0}b_{4}\right)+\mathcal{X}_{\vec{\mathcal{F}}\!\!_{\mathcal{L}};4}+\frac{{\rm d}}{{\rm d}r}\left(r\mathcal{Y}_{\vec{\mathcal{F}}\!\!_{\mathcal{L}};4}\right)\right]\hbox{ }\hbox{ }\hbox{ }\hbox{(cf. eq. \ref{poisson-pert})}.
\end{equation}
We proceed as for $l=2$ to calculate the Ohmic heating:
\begin{eqnarray}
\mathcal{J}_{4}&=&\frac{1}{4\pi}\left\{4\sqrt{\frac{3}{7}}\left[\mathcal{A}_{0}^{3}\mathcal{D}_{0}^{1}+\mathcal{A}_{0}^{1}\mathcal{D}_{0}^{3}-\frac{3}{5}\left(\mathcal{B}_{0}^{3}\mathcal{E}_{0}^{1}+\mathcal{C}_{0}^{3}\mathcal{F}_{0}^{1}+\mathcal{B}_{0}^{1}\mathcal{E}_{0}^{3}+\mathcal{C}_{0}^{1}\mathcal{F}_{0}^{3}\right)\right]+\frac{18}{7}\left[\mathcal{A}_{0}^{2}\mathcal{D}_{0}^{2}-\frac{4}{5}\left(\mathcal{B}_{0}^{2}\mathcal{E}_{0}^{2}+\mathcal{C}_{0}^{2}\mathcal{F}_{0}^{2}\right)\right]\right.\nonumber\\
&+&{\left. \frac{2}{11}\left[9\mathcal{A}_{0}^{3}\mathcal{D}_{0}^{3}+\frac{28898}{105}\left(\mathcal{B}_{0}^{3}\mathcal{E}_{0}^{3}+\mathcal{C}_{0}^{3}\mathcal{F}_{0}^{3}\right)\right]\right\}}
\end{eqnarray}
where the $\mathcal{A}_{0}^{l}$, $\mathcal{B}_{0}^{l}$, $\mathcal{C}_{0}^{l}$, $\mathcal{D}_{0}^{l}$, $\mathcal{E}_{0}^{l}$ and $\mathcal{F}_{0}^{l}$ with $l=\left\{1,2,3\right\}$ have been given in the previous section in (\ref{OhmCoeff}).

\subsubsection{Baroclinic relation}
We apply (\ref{baro1}) to $l=4$. We get:
\begin{equation}
\varphi\Lambda_{4}-\delta\Psi_{4}=\frac{r}{\overline{g}}\left[\mathcal{D}_{4}+\frac{\mathcal{X}_{\vec{\mathcal{F}}\!\!_{\mathcal{L}};4}}{r\overline{\rho}}+\frac{1}{r}\frac{{\rm d}}{{\rm d}r}\left(r\frac{\mathcal{Y}_{\vec{\mathcal{F}}\!\!_{\mathcal{L}};4}}{\overline{\rho}}\right)\right]
\end{equation}
where $\mathcal{D}_{4}=\frac{6}{35}\left[r\partial_{r}\left(\overline{\Omega}\Omega_{2}\right)-2\overline{\Omega}\Omega_{2}\right]$ and where $\mathcal{X}_{\vec{\mathcal{F}}\!\!_{\mathcal{L}};4}$ and $\mathcal{Y}_{\vec{\mathcal{F}}\!\!_{\mathcal{L}};4}$ have been given in (\ref{XYFL4}).

\subsubsection{Horizontal fluctuation of the molecular weight}
This equation is not affected by the introduction of the magnetic field:
\begin{equation}
\frac{{\rm d}\Lambda_{4}}{{\rm d}t}-\frac{{\rm d}\ln \overline{\mu}}{{\rm d}t}\Lambda_{4}=\frac{U_4}{H_p}\nabla_{\mu}-\frac{20}{r^2}D_{h}\Lambda_{4}.
\end{equation}
These equations are ready to be implemented in a stellar evolution code, together with the boundary conditions discussed in \S7.

\end{document}